\documentclass[useAMS,usenatbib]{mn2e}
\usepackage{epsfig,lscape,amsmath,amssymb}
\usepackage{graphicx,longtable,times,threeparttable,multirow}
\usepackage{rotating,color}
\usepackage[T1]{fontenc}
\usepackage{aecompl}
\usepackage{bm}

\DeclareGraphicsExtensions{.ps, .eps,.eps.gz,.epsi}

\begin{document}

\title[]{LAMOST Spectroscopic Survey of the Galactic Anticentre (LSS-GAC): the second release of value-added catalogues}
\author[Xiang et al.]{M.-S. Xiang$^{1}$\thanks{LAMOST Fellow}, 
        X.-W. Liu$^{2, 3}$\thanks{E-mail: x.liu@pku.edu.cn}, H.-B. Yuan$^{4}$,
        Z.-Y. Huo$^{1}$, Y. Huang$^{2}$\footnotemark[1], C. Wang$^{2}$, 
        \newauthor B.-Q. Chen$^{2}$\footnotemark[1], J.-J. Ren$^{2}$\footnotemark[1], 
        H.-W. Zhang$^{2}$, Z.-J. Tian$^{2}$\footnotemark[1], Y. Yang$^{2, 4}$, 
        J.-R. Shi$^{1}$, \newauthor J.-K. Zhao$^{1}$, J. Li$^{5}$, Y.-H. Zhao$^{1}$, X.-Q. Cui$^{6}$, G.-P. Li$^{6}$, 
         Y.-H. Hou$^{6}$, Y. Zhang$^{6}$, \newauthor W. Zhang$^{1}$, J.-L. Wang$^{1}$, Y.-Z. Wu$^{1}$, 
         Z.-H. Cao$^{1}$, H.-L. Yan$^{1}$, T.-S. Yan$^{1}$, 
         A.-L. Luo$^{1}$, \newauthor H.-T. Zhang$^{1}$, Z.-R. Bai$^{1}$, 
         H.-L. Yuan$^{1}$, Y.-Q. Dong$^{1}$, Y.-J. Lei$^{1}$,  G.-W. Li$^{1}$ 
\\ \\ $1$ Key Laboratory of Optical Astronomy, National Astronomical Observatories, 
    Chinese Academy of Sciences, Beijing 100012, P. R. China \\
$2$ Department of Astronomy, Peking University, Beijing 100871, P. R. China \\
$3$ Kavli Institute for Astronomy and Astrophysics, Peking University, Beijing 100871, P. R. China \\ 
$4$ Department of Astronomy, Beijing Normal University, Beijing 100875, P. R. China \\
$5$ Department of Space Science and Astronomy, Hebei Normal University, Shijiazhuang 050024,China \\
$6$ Nanjing Institute of Astronomical Optics \& Technology, National Astronomical Observatories, 
    Chinese Academy of Sciences, Nanjing 210042, P. R. China \\}

\date{Received:}

\pagerange{\pageref{firstpage}--\pageref{lastpage}} \pubyear{2015}

\maketitle

\label{firstpage}

\begin{abstract}{
We present the second release of value-added catalogues of the LAMOST Spectroscopic
Survey of the Galactic Anticentre (LSS-GAC DR2). The catalogues present values of radial velocity $V_{\rm r}$,
atmospheric parameters --- effective temperature $T_{\rm eff}$, surface gravity log\,$g$,
metallicity [Fe/H], $\alpha$-element to iron (metal) abundance ratio [$\alpha$/Fe] ([$\alpha$/M]),
elemental abundances [C/H] and [N/H], and absolute magnitudes ${\rm M}_V$ and ${\rm M}_{K_{\rm s}}$
deduced from 1.8 million spectra of 1.4 million unique stars targeted by the LSS-GAC since
September 2011 until June 2014. The catalogues also give values of interstellar reddening, distance
and orbital parameters determined with a variety of techniques, as well as                
proper motions and multi-band photometry from the far-UV to the mid-IR collected from the
literature and various surveys. Accuracies of radial velocities reach 5\,km\,s$^{-1}$
for late-type stars, and those of distance estimates range between 10 -- 30 per cent, depending on
the spectral signal-to-noise ratios. Precisions of [Fe/H], [C/H] and [N/H] estimates   
reach 0.1\,dex, and those of [$\alpha$/Fe] and [$\alpha$/M] reach 0.05\,dex.  
The large number of stars, the contiguous sky coverage, the simple yet non-trivial target 
selection function and the robust estimates of stellar radial velocities and
atmospheric parameters, distances and elemental abundances, make the catalogues a valuable
data set to study the structure and evolution of the Galaxy, especially the solar-neighbourhood
and the outer disk.}
\end{abstract}

\section{Introduction}
\label{sect:intro}
Better understanding the structure, stellar populations, and the chemical and dynamic evolution of the Milky Way 
is both a challenge and an opportunity of modern galactic astronomy. The Milky Way is the only galaxy 
whose distribution of stellar populations can be mapped out in full dimensionality --- three-dimensional position 
and velocity, age, as well as photospheric elemental abundances. However, owing to our location inside the Milky Way disk, 
the hundreds of billions of Galactic stars are distributed over the whole sky, and our views in the Galactic disk 
are seriously limited by the interstellar dust extinction. Obtaining the full dimensional distribution 
of a complete stellar sample is thus a great challenge. It is only recently that comprehensive 
surveys of Galactic stars become feasible, thanks to the implementation of a number of large-scale 
photometric and spectroscopic surveys, such as the Sloan Digital Sky Survey \citep[SDSS;][]{York+2000}, 
the Two Micron All Sky Survey \citep[2MASS;][]{Skrutskie+2006}, the Apache Point Observatory Galactic Evolution
Experiment \citep[APOGEE;][]{Majewski+2010}, the LAMOST Experiment for Galactic Understanding and 
Exploration \citep[LEGUE;][]{Deng+2012, Zhao+2012} and the $Gaia$ astrometric survey \citep{Perryman+2001}.

As a major component of the LEGUE project, the LAMOST Spectroscopic Survey of the Galactic Anticentre
\citep[LSS-GAC;][]{Liu+2014, Yuan+2015a} is being carried out with the aim to obtain a statistically complete stellar spectroscopic 
sample in a contiguous sky area around the Galactic anticentre, taking full advantage of the 
large number of fibers (4,000) and field of view (20\,sq.deg.) offered by LAMOST \citep{Cui+2012}. 
The survey will allow us to acquire a deeper and more comprehensive understanding of the structure, 
origin and evolution of the Galactic disk and halo, 
as well as the transition region between them. The main scientific goals of LSS-GAC include:  
(a) to characterize the stellar populations, chemical composition, kinematics and 
structure of the thin and thick disks and their interface with the halo; 
(b) to understand the temporal and secular evolution of the disk(s); 
(c) to probe the gravitational potential and dark matter distribution;
(d) to identify star aggregates and substructures in the multi-dimensional phase space;
(e) to map the interstellar extinction as a function of distance;
(f) to search for and study rare objects (e.g. stars of peculiar chemical composition
or hypervelocities);
(g) to study variable stars and binaries with multi-epoch spectroscopy.

LSS-GAC plans to collect low-resolution ($R\sim1800$) optical spectra ($\lambda\lambda$3700 -- 9000\AA) 
of more than 3 million stars down to a limiting magnitude of $r\sim$17.8\,mag (to 18.5\,mag for selected fields) 
in a contiguous sky area of over 3400\,sq.deg. 
centred on the Galactic anticentre ($|b| \leq 30^\circ$, $150 \leq l \leq 210^\circ$), 
and deliver spectral classifications, stellar parameters (radial velocity $V_{\rm r}$, effective temperature $T_{\rm eff}$, 
surface gravity log\,$g$, metallicity [Fe/H], $\alpha$-element to iron abundance ratio [$\alpha$/Fe], and 
individual elemental abundances), as well as values of interstellar extinction and distance of the surveyed stars, 
so as to build-up an unprecedented, statistically representative multi-dimensional database for the Galactic (disk) studies. 
The targets of LSS-GAC are selected uniformly in the planes of ($g-r$, $r$) and ($r-i$, $r$) Hess diagrams 
and in the (RA, Dec) space with a Monte Carlo method, based on the Xuyi Schmidt Telescope Photometric Survey of 
the Galactic Anticentre \cite[XSTPS-GAC;][]{Zhang+2013, Zhang+2014, Liu+2014, Yuan+2015a},  
a CCD imaging photometric survey of $\sim 7000$\,sq.deg. with the Xuyi 1.04/1.20m Schmidt Telescope. 
Stars of all colours are sampled by LSS-GAC. The sampling rates are higher for stars of rare colours, 
without losing the representation of bulk stars, given the high sampling density ($\gtrsim1000$ stars per sq.deg.). 
This simple yet non-trivial target selection strategy allows for a statistically 
meaningful study of the underlying stellar populations for a wide range of object class, from  
white dwarfs \citep[e.g.][]{Rebassa-Mansergas+2015}, main sequence turn-off stars \citep[e.g.][]{Xiang+2015c} 
to red clump giants \citep[e.g.][]{Huang+2015a}, 
after the selection function has been properly taken into account.    

As an extension, LSS-GAC also surveys objects in a contiguous area of a few hundred 
sq.deg. around M31 and M33. The targets include background quasars, planetary nebulae (PNe), H\,{\sc ii} 
regions, globular clusters, supergiant stars, as well as foreground Galactic stars. 
In addition, to make full use of all available observing time, LSS-GAC 
targets very bright (VB) stars of $r<14$\,mag in sky areas accessible to LAMOST 
$(-10^\circ \leq {\rm Dec}  \leq 60^\circ)$ in poor observing conditions (bright/grey lunar nights, 
or nights of poor transparency). Those very bright stars comprise an excellent sample supplementary 
to the main one. Given their relatively low surface densities, at least for the areas outside the disk, 
LSS-GAC has achieved a very high sampling completeness (50 per cent) for those very bright stars, 
making the sample a golden mine to study the solar-neighbourhood. 
   
 Radial velocities and atmospheric parameters (effective temperature $T_{\rm eff}$, 
 surface gravity log\,$g$, metallicity [Fe/H]) have been deduced from the LSS-GAC spectra for A/F/G/K-type 
 stars using both the official LAMOST Stellar parameter Pipeline \citep[LASP;][]{Wu+2011, Wu+2014} and the 
 LAMOST Stellar Parameter Pipeline at Peking University \citep[LSP3;][]{Xiang+2015a}. 
 Typical precisions of the results, depending on the spectral signal-to-noise ratio (SNR) 
 and the spectral type, are a few (5 -- 10) km~s$^{-1}$ for radial velocity $V_{\rm r}$, 100 -- 200\,K for 
 $T_{\rm eff}$, 0.1 -- 0.3\,dex for log\,$g$, 0.1 -- 0.2\,dex for [Fe/H] \citep{Xiang+2015a, Luo+2015, 
 Gao+2015, Ren+2016, Wang+2016}. Values of $\alpha$-element to iron abundance ratio [$\alpha$/Fe], as well as 
 abundances of individual elements (e.g. [C/H] and [N/H]) have also been derived with 
 LSP3 \citep{Liji+2016, Xiang+2016}, with precisions 
 similar to those achieved by the APOGEE survey for giant stars \citep{Xiang+2016}.       
 Efforts have also been made to derive stellar parameters from LAMOST spectra with  
 other pipelines, such as the SSPP \citep{Lee+2015} and the $Cannon$ \citep{Ho+2016}. 
 \citet{Liuchao+2015} refine the LASP estimates of $\log\,g$ using a support vector regression 
 (SVR) model based on {\em Kepler}\, asteroseismic measurements of giant stars. 
 Stellar extinction and distances have been deduced for LSS-GAC sample stars using a variety of methods 
 \citep{Chen+2014, Yuan+2015a, Carlin+2015, WangJL+2016}, with typical uncertainties of $E_{B-V}$ 
 of about 0.04\,mag, and of distance between 10 -- 30 per cent, 
 depending on the stellar spectral type \citep{Yuan+2015a}. 
 
Following a year-long Pilot Survey, LSS-GAC was initiated in October, 2012, and is expected to  
complete in the summer of 2017. The LSS-GAC data collected up to the end of the first year 
of the Regular Survey are public available from 
two formal official data releases, namely the early \citep[LAMOST EDR;][]{Luo+2012} 
and first \citep[LAMOST DR1;][]{Luo+2015} data release\footnote{http://dr1.lamost.org}. 
The LAMOST EDR includes spectra and stellar parameters derived with the LASP 
for stars observed during the Pilot Survey, while the LAMOST DR1 includes stars observed 
by June, 2013. In addition, there is a public release of LSS-GAC value-added catalogues  
for stars observed by June, 2013, the LSS-GAC DR1\footnote{http://lamost973.pku.edu.cn/site/data} 
\citep{Yuan+2015a}. LSS-GAC DR1 presents stellar parameters derived with LSP3, 
values of interstellar extinction and stellar distance deduced with a variety of methods, 
as well as magnitudes of broadband photometry compiled from various  
photometric catalogues (e.g. GALEX, SDSS, XSTPS-GAC, UCAC4, 2MASS and WISE),  
and values of proper motions from the UCAC4 and PPMXL catalogues and those derived by 
combing the XSTPS-GAC and 2MASS astrometric measurements, and, finally, 
stellar orbital parameters (e.g. eccentricity) computed assuming specific Galactic potentials.    
 
This paper presents the second release of value-added catalogues of LSS-GAC (LSS-GAC DR2). 
LSS-GAC DR2 presents the aforementioned multi-dimensional parameters deduced from 
1,796,819 spectra of 1,408,737 unique stars observed by June, 2014. 
Compared to LSS-GAC DR1, in addition to a significant  
increase in stellar number, several improvements to the data reduction and 
stellar parameter determinations have been implemented, including: (1) An upgraded LAMOST 2D pipeline 
has been used to process the spectra; (2) The spectral template library used by LSP3 has been updated,  
adding more than 200 new templates. The atmospheric parameters for all template stars 
have also been re-determined/calibrated; (3) Values of $\alpha$-element to iron abundance 
ratio [$\alpha$/Fe] have been estimated with LSP3; (4) Accurate values of stellar atmospheric 
parameters ($T_{\rm eff}$, log\,$g$, [Fe/H], [$\alpha$/Fe]), absolute magnitudes ${\rm M}_V$ 
and ${\rm M}_{K_{\rm s}}$, as well as elemental abundances [C/H] and [N/H],  
have also been estimated from the spectra using a multivariate regression method based on 
kernel-based principal component analysis (KPCA).  

The paper is organized as follows. Section\,2 describes the observations included in the LSS-GAC DR2, 
including a brief review of the target selection algorithm and the observational footprint. 
Section\,3 introduces the data reduction briefly. Section\,4 presents a detailed description 
of the improvements in stellar parameter determinations incorporated in LSS-GAC DR2. 
Section\,5 briefly discusses the duplicate observations, which accounts for nearly 30 per cent of 
all observations. Section\,6 introduces the determinations of extinction and distance. 
Section\,7 presents proper motions and derivation of stellar orbital parameters. 
The format of value-added catalogues is described in Section\,8,  
followed by a summary in Section\,9.
 
\section{Observations}
To make good use of observing time of different qualities as well as to avoid fibre cross-talking, 
LSS-GAC stars are targeted by 4 types of survey plates defined in $r$-band magnitude \citep{Liu+2014, Yuan+2015a}. 
Stars of $r<14.0$\,mag are targeted 
by very bright (VB) plates, and observed in grey/lunar nights, with typical exposure time of (2 -- 3) $\times$ 600\,s. 
Stars of $14.0<r\lesssim16.3$\,mag are targeted by bright (B) plates, and observed in grey/dark 
nights, with typical exposure time of 2 $\times$ 1200\,s, whereas stars of $16.3\lesssim r \lesssim17.8$\,mag, 
and of $17.8\lesssim r<18.5$\,mag are targeted respectively by medium-bright (M) and faint (F) plates, 
and observed in dark nights of excellent observing conditions (in term of seeing and transparency), 
with typical exposure time of 3$\times$1800\,s. 

By June, 2014, 314 plates (194 B + 103 M + 17 F) for the LSS-GAC 
main survey,  59 plates (38 B + 17 M + 4 F) for the M31/M33 survey and 682 plates for the VB survey 
have been observed. Note that spectra of a few plates could not be successfully processed with the 
2D pipeline and are therefore not included in the above statistics. During the survey, some observing 
time, on the level of one or two grey nights per month has been reserved for monitoring 
the instrument performance \citep[e.g. throughput and accuracy of fibre positioning;][]{Liu+2014, Yuan+2015a}. 
Some of those reserved nights have been used to target some of the LSS-GAC plates, yielding 
another 94 observed plates (79 B/M/F + 15 V). Finally, 43 B or M or F 
LSS-GAC plates were observed from September to October, 2011, when the LAMOST 
was being commissioned before the start of Pilot Survey. For those 43 plates, 
the magnitude limits of assigning stars in B/M/F plates are not exactly the same as those 
adopted during the Pilot and Formal surveys. For convenience, all plates observed 
using reserved time as well as those collected during the commissioning phase have been grouped into, 
as appropriately, the LSS-GAC main, M31/M33 and VB survey, respectively, leading to a total of 395 
plates for the LSS-GAC main, 100 plates for the M31/M33 and 697 plates of the VB survey, respectively. 

The total numbers of spectra collected and unique stars targeted by those plates are listed in Table\,1. 
Table\,1 also lists the numbers of spectra and unique stars that are successfully observed, 
defined by a spectral SNR of higher than 10 either in the blue ($\sim$4650\,{\AA}) or the ($\sim$7450\,{\AA}) 
part of the spectrum \citep{Liu+2014}. The numbers of spectra and stars have increased 
significantly compared to those of LSS-GAC DR1, which contains, for example, 225,522 spectra of 189,042  
unique stars of SNR(4650{\AA}) $>10$ for the main survey, and 457,906 spectra 
of 385,672 unique stars of SNR(4650{\AA}) $>10$ for the VB survey \citep[cf.][]{Yuan+2015a}. 
Due to the overlapping of LAMOST fields of view of adjacent plates and the repeating of 
observations failed to meet the quality control, there is a considerable fraction of stars 
that have been repeatedly targeted several times. For LSS-GAC DR2, amongst the 948,361 
unique stars targeted by the main survey, 71.0, 20.6, 6.0, 1.7, 0.5, 0.1 per cent of the stars 
are observed by one to six times, respectively.
The corresponding fractions for the M31/M33 survey are 60.1, 20.4, 10.1, 4.4, 2.5, 1.3 per cent, 
and those for the VB survey are 73.5, 20.5, 3.7, 1.7, 0.3, 0.2 per cent. 
For those unique stars with SNR(4650{\AA}) $>10$, the corresponding fractions are 
83.6, 13.2, 2.4, 0.6, 0.1, 0.03 per cent  
for the LSS-GAC main survey, 72.6, 19.2, 5.6, 1.7, 0.7, 0.2 per cent for the M31/M33 survey, 
and 78.3, 17.3, 2.9, 1.2, 0.2, 0.1 per cent for the VB survey. 
Note that there are also a small fraction ($\sim0.8$ per cent) of stars that are targeted 
by both the LSS-GAC main (or M31/M33) survey and the VB survey. These duplicates were not considered 
when calculating the above percentage numbers.

Figs.\,1 and 2 plot the footprints of stars with either a blue or red spectral SNR higher than 10 
for LSS-GAC DR2. The footprints of stars targeted by VB, B and M plates are plotted in separate panels. 
For the main survey, the strategy is to extend the observations of a stripe of ${\rm Dec}\sim29^\circ$ 
to both higher and lower Declinations \citep{Yuan+2015a}. 
Compared to LSS-GAC DR1, LSS-GAC DR2 has completed two more stripes for B and M 
plates, namely those of ${\rm Dec}\sim34^\circ$ and $\sim24^\circ$, respectively. A few B and M plates 
of Galactic latitude $b > 35^\circ$ were observed using either the reserved time or during the commissioning phase.
For the M31/M33 survey, 7 B and 9 M plates were collected in the second year of the Regular  
Survey (September 2013 -- June 2014), leading to a much larger sky coverage compared to LSS-GAC DR1.
Significant progress in the observation of VB plates is seen in LSS-GAC DR2, in 
term of both the area and continuity of the sky coverage. 
 
\begin{table*}
\caption{Numbers of spectra and unique stars (in parentheses) observed by LSS-GAC by June, 2014.}
\label{}
\begin{tabular}{ccccc}
\hline
             &                       LSS-GAC Main Survey & M31/M33 Survey  & VB Survey \\
 All SNRs &  1,332,812 (948,361) &  305,226 (171,259) & 1,944,525 (1,431,219)  \\
 SNR (4650{\AA}) $> 10$     &  510,531 (423,503) & 91,921 (65,841) & 1,194,367 (922,935)  \\ 
 SNR (7450{\AA}) $> 10$    & 688,459 (572,438)  & 68,380 (58,197)  & 1,358,618 (1,050,143)  \\ 
SNR (4650{\AA}) $> 10$  or  SNR(7450) $> 10$  & 763,723 (618,924) &  113,597 (80,452) & 1,397,538 (1,075,677) \\
\hline
 \hline
\end{tabular}
\end{table*}

 \begin{figure*}
\centering
\includegraphics[width=180mm]{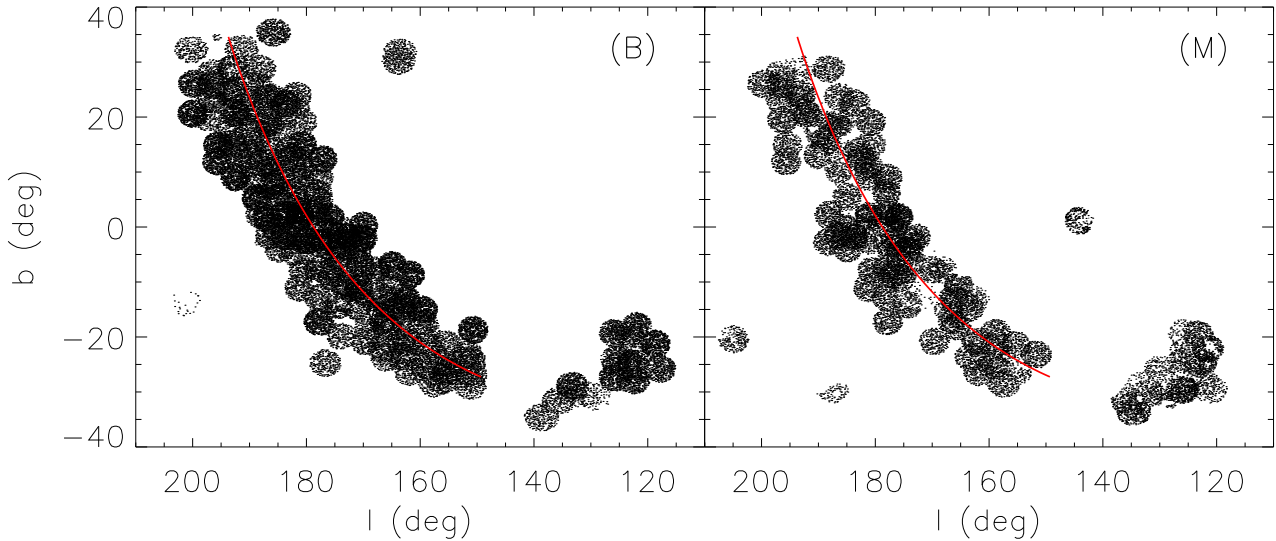}
\caption{LSS-GAC DR2 footprints in a Galactic coordinate system of stars observed respectively with 
bright (B; left panel) and medium (M; right panel) plates. The red line denotes a constant 
Declination of 30$^\circ$.
To reduce the figure file size, only 1 in 10 observed stars are plotted.}
\label{Fig1}
\end{figure*}

\begin{figure}
\centering
\includegraphics[width=85mm]{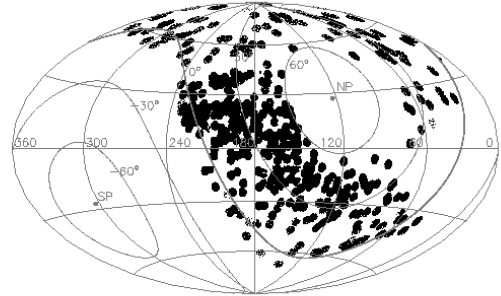}
\caption{LSS-GAC DR2 footprint for the very bright (VB) plates. 
To reduce the figure file size, only 1 in 10 observed stars are plotted.}
\label{Fig2}
\end{figure}

\begin{figure}
\centering
\includegraphics[width=85mm]{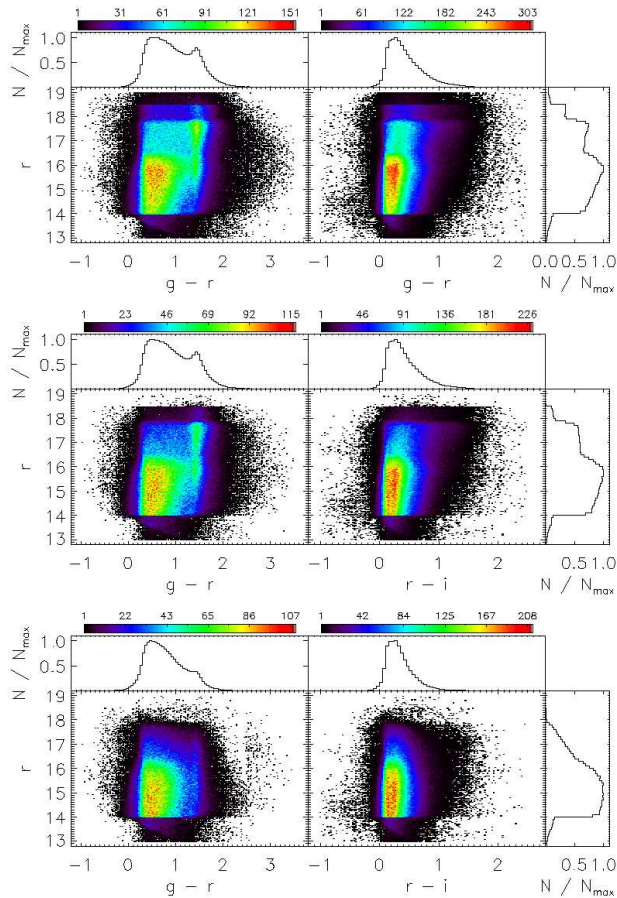}
\caption{Colour-coded stellar density distributions in the colour-magnitude ($g-r$, $r$)  and ($r-i$, $r$) 
diagrams for the LSS-GAC main survey. The upper panels show all observed stars, while the middle 
panels show those with either SNR(4650{\AA}) $>10$ or SNR(7450{\AA}) $>10$, and the lower panels 
show those with SNR(4650{\AA}) $>10$. The histograms show the one dimensional distributions of 
stars in colours $(g-r)$ and $(r-i)$ or in magnitude $r$, respectively, 
normalized to the maximum value.}
\label{Fig3}
\end{figure}

\begin{figure}
\centering
\includegraphics[width=85mm]{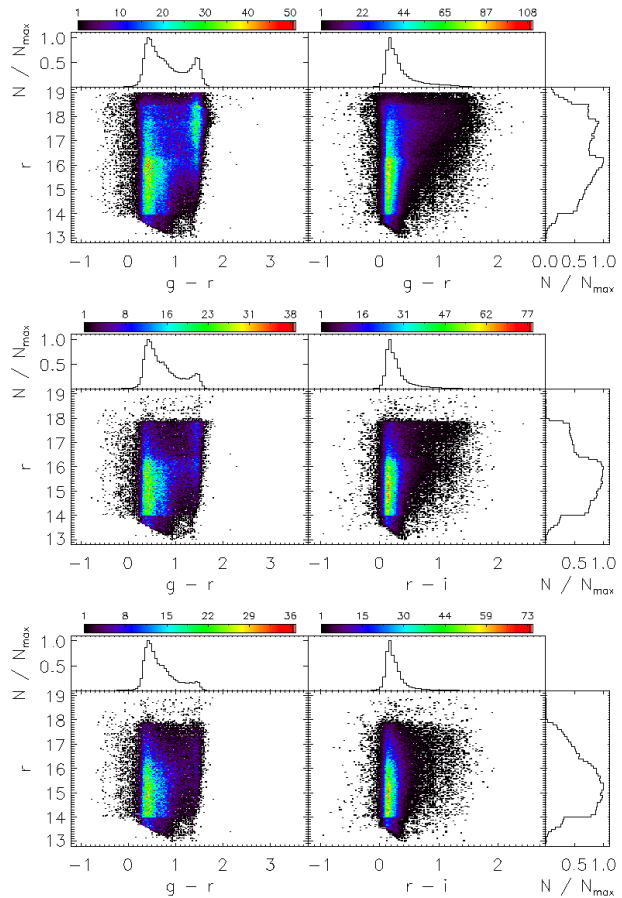}
\caption{Same as Fig.\,3 but for the M31/M33 survey.}
\label{Fig4}
\end{figure}

\begin{figure}
\centering
\includegraphics[width=85mm]{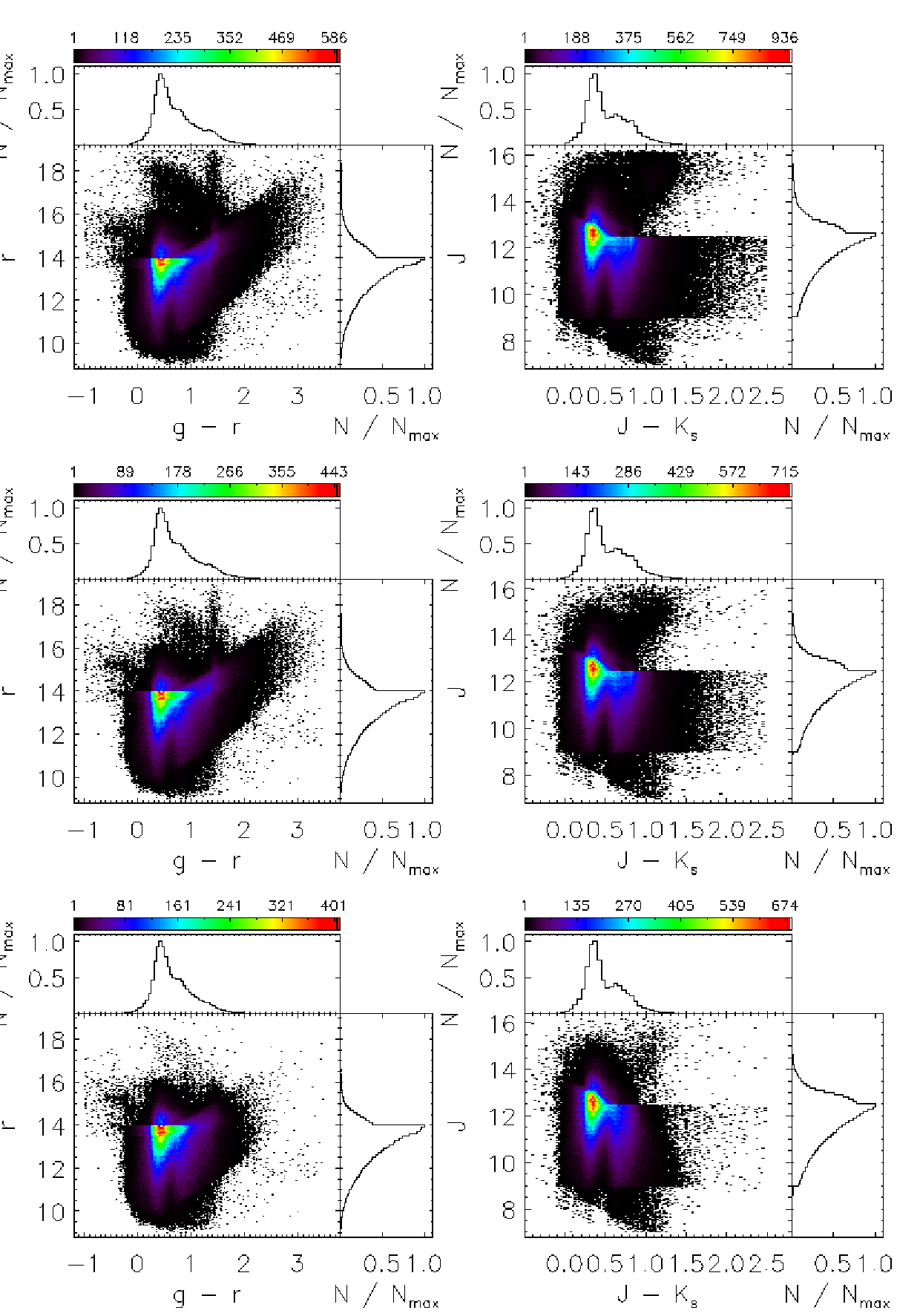}
\caption{Same as Fig\,3 but for the very bright survey. Colour-magnitude diagrams of ($g-r$, $r$) and ($J-K_{\rm s}$, $J$) are shown.}
\label{Fig5}
\end{figure}

Figs.\,3 -- 5 plot the density distributions of stars targeted, as well as those successfully observed 
(i.e. with a spectral SNR higher than 10, either in the blue or red), 
in the colour -- magnitude diagrams (CMDs) for the main, M31/M33 
and VB surveys, respectively. For the main and M31/M33 surveys, 
the diagrams are for colour -- magnitude combinations $(g-r, r)$ and $(r-i, r)$ 
used to select targets \citep{Yuan+2015a}. 
Magnitudes of $g,r,i$ bands are from the XSTPS-GAC  survey, except for a few 
plates of high Galactic latitudes, for which magnitudes from the SDSS photometric 
survey are used. The figures show that, as planned, stars of all colours have been observed. 
Figs.\,3 -- 5 also show that the distributions of stars that have either a blue or red spectral SNR 
higher than 10, as plotted in the middle panels of those three figures, are quite similar to 
those targeted, plotted in the top panels of the three figures, except for faint ones ($r>17.8$\,mag). 
Note that for the main and M31/M33 surveys, some stars of either $r > 18.5$ or $r<14.0$\,mag 
were observed during the commissioning phase. 
In contrast, the distributions of stars of ${\rm SNR (4650\AA)} > 10$ are quite 
different -- there are fewer faint stars of red colours. 
This is caused by a combination of the effects of low intrinsic fluxes in the blue of red stars 
and the lower throughputs of the spectrographs in the blue compared to those in the red \citep{Cui+2012}.   
For the VB survey, $(g-r, r)$ and $(J-K_{\rm s}, J)$ diagrams are plotted. 
The $g,r,i$ magnitudes are taken from the AAVSO Photometric All-Sky Survey \citep[APASS;][]{Munari+2014}, 
which have a bright limiting magnitude of about 10\,mag and a faint 
limiting magnitude (10$\sigma$) of about 16.5\,mag in $g,r,i$-bands. For stars of $r > 14.0$\,mag, 
magnitudes from the XSTPS-GAC or SDSS surveys are adopted if available. 
A comparison of stars common to XSTPS-GAC and APASS surveys shows 
good agreement in both magnitudes and colours for stars of $14.0<r<15.0$\,mag, 
with differences of just a few ($<5$) per cent.  
Due to the heterogeneous input catalogs and magnitude cut criteria used for the VB survey \citep{Yuan+2015a}, 
the morphologies of CMD distributions of VB plates are more complicated than 
those of the main and M31/M33 plates. Nevertheless, due to the high sampling rates 
of very bright stars \citep[e.g.][]{Xiang+2015c}, and the fact that stars of all colours have been 
observed without a strong colour bias, the selection function can still be well accounted for 
with some effort and care, if not straightforwardly.

\begin{figure}
\centering
\includegraphics[width=85mm]{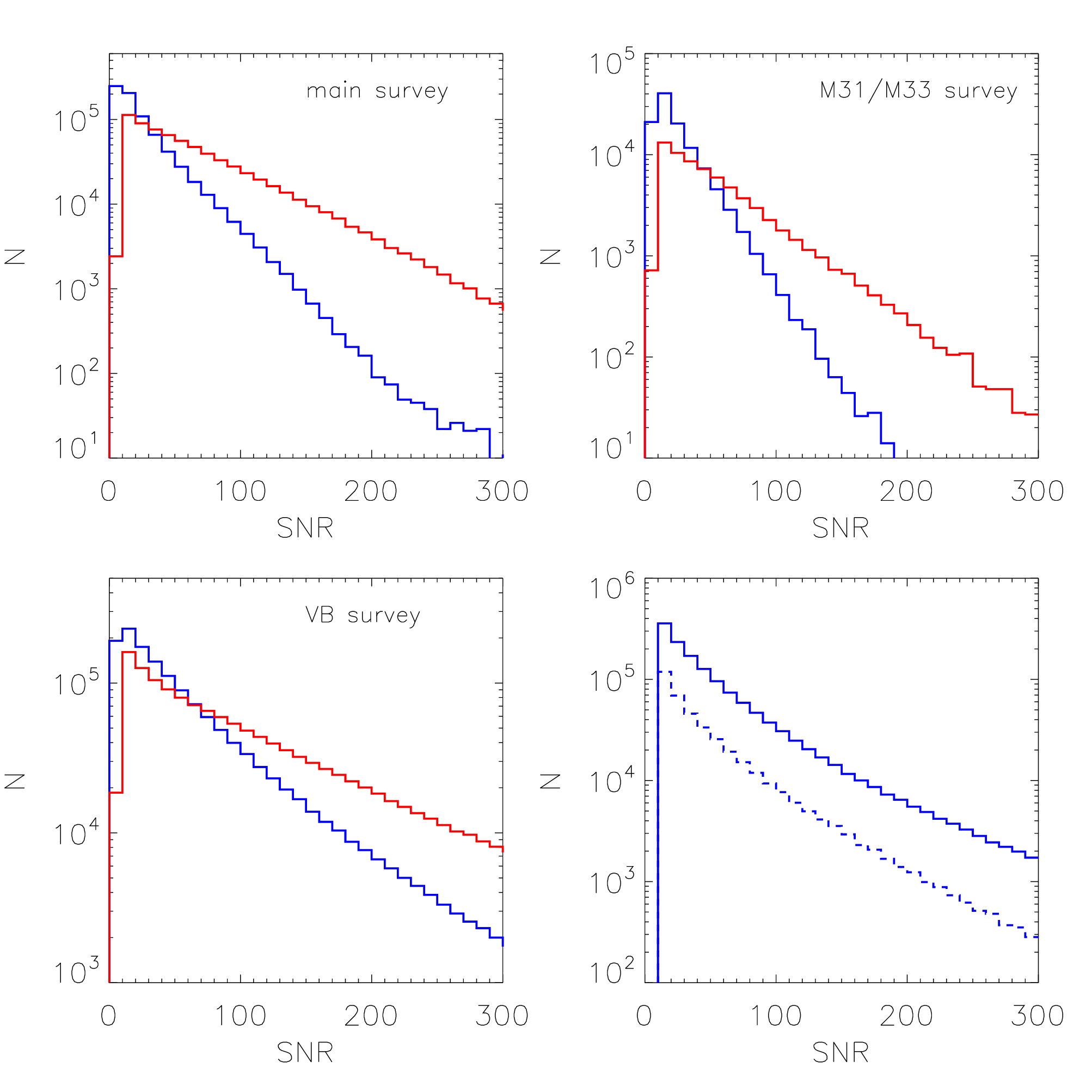}
\caption{Distribution of spectral SNRs for the main (top left), M31/M33 (top right) and VB (bottom left) surveys. 
Blue and red lines represent spectral SNR at 4650 and 7450\,{\AA}, respectively.
The bottom right panel plots SNR (4650{\AA}) for the whole sample that have a spectral SNR (4650{\AA}) 
higher than 10 for dwarfs (solid line) and giants (dashed line), respectively.}
\label{Fig6}
\end{figure}
Fig.\,6 plots the SNR distribution of spectra for the main, M31/M33 and VB surveys, 
as well as those of the whole sample for dwarf and giant stars. 
Only spectra with either SNR(4650{\AA}) or SNR(7450{\AA}) higher than 10 are plotted. 
The number of spectra in logarithmic scale decreases approximately linearly with increasing SNR. 
SNRs of the red part of the spectra are generally higher than those of the blue part.
Also, spectra of the VB survey have generally higher SNRs than those of the main and M31/M33 surveys. 
Distribution of SNRs for the giants are similar to those of the dwarfs.  
Here the classification of dwarfs and giants is based on the results of LSP3 (cf. Section 4). 
For the whole sample, about 36, 57 and 73 per cent of the spectra have a SNR higher than 50, 
30 and 20, respectively, in the blue part of the spectrum.

\section{data reduction}
The raw spectra of LSS-GAC used to generate the value-added catalogues were processed 
at Peking University with the LAMOST 2D reduction pipeline \citep{Luo+2012, Luo+2015} 
to extract the 1D spectra. This process includes several basic reduction steps, including bias subtraction, 
fibre tracing, fibre flat-fielding, wavelength calibration and sky subtraction. 
Both fibre tracing and flat-fielding were first carried out using twilight flat-fields. The results were  
further revised using sky emission lines when processing the object spectra to 
account for the potential shifts of fibre traces and the variations in fibre throughput.  
Typical precision of fibre flat-fielding, as indicated by the dispersions of flat-fields acquired 
in different days, is better than 1 per cent. Wavelength calibration was carried out using 
exposures of a Cd-Hg arc lamp for the blue-arm spectra and an Ar-Ne arc lamp for the red-arm spectra.   
Typically, the residuals of wavelength calibration for the individual arc lines have a mean value close to zero and 
a standard deviation of $\sim$\,0.02{\AA}, which corresponds to $\sim$1\,km\,s$^{-1}$ in velocity space. 
When processing the object spectra, sky emission lines are used to adjust the 
wavelengths to account for any residual systematic errors in the wavelength calibration and/or 
potential wavelength drifts between the arc-lamp and object spectra. 
A comparison of stellar radial velocities with the APOGEE measurements for LAMOST-APOGEE 
common stars yields an offset of $-3$ to $-4$\,km\,s$^{-1}$ \citep{Xiang+2015b, Luo+2015}, 
and the offset is found to be stable in the past few years, with typical night-to-night variations 
of about 2\,km\,s$^{-1}$. 
For each of the 16 spectrographs, about 20 fibres are typically assigned to target sky background 
for sky subtraction. The numbers of sky fibres allocated for sky measurement are higher  
for plates of low source surface density, e.g. VB plates of $|b|>10^\circ$. To subtract the sky background, 
the 2D pipeline creates a super-sky by B-spline fitting of fluxes measured by the individual sky fibres. 
The measured fluxes of sky emission lines in the object spectra are normalized to those of 
the super-sky spectrum to correct for potential differences of throughputs of fibres used to 
measure the sky and the objects. The correction assumes that the strengths of the sky 
emission lines are constant across the sky area covered by a given spectrograph, which is about one square degree, 
and any differences in sky emission line fluxes as measured by the sky and object fibres, 
are caused by errors in flat-fielding. The correction also assumes that 
for the sky background, the continuum scales with emission line flux.    

The resultant 1D spectra were then processed with the flux calibration pipeline developed specifically for 
LSS-GAC to deal with fields of low Galactic latitudes that may suffer from substantial interstellar extinction. 
The pipeline generates flux-calibrated spectra as well as co-adds the individual exposures of a given 
plate \citep{Xiang+2015a, Yuan+2015a}. 
To deal with the interstellar extinction, the pipeline calibrates the spectra in an iterative way, 
using F-type stars selected based on the stellar atmospheric parameters yielded by LSP3 
as flux standards. The theoretical synthetic spectra from \citet{Munari+2005} of 
the same atmospheric parameters are adopted as the intrinsic spectral energy distributions (SEDs) of the standards. 
Typical (relative) uncertainties of the spectral response curves (SRCs) thus derived are about 
10 per cent for the wavelength range 4000 -- 9000\,{\AA} \citep{Xiang+2015b}.    

Compared to the pipelines used to generate LSS-GAC DR1, a few improvements 
have been implemented: (1) Several fibres of Spectrograph \#4 are found to be mis-identified 
before June, 2013 (Luo A.-L., private communication). 
As a result, the coordinates, magnitudes, spectra and stellar parameters of stars targeted by those fibres
were wrongly assigned in LAMOST DR1 \citep{Luo+2015} as well as LSS-GAC DR1 \citep{Yuan+2015a}. 
The fibres are: \#76 (87), 87 (79), 79 (95), 95 (84), 84 (76), 44 (31), 
31 (46), 46 (26), 26 (44), where the numbers in the brackets are the correct ones. 
The errors have been corrected;
(2) The values of interstellar reddening of flux standard stars are now derived with 
the star-pair method \citep{Yuan+2015a}, replacing those deduced by comparing the 
observed and synthetic colours, as adopted in LSS-GAC DR1. 
The change is based on the consideration that the star-pair method for extinction determination 
is model-free, and yields in general more robust results than the method adopted for LSS-GAC DR1 \citep{Yuan+2015a}; 
(3) For the flux calibration of VB plates, $g, r, i$ magnitudes from the APASS survey \citep{Munari+2014} 
are combined with 2MASS $J, H, K_{\rm s}$ magnitudes to derive values of extinction of  
flux-calibration standard stars. In LSS-GAC DR1, only 2MASS $J, H, K_{\rm s}$ magnitudes were used. 
The change should significantly improve the accuracy of extinction estimates for standards used to flux-calibrate VB plates.

Considering that the SNR alone does not give a fully description of the quality of a spectrum, 
a few flags are now added to image header of a processed spectrum.  
The first flag is the ratio of (sky-subtracted) stellar flux to the flux of (super-) sky 
adopted for sky-subtraction. Due to the uncertainties in sky-subtraction, the spectra of some stars, especially
those observed under bright lunar conditions, may have artificially high SNRs, yet this ratio can be 
in fact quite small for those spectra. The flag is denoted by  
`OBJECT\_SKY\_RATIO' in the value-added catalogues. A second flag is used to mark fibres that maybe potentially 
affected by the nearby saturated fibres. Saturation occurs for very bright stars. When a fibre saturates, 
spectra of nearby fibres, especially those of faint stars, can be seriously contaminated 
by flux crosstalk, leading to incorrect SNRs and wrongly estimated stellar parameters. 
When a fibre saturates, stars observed by the adjacent 50 fibers ($\pm25$) are now marked   
by flag `SATFLAG'  in the value-added catalogues.
Even when a fiber is not saturated, crosstalk may still occur if the flux of that fibre is very high. 
To account for such situation, a third flag is introduced. If the spectrum from a given fibre 
has a SNR higher than 300, then the adjacent 4 ($\pm2$) fibres are assigned a `BRIGHTFLAG' 
value of 1; otherwise the flag has a value of 0. For each fibre, the value of maximum SNR of the nearest 5 fibres 
(the adjacent 4 plus the fibre of concern itself), is also listed as `BRIGHTSNR' in the value-added catalogues. 
Finally, a flag has been created to mark bad fibres. Among the 4000 fibres of LAMOST,
some have very low throughputs or suffer from serious positioning errors. 
Spectra yielded by those ``bad'' fibers cannot be trusted. The number of bad fibres continuously increases 
with time, and reaches about 200 by June, 2014. Those fibers are marked by `BADFIBRE' in the value-added catalogues.
In addition to those newly created flags, information of observing conditions 
with regard to the moon (phase, angular distance), the airmass and 
the pointing position of the telescope are now also included in LSS-GAC DR2.  

\section{Stellar parameter determination: improvements of LSP3}
LSS-GAC DR1 presents values of radial velocity $V_{\rm r}$ and stellar atmospheric parameters (effective temperature $T_{\rm eff}$, 
surface gravity log\,$g$, metallicity [Fe/H]) derived from LSS-GAC spectra with LSP3. 
Since then, a few improvements of LSP3 have been implemented and are included in LSS-GAC DR2, 
including (1) A number of new spectral templates have been added to the 
MILES library, and atmospheric parameters of the template stars have been re-determined/calibrated; 
(2) Several flags are now assigned to describe the best-matching templates that has the characteristics 
of a, e.g., variable star, binary, double/multiple star or supergiant etc. 
of a given target spectrum; (3) Values of $\alpha$-element to iron abundance ratio [$\alpha$/Fe] have been  
estimated by template matching with a synthetic spectral library; 
(4)  A multivariate regression method based on kernel-based principal component analysis (KPCA) has been  
used to obtain an independent set of estimates of stellar atmospheric parameters, including $T_{\rm eff}$, 
log\,$g$, [M/H], [Fe/H], [$\alpha$/M], [$\alpha$/Fe], absolute magnitudes ${\rm M}_V$ and ${\rm M}_{K_{\rm s}}$ 
as well as individual elemental abundances including [C/H] and [N/H].

\subsection{Updates to the MILES library}
The original MILES spectral library contains medium-to-low resolution (full-width-at-half-maximum FWHM $\sim2.5${\AA}) 
long-slit spectra of wavelength range 3525 -- 7410\,{\AA} for 985 stars that have robust stellar parameter 
estimates in the literature, mostly determined with high-resolution spectroscopy \citep{Sanchez-Blazquez+2006, Falcon-Barroso+2011}. 
The MILES library is adopted by LSP3 as templates for estimation of atmospheric 
parameters from LAMOST spectra. Compared to other template libraries available in the literature, 
MILES has  two advantages. Firstly, the MILES spectra 
have robust flux calibration and the spectral resolution matches well that of  
LAMOST spectra. Secondly,  
the template stars cover a large volume of parameter space ($3000<T_{\rm eff}<40,000$\,K, 
$0<{\rm log}\,g<5$\,dex and $-3.0<{\rm [Fe/H]}<0.5$\,dex). 

Still, for the purpose of accurate stellar atmospheric parameter estimation, 
the MILES library has a few defects in want of improvement. 
One is the limited spectral wavelength coverage. LAMOST spectra cover the full optical wavelength 
range of 3700 -- 9000\,{\AA}, whereas MILES spectra stop at 7410{\AA} in 
the red. As a result, LAMOST spectra in the 7400 -- 9000\,{\AA} wavelength range 
have hitherto not been utilized for parameter estimation. There are a few 
prominent features in this wavelength range that are sensitive abundance indicators, 
including, e.g. the Ca\,{\sc ii} $\lambda$8498, 8542, 8664 triplet and  
the Na\,{\sc i} $\lambda$8193, 8197 doublet. In addition, since LSS-GAC targets stars 
of all colours, especially those in the disk, about 30 per cent spectra collected have poor SNRs in the blue 
but good SNRs in the red. Those stars are either intrinsically red or suffer from heavy interstellar extinction. 
Stellar parameters for those stars have currently not been derived from LAMOST spectra,  
by either LSP3 or LASP, due to the fact that both the MILES and ELODIE (adopted by LASP) libraries, 
do not have wavelength coverage long enough in the red. Another defect of the MILES library is the inhomogenous 
coverage of stars in the parameter space. As shown in \citet{Xiang+2015b}, 
there are holes and clusters in the distribution of MILES stars in the parameter space, 
leading to some significant artifacts in the resultant parameters. 
Finally, stellar atmospheric parameters of the original MILES library are 
collected from various sources in the literature thus suffer from systematic errors. 
Although \citet{Cenarro+2007} have taken effort to homogenize the parameters in order to 
account for the systematics amongst the 
values from the different sources, the homogenization 
was carried out for only a limited temperature range of $4000<T_{\rm eff}<6300$\,K. 
With more data available, there is room of considerable improvement.

To deal with the limited wavelength coverage of the MILES spectra and to improve the 
coverage and distribution of template stars in the parameter space, an observational 
campaign is being carried out to observe additional template stars that fill the holes 
in parameter space, to enlarge the coverage of parameter space, as well as to extend 
the spectral wavelength coverage to 9200\,{\AA}, using the NAOC 2.16\,m 
telescope and the YAO 2.4\,m telescope (Wang et al. in preparation). 
The project plans to obtain long-slit spectra covering 3600 -- 9200{\AA} for some 900 
template stars newly selected from the PASTEL catalog \citep{Soubiran+2010}, 
a compilation of stars with robust stellar parameters, mostly inferred from high resolution spectroscopy. 
The project will also extend the wavelength coverage of all the original MILES spectra to 9200\,{\AA}. 
In the current work, 267 new template spectra obtained by the campaign by October, 2015 have 
been added to the MILES library to generate parameter estimates presented in LSS-GAC DR2. 
In addition, to reduce the systematic and random errors of the atmospheric parameters of the template stars,  
Huang et al. (in preparation) have re-calibrated/determined the atmospheric parameters 
of all template stars, both old and newly selected. 
In doing so, the recent determinations of parameters of template stars available  
from the PASTEL catalogues have been adopted, replacing the older values used by the original MILES library. 
The updated values of metallicity are then used to calculate effective temperatures using 
the newly published metallicity-dependent colour-temperature relations of \citet{Huang+2015b}, 
deduced from more than two hundred nearby stars with direct, interferometric angular 
size measurements and Hipparcos parallaxes. Values of surface gravity are 
re-determined using Hipparcos parallaxes \citep{Perryman+1997, Anderson+2012} 
and stellar isochrones from the Dartmouth Stellar Evolution Database \citep{Dotter+2008}.   
Values of [Fe/H] are then re-calibrated to the standard scale of Gaia-ESO survey \citep{Jofre+2015}.
Fig.\,6 plots the newly calibrated/determined parameters of the template stars in the 
$T_{\rm eff}$ -- log\,$g$ and $T_{\rm eff}$ -- [Fe/H] planes.  
Note that in generating LSS-GAC DR1, 85 of the original MILES stars were abandoned 
because there were no complete parameter estimates in the literatures. Those stars are now included in 
the library as their parameters have been re-determined. Also, with the  
newly selected and observed templates added to the library, we have now discarded the 400
interpolated spectra used when generating LSS-GAC DR1.

\begin{figure}
\centering
\includegraphics[width=80mm]{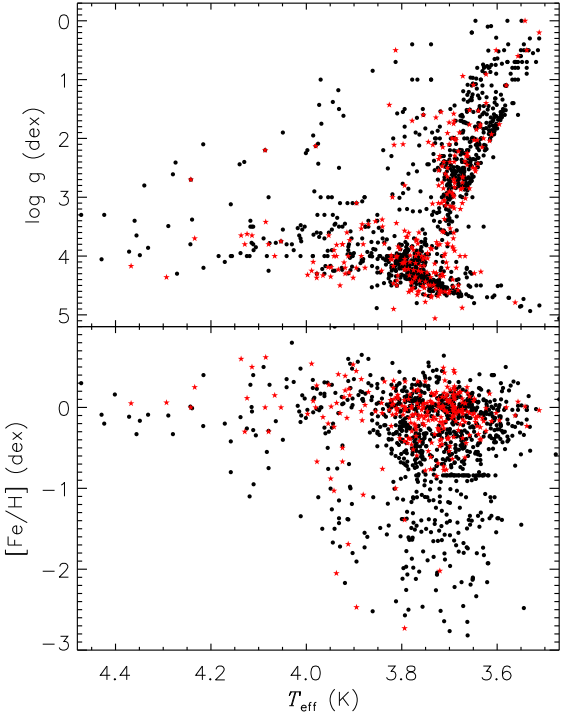}
\caption{Distributions of template stars used by LSP3 in the $T_{\rm eff}$ -- log\,$g$ and 
$T_{\rm eff}$ -- [Fe/H] planes. Black dots represent stars in the original MILES library, while red dots represent 
267 newly observed templates employed in the generation of LSS-GAC DR2 (see the text).}
\label{Fig7}
\end{figure}

\subsection{[$\alpha$/Fe] estimation by matching with synthetic spectra}
$\alpha$-element to iron abundance ratio [$\alpha$/Fe] is a good indicator of the Galactic 
chemical enrichment history \citep[e.g.][]{Lee+2011b}, and thus valuable to derive. To estimate ratios [$\alpha$/Fe] 
from LAMOST spectra, a method of template matching based on $\chi^2$-fitting with synthetic spectra 
is developed for LSP3. Details about the method and robustness tests of the deduced [$\alpha$/Fe] 
values are described in \citet{Liji+2016}. Here we briefly summarize the method and 
point out a few improvements that may lead to better results.    

The synthetic library used to estimate [$\alpha$/Fe]  was generated with the SPECTRUM code \citep{Gray+1999} 
of version 2.76, utilizing the Kurucz stellar model atmospheres of \citet{Castelli+2004}. The solar [$\alpha$/Fe] ratio 
is set to zero, and the $\alpha$-enhanced grids are generated by scaling the elemental 
abundances of O, Mg, Si, S, Ca and Ti, and those of C and N abundances 
are also enhanced by the amount of $\alpha$-elements. Lines from more then 15 diatomic molecule species, 
including H$_2$, CH, NH, OH, MgH, AlH, SiH, CaH, C$_2$, CN, CO, AlO, SiO, TiO and ZrO, 
are taken into account in the calculation of the opacity. Isotope lines are also taken into account. 
In total, 320,000 synthetic spectra are generated. Table\,2 lists the parameter ranges and
steps of the grids. Note that the grids adopted here have a step of 0.1\,dex in [$\alpha$/Fe],  
half of the value used in \citet{Liji+2016}.  
All the computed synthetic spectra have a resolution of 2.5\,{\AA} FWHM, and is invariant with wavelength. 
\begin{table}
\caption{Grids of KURUCZ synthetic spectra for [$\alpha$/Fe] estimation.}
\label{}
\begin{tabular}{ccc}
\hline
 Parameter & Range                                             & Step            \\ 
\hline
  $T_{\rm eff}$   & [4000, 8000]\,K                          &  100 \,K       \\ 
  log\,$g$   & [0.0, 5.0]\,dex                                       &   0.25\,dex   \\  
  ${\rm [Fe/H]}$    & [$-4.0$, 0.5]\,dex                                      & 0.2\,dex for ${\rm [Fe/H]} < -1.0$\,dex \\
         &   & 0.1\,dex for ${\rm [Fe/H]} > -1.0$\,dex \\
  ${\rm [\alpha/Fe]}$   & [$-$0.4, 1.0]\,dex                               & 0.1\,dex       \\
\hline
\end{tabular}
\end{table}

For a target spectrum with atmospheric parameters $T_{\rm eff}$, log\,$g$ and [Fe/H] 
yielded by LSP3, the synthetic spectra are interpolated to generate a set of spectra that 
have the same atmospheric parameters as the target 
for all grid values of [$\alpha$/Fe].
Values of $\chi^2$ between the target and the individual interpolated synthetic spectra 
are then calculated. 
A Gaussian plus a second order polynomial is then used to fit the deduced $\chi^2$ 
as a function of [$\alpha$/Fe]. The value of [$\alpha$/Fe] that yields  
minimum $\chi^2$ is taken to be the [$\alpha$/Fe] ratio of the target spectrum. 
To compute $\chi^2$, \citet{Liji+2016} use spectral 
segments 4400 -- 4600\,{\AA} and 5000 -- 5300\,{\AA}. The 4400 -- 4600\,{\AA} segment
contains mainly Ti features, while that of 5000 -- 5300\,{\AA} contains mainly Mg\,{\sc i} 
features, as well as a few features of Ca, Ti and Si. 
Note that given the low resolution as well as limited SNRs of LAMOST spectra, 
Ca, Ti and Si features within those two spectral segments contribute in fact only a small fraction 
of the calculated values of $\chi^2$, and are therefore not very useful for the determination of [$\alpha$/Fe].
In metal-rich stars, the Mg\,{\sc i}\,b features are in general prominent enough for a 
robust determination of [$\alpha$/Fe]. However, in metal-poor (${\rm [Fe/H]}<-1.0$\,dex) stars, 
the Mg\,{\sc i}\,b lines become less prominent so that the [$\alpha$/Fe] have 
larger uncertainties. In the current work, in order to improve precision of [$\alpha$/Fe] 
estimates, especially for metal-poor stars, we have opted to include the 3910 -- 3980\,{\AA} spectral 
segment that contains the Ca~{\sc ii} HK lines in the calculation of $\chi^2$ values.
Meanwhile, as an option, we also provide results yielded using the exactly same spectral segments 
as \citet{Liji+2016}. This is useful considering that the strong Ca~{\sc ii} HK lines 
in model spectra for metal-rich stars maybe not accurately synthesized. 

The resolution of LAMOST spectra from individual fibers varies from one to another, 
as well as with wavelength \citep{Xiang+2015b}. To account for this in template matching, 
the synthetic spectra are degraded in resolution to match that of the target spectrum. 
The latter is derived utilizing the arc spectrum. The deduced resolution 
as a function of wavelength is further scaled to match the resolution yielded 
by sky emission lines detected in the target spectrum in order to account for 
systematic variations of spectral resolution between the arc and target exposures. 
Typically, for a given spectrograph, fibre to fibre variations of spectral resolution amount to 
0.3\,{\AA}, rising to 0.5 -- 1.0\,{\AA} among the different spectrographs. 
Systematic variations of spectral resolution between the 
arc and target exposures are found to be typically 0.2\,{\AA}. 

In addition, in order to allow for possible uncertainties in the input atmospheric parameters 
$T_{\rm eff}$, $\log\,g$ and [Fe/H] yielded by LSP3 as well as any possible mismatch 
between the LSP3 atmospheric parameters and those of the Kurucz stellar model 
atmospheres, in the current work, we have opted not to fix the input values of 
$T_{\rm eff}$, log\,$g$ and [Fe/H] as yielded by LSP3, but allow them to vary in limited ranges
 around the initial values. The ranges are set to 2$\sigma$ uncertainties of the parameters 
 concerned, with lower limits of 500\,K, 0.5\,dex and 0.5\,dex and upper limits of 
 1000\,K, 1.0\,dex and 1.0\,dex for $T_{\rm eff}$, $\log\,g$ and [Fe/H], respectively.  
For each grid value of [$\alpha$/Fe], the synthetic spectrum that has atmospheric 
parameters $T_{\rm eff}$, $\log\,g$ and [Fe/H] within the above ranges and fits the target 
spectrum best (i.e. yielding the smallest $\chi^2$) is taken as the choice of synthetic 
spectrum when fitting and deriving [$\alpha$/Fe] using the technique described in \citet{Liji+2016}.  

Values of [$\alpha$/Fe] are derived with the above algorithm for all 
LSS-GAC DR2 stars of a spectral SNR higher than 15. 
The left panel in the third row of Fig.\,9 plots the density distribution of 
LSS-GAC DR2 dwarf stars in the [Fe/H] -- [$\alpha$/Fe] plane. 
Here the [$\alpha$/Fe] are those derived using spectra including the 3910 -- 3980\,{\AA} segments. 
The figure shows an [$\alpha$/Fe] plateau for metal-poor (${\rm [Fe/H]} < -1.0$\,dex) 
stars, at a median value about 0.4\,dex. For more metal-rich stars, [$\alpha$/Fe] decreases 
with increasing [Fe/H], reaching a median value of zero near the solar metallicity, 
which means that the zero-point offset, i.e. deviations of [$\alpha$/Fe] values from zero 
at solar metallicity, is small, which is in contrast to \citet{Liji+2016} who find a zero-point 
offset of $-0.12$\,dex. However, we indeed find a zero-point offset of about $-0.1$\,dex 
for [$\alpha$/Fe] derived using only the 4400 -- 4600\,{\AA} and 5000 -- 5300\,{\AA} segments, 
which is basically consistent with \citet{Liji+2016}. The offset is found to be mainly contributed 
by the 4400 -- 4600\,{\AA} segment. The causes of this difference are not 
fully understand yet. We suspect there may be some unrealistic inputs in either 
the atmosphere model or the atomic and molecular data used to generate the synthetic spectra. 
Note that here we opt not to introduce any external corrections on the estimated [$\alpha$/Fe].
Random errors of [$\alpha$/Fe] induced by spectral noises, as estimated by comparing the results 
deduced from duplicate observations, are a function of spectral SNR and atmospheric parameters 
$T_{\rm eff}$, log\,$g$ and [Fe/H], and have typical values that decrease from $\sim$0.1 to $\sim$0.05\,dex 
as the spectral SNR increases from 20 to a value higher than 50. 
To provide a realistic error estimate for [$\alpha$/Fe], the  
random error induced by spectral noises is combined with the method error, 
which is assumed to have a constant value of 0.09\,dex, estimated by a comparison with 
high-resolution measurements \citep[cf.][]{Liji+2016}. 

For giant stars, [$\alpha$/Fe] estimated with the above algorithm exhibits a 
zero-point offsets between 0.1 and 0.2\,dex, probably caused by inadequacies 
of the synthetic spectra for giant stars. In this work, no corrections for those offsets are applied. 
Note that for giant stars, [$\alpha$/Fe] values are also estimated with the KPCA regression 
method using the LAMOST-APOGEE common stars as the training data set (cf.\S{4.3}), 
and the resultant values are presented as the recommended ones (cf.\S{4.6}). 

\subsection{Stellar atmospheric parameters estimated with KPCA method}
The LSP3 version used to generate LSS-GAC DR1 estimates stellar atmospheric parameters 
$T_{\rm eff}$, log\,$g$ and [Fe/H] with a $\chi^2$-based weighted-mean algorithm. 
The algorithm achieves a high precision in the sense that random errors of the deduced 
parameters induced by spectral noises are well controlled, even for stars of SNRs as low as 10. 
Nevertheless, parameters estimated with the 
weighted-mean algorithm suffer from several artifacts. One is the so-called `suppression effect' -- values of 
the derived log\,$g$ are narrowed down to an artificially small range. This is partly 
caused by the fact that $\chi^2$ calculated from a LAMOST spectrum with respect to 
a template spectrum is only moderately sensitive to log\,$g$, 
thus the sets of templates used to calculate the weighted-mean 
values of $\log\,g$ for the individual target sources often have similar distributions in log\,$g$. 
Another artifact is the so-called `boundary effect' -- parameters of stars with true parameters 
that pass or are close to the boundary of parameter space covered by the templates are 
often either underestimated or overestimated systematically by the weighted-mean algorithm. 
This effect is especially serious for [Fe/H] and log\,$g$ estimation. 
Finally, due to the inhomogeneous distribution of templates in 
the parameter space, moderate clustering effect is also seen in the 
deduced parameter values. 

To overcome the above defects of the weighted-mean algorithm, a regression method based on  
Kernel-based Principal Component Analysis (KPCA) has recently been incorporated into LSP3 \citep{Xiang+2016}. 
The method is implemented in a machine learning scheme. A training data set is first 
defined to extract non-linear principal components (and the loading vectors) as well as to build-up 
regression relations between the principal components and the target parameters. 
Four training data sets are defined for the determination of specific sets of parameters  
of specific types of stars. They are: (1) The MILES library for the estimation of  
$T_{\rm eff}$ and [Fe/H] of A/F/G/K stars, and for the estimation of log\,$g$ of stars 
that have log\,$g$ values larger than 3.0\,dex as given by the weighted-mean algorithm. 
The latter stars are mainly dwarfs and sub-giants;
(2) A sample of LAMOST-Hipparcos common stars with accurate parallax (thus distance 
and absolute magnitude) measurements for the estimation of absolute magnitudes 
(${\rm M}_V$ and ${\rm M}_{K_{\rm s}}$) directly from LAMOST spectra; 
(3) A sample of LAMOST-$Kepler$ common stars with accurate asteroseismic $\log\,g$ 
measurements for the estimation of log\,$g$ of giant stars; and (4) A sample of LAMOST-APOGEE 
common stars for the estimation of metal abundance [M/H], $\alpha$-element to 
iron (metal) abundance ratio [$\alpha$/Fe] ([$\alpha$/M]), and of individual elemental abundances 
including [Fe/H], [C/H] and [N/H] for giant stars.
A detailed description of the implementation and test of the method is presented in \citet{Xiang+2016}. 
Here we note two updates of the implementation with respect to those of \citet{Xiang+2016}.
One is that a unified number of principal components (PCs) of 100 is adopted except for 
the estimation of absolute magnitudes with the LAMOST-Hipparcos training set, for which 
both 100 and 300 PCs are adopted. Another modification is that additional training stars 
have been added to the LAMOST-$Kepler$ and LAMOST-APOGEE training sets as a result that  
more common stars become available as the surveys progress. Both training sets now contain 
(exactly) 3000 stars. 

Compared to the weighted-mean results, log\,$g$ and [Fe/H] derived with the KPCA 
regression method using the MILES training set have been found to suffer from  
less systematics, and thus better for statistical analyses. A comparison with asteroseismic 
measurements shows that for stars of $T_{\rm eff}<6000$\,K and ${\rm [Fe/H]}>-1.5$\,dex, 
uncertainties of KPCA log\,$g$ estimates can be as small as 0.1\,dex for spectral SNRs higher than 50.  
The uncertainties increase substantially as the SNR deteriorates, reaching $\sim$0.2\,dex 
at a SNR of 20 \citep{Xiang+2016}. Similarly, comparisons with 
[Fe/H] determinations from high resolution spectroscopy show that uncertainties of KPCA [Fe/H] 
estimates are $\sim$0.1\,dex for good SNRs. 
For metal-poor (${\rm [Fe/H]}<-1.5$\,dex) or hot ($T_{\rm eff}>8000$\,K) stars, 
KPCA estimates of both log\,$g$ and [Fe/H] become less reliable, and should be used with caution.    
Systematic differences between KPCA and weighted-mean estimates of $T_{\rm eff}$ exist  
for stars hotter than $\sim6500$\,K, with deviations reach as much as $\sim$300\,K for stars 
of $T_{\rm eff}\sim7500$\,K. 
More studies are needed to understand the cause of deviations. For the time being, we recommend  
weighted-mean $T_{\rm eff}$ values as they are estimated in a straightforward way 
and well validated \citep{Xiang+2015b}. However, note that weighted-mean estimates of $T_{\rm eff}$ 
suffer from clustering effect so that they may weakly clump in the parameter space 
on the scale of a few tens Kelvin, while the KPCA estimates of $T_{\rm eff}$ do not have such problem. 

Surface gravities of giant stars of $T_{\rm eff}<5600$ and log\,$g < 3.8$\,dex are 
estimated with the KPCA method using the LAMOST-$Kepler$ training set. 
Here the cuts of $T_{\rm eff}$ and log\,$g$ are based on values yielded by the weighted-mean method. 
The KPCA estimate of log\,$g$ is likely to be accurate to 0.1\,dex given a spectral SNR 
higher than 50. The uncertainty increases to $\sim$0.2\,dex at a SNR of 20 \citep{Huang+2015c, Xiang+2016}. 

Absolute magnitudes ${\rm M}_V$ and ${\rm M}_{K_{\rm s}}$ of all stars with a weighted-mean 
effective temperature lower than 12,000\,K are estimated directly from LAMOST spectra 
using the LAMOST-Hipparcos training set. Two sets of absolute magnitudes are given, 
corresponding to respectively 100 and 300 PCs adopted for parameter estimation. 
For absolute magnitudes estimated using 100 PCs, typical uncertainties are 0.3--0.4\,mag 
in both ${\rm M}_V$ and ${\rm M}_{K_{\rm s}}$ for a spectral SNR higher than 50, 
and the values increase to $\sim$0.6\,mag at a SNR of 20. For absolute magnitudes estimated 
using 300 PCs, uncertainties are only 0.2--0.3\,mag for a spectral SNR higher than 50, 
but the results are more sensitive to SNR, with a typical uncertainty of $\sim$0.7\,mag at a SNR of 20.    

\begin{figure}
\centering
\includegraphics[width=85mm]{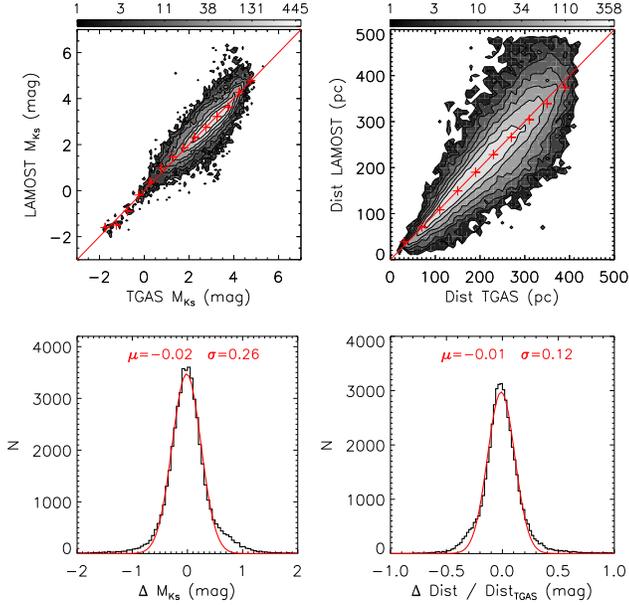}
\caption{Comparison of absolute magnitudes and distances with those inferred from Gaia TGAS 
parallax for 50,000 LAMOST-TGAS common stars that have a TGAS-based magnitude error 
smaller than 0.2\,mag. The upper panels shows colour-coded contours of stellar number 
density in logarithmic scale. Crosses in red are median values of our estimates calculated 
in bins of the TGAS-based values. The lower panels plot distribution of differences of magnitudes 
and distances between our estimates and the TGAS-based values. Red lines are Gaussian 
fits to the distribution, with the mean and dispersion of the Gaussian marked in the plot.}
\label{Fig8}
\end{figure}
A significant advantage of estimating absolute magnitudes directly from the observed spectra 
is that the magnitudes as well as the resultant distance moduli, are model independent. 
Fig.\,8 plots a comparison of the estimates of ${\rm M}_{K_{\rm s}}$ for 300 PCs as well as 
the distances thus estimated with results inferred from the Gaia TGAS parallaxes \citep{Lindegren2016} 
for a sample of 50,000 LAMOST-TGAS common stars that have a TGAS-based magnitude 
error smaller than 0.2\,mag. Here the TGAS distances and ${\rm M}_{K_{\rm s}}$ values are derived 
using values of interstellar extinction derived with the star pair method (cf. Section 6.1). The figure shows very 
good agreement between our estimates and the TGAS results. Systematics in both ${\rm M}_{K_{\rm s}}$ 
and distance estimates are negligible, and the dispersion is only 0.26\,mag for ${\rm M}_{K_{\rm s}}$,
12 per cent for distance. A similar comparison of results yielded using 100 PCs yields a mean 
difference of 0.04\,mag and a dispersion of 0.29\,mag in ${\rm M}_{K_{\rm s}}$, and 
a mean difference of $-2$ per cent and a dispersion of 13 per cent in distance estimates. 
Of course, spectral SNRs for the LAMOST-TGAS common stars 
are very high, which have a median value of 150, as the stars are very bright. 
Note that Fig.\,8 shows also a positive non-Gaussian tail in the difference of ${\rm M}_{K_{\rm s}}$, 
with a corresponding negative tail in the difference of distance. This non-Gaussian tail is likely 
caused by binary stars, for which the estimates of absolute magnitudes from LAMOST 
spectra are only marginally affected, whereas the photometric magnitudes are underestimated 
respect to those assuming single stars. 
Note that not all the LAMOST-TGAS stars used for the comparison can be found in the LSS-GAC 
value-added catalogues as many of them are targeted by other survey projects of LAMOST. 
The spectra of those stars have been processed with LSP3 and the data are only internally available 
for the moment.

Metal (iron) abundance [M/H] ([Fe/H]), $\alpha$-element to metal (iron) 
abundance ratio [$\alpha$/M] ([$\alpha$/Fe]), as well as abundances of carbon and nitrogen,   
[C/H] and [N/H], are estimated using the LAMOST-APOGEE training set. 
\citet{Xiang+2016} have demonstrated that the results have a precision 
comparable to those deduced from APOGEE spectra with the ASPCAP pipeline \citep{Garcia_Perez+2015, Holtzman+2015}. 
Specifically, estimates of [M/H], [Fe/H], [C/H] and [N/H] have a precision better than 0.1\,dex 
given a spectral SNR high than 30, and $\sim$0.15\,dex for a SNR of 20. 
We note however, since the APOGEE stellar parameters are 
not externally calibrated except for [M/H], any systematics in the APOGEE results 
propagated into ours through the training set. As pointed out by \citet{Holtzman+2015}, 
APOGEE estimates of elemental abundances may suffer from systematic biases of about 0.1--0.2\,dex. 
For [M/H], the APOGEE values are calibrated to [Fe/H] measurements of star clusters for [Fe/H] range [$-$2.5, 0.5]\,dex. 
Nevertheless, it is found that [Fe/H] ([M/H]) estimated with the LAMOST-APOGEE training set  
are 0.1\,dex systematically higher than those estimated with the MILES library. 
Such an overestimation is also confirmed by examining common stars with high resolution 
spectroscopic [Fe/H] measurements available from the PASTEL catalogue. The discrepancy 
is likely due to an offset in absolute value between the APOGEE and MILES (PASTEL) metallicities. 
The estimated [$\alpha$/M] ([$\alpha$/Fe]) values have a typical precision of 0.03 -- 0.06\,dex given 
a spectral SNR higher than 20. \citet{Xiang+2016} have demonstrated that, 
as a consequence of such a high precision, a clear distinction between the sequences of thick  
and thin disk stars is seen in the [M/H] ([Fe/H]) -- [$\alpha$/M] ([$\alpha$/Fe]) plane, 
quite similar to that revealed by results from high-resolution spectroscopy.
 Note however that, for stars with [$\alpha$/M] higher than 0.30\,dex or 
lower than 0.0\,dex, the KPCA values are probably systematically underestimated or overestimated 
due to a lack of training stars of such abundance ratios. In addition, given that APOGEE [$\alpha$/M] ([$\alpha$/Fe])  
values are not externally calibrated, there may also be some systematics  
hided in our results. 

\subsection{Estimation of parameter errors} 
Errors of the deduced stellar parameters are estimated in a statistical way. 
Parameter errors contributed by both spectral noises and inadequacies of the method are taken into account.  
For $T_{\rm eff}$, log\,$g$ and [Fe/H] estimated by the weighted-mean algorithm, 
or those by the KPCA method using the MILES training set, 
as well as ${\rm M}_V$ and ${\rm M}_{K_{\rm s}}$ estimated by the KPCA method 
using the LAMOST-Hipparcos training set, parameter errors induced by 
spectral noises and those by the method are estimated separately. 
Errors induced by spectral noises, as in the case of LSS-GAC DR1, are estimated by comparing 
results derived from duplicate observations made in different nights, and 
are estimated separately for giants (log\,$g < 3.5$\,dex) and dwarfs (log\,$g > 3.5$\,dex). 
The results are fitted with a second-order polynomial function of SNR, $T_{\rm eff}$ and [Fe/H],  
\begin{equation}
\begin{aligned}
\sigma =  & c_0 + c_1\times {\rm SNR} + c_2 \times T_{\rm eff} + c_3 \times {\rm [Fe/H]} + c_4 \times {\rm SNR}^2 \\
  &     + c_5 \times T_{\rm eff}^2 + c_6 \times {\rm [Fe/H]}^2 +  c_7 \times {\rm SNR} \times T_{\rm eff} \\
  &    + c_8 \times {\rm SNR} \times {\rm [Fe/H]} + c_9 \times T_{\rm eff} \times {\rm [Fe/H]}  
\end{aligned}
\end{equation}
Here log\,$g$ estimated with the weighted-mean algorithm is used to group stars 
into giants and dwarfs. 
Generally, the errors decrease significantly with increasing SNR, and hot or metal-poor 
stars have larger errors than cool or metal-rich ones. 

The method errors are deduced from the residuals obtained by applying the  
method to spectra of the template library or the training sets themselves. Both the mean and dispersion 
of the residuals are calculated and fitted as functions of $T_{\rm eff}$ and [Fe/H], for dwarfs and giants 
separately. The mean reflects the bias induced by the method, 
and is thus corrected for for all target stars. While the dispersion is combined with the error 
induced by the spectral noises to yield the final value of error of the estimated parameter of concern. 
In doing so, a grid of method errors (mean and dispersion) is first 
created in the $T_{\rm eff}$ -- [Fe/H] plane, and for a given set of atmospheric parameters of 
a target star, the corresponding value of method error is interpolated from the grid.  
   
For log\,$g$ estimated with the KPCA method using the LAMOST-$Kepler$ training set, 
as well as metal/elemental abundances ([M/H], [Fe/H], [C/H], [N/H], [$\alpha$/M], [$\alpha$/Fe]) 
estimated using the LAMOST-APOGEE training set, since there are sufficiently large number 
of stars in common with the $Kepler$ and APOGEE surveys that have not been included 
in the training samples, they can be used as the test samples \citep{Xiang+2016} to 
directly estimate the parameter errors. 
Given that no obvious trends with $T_{\rm eff}$, log\,$g$ and [Fe/H] are seen, 
the errors are estimated as a function of spectral SNR only.   

\subsection{Specific flags}
Specific flags are assigned to each star to better describe the quality of estimated parameters. 
A flag is assigned to describe the type of the best-matching template star.  
Based on the SIMBAD database \citep{Wenger+2000}, LSP3 template stars are divided into 38 groups  
as listed in Table\,3. While the majority  
template stars are normal (single) stars of A/F/G/K/M spectral types, there are also considerable 
numbers of spectroscopic binaries, double or multiple stars, variable stars, as well as of other rare types. 
The flag is set to help identify stars of specific type, 
although a careful analysis is essential to validate the results.
The flag is an integer of value from 1 to 38, and is labeled respectively by `TYPEFLAG\_CHI2' and 
`TYPEFLAG\_CORR' for results based on the minimum $\chi^2$ and correlation 
matching algorithms \citep{Xiang+2015b}.

\begin{table}
\caption{Types of template stars.}
\label{}
\begin{tabular}{lll}
\hline
Flag & Type                                             & N            \\ 
\hline
  1   & Normal star  (single; AFGKM)                 &  757 \\ 
  2   & Pre-main sequence star               &   1 \\ 
  3   & L/T-type star                                 & 9\\
  4   & O/B-type star                                 & 19\\
  5   & S star                                            & 6 \\
  6   & Carbon star                                   &5 \\
  7   & Blue supergiant star                     &6\\ 
  8   & Red super giant  star                                    &3\\ 
  9   & Evolved supergiant star                                &1 \\
 10   & Horizontal branch (HB) Star (not include RCs)                         &27\\ 
 11   & Asymptotic giant branch (AGB) Star            &6\\
 12   & Post-AGB star (proto-PN)                            &4\\ 
 13  & Planetary nebula (PN)                                  &1 \\
 14  & Star in nebula                                                &3 \\
 15  & Emission line star                                          &2 \\
 16  & Pulsating variable star                                   &5\\ 
 17  & Semi-regular pulsating star                           &22 \\
 18  & Classical Cepheid (delta Cep type)               &10 \\
 19  & Variable star of beta Cep type                       &4\\
 20  & Variable star of RR Lyr type                            &5\\
 21  & Variable star of BY Dra type                           &28\\
 22  & Variable star of RS CVn type                         &12\\
 23  & Variable star of alpha2 CVn type                   &23\\
 24  & Variable star of delta Sct type                       &21\\
 25  & Variable star of RV Tau type                         &1\\
 26  & Rotationally variable star                               &4\\
 27  & T Tau-type star                                              & 1 \\
 28  & Flare star                                                       &10 \\
 29  & Peculiar star                                                   &4\\
 30  & Long-period variable star                               &13\\
 31  & Spectroscopic binary                                  & 80 \\
 32  & Eclipsing binary of Algol type (detached)       &7\\
 33  & Eclipsing binary of beta Lyr type (semi-detached) &2\\
 34  & Symbiotic star                                             & 2\\ 
 35  & Star with envelope of CH type                        &7\\
 36 & Variable star (unclassfied)                                    &78\\
 37  & Double or multiple star                               &79\\
 38  & WD  (DA)                                                       &3 \\
\hline
 \hline
\end{tabular}
\end{table}

The second flag describes the correlation coefficient for radial velocity estimation. 
Although LSS-GAC intends to target stars that are identified as point sources in 
the photometry catalogues, there are still some contaminations from extragalactic 
sources (e.g. galaxies, QSOs), for which LSP3 gives problematic radial velocities 
because the pipeline treats all input spectra as stellar.  
In addition, radial velocity estimates can be problematic for stars of unusual spectra 
(e.g. of emission line stars) or defective spectra (e.g. those seriously affected by cosmic rays 
and/or scatter light). In such cases, it is found that the peak correlation coefficient   
for radial velocity estimation is small compared to those of bulk stars. 
To mark those objects, a flag is assigned to each star in a way similar to 
the third flag in \citet{Xiang+2015b}, with the only difference that
the current flag is set to be a positive float number, with negative ones replaced by 0.0.
Experience suggests that one should treat the radial velocity cautiously  
if the flag has a value larger than 6.0. The flag is denoted by `VR\_FLAG' in the value-added 
catalogues. The peak correlation coefficient is also given by flag
`PEAK\_CORR\_COEFF'. 

Similar to the second flag of \citet{Xiang+2015b}, a flag is used to describe 
anomalies in the minimum $\chi^2$ of the best-matching spectral template. 
This flag is designed for parameters estimated with the weighted-mean 
algorithm. Again, the current flag is set to be a positive float, with negative 
values replaced by 0.0. The flag is denoted by `CHI2\_FLAG'.  
The minimum $\chi^2$ of the best-matching template is presented as `MIN\_CHI2'. 

Another flag is assigned to indicate the quality of parameters estimated with the KPCA method. 
It is defined as the maximum value of the kernel function, $d_{\rm g}$, which reflects 
the minimum distance between the target and training spectra \citep{Xiang+2016}.   
The flag is a float of value between 0.0 and 1.0, with larger values indicating better 
parameter estimates. All KPCA estimates that have  
a $d_{\rm g}$ value smaller than 0.2 are not provided in the value-added catalogues due to their low reliability. 
The flag is labeled respectively by `DG\_MILES\_TM',  `DG\_MILES\_G', `DG\_HIP',  `DG\_KEP', 
`DG\_APO', for parameters estimated with the aforementioned different training sets, (cf. Table\,5).

\subsection{Recommended parameter values}
\begin{figure*}
\centering
\includegraphics[width=160mm]{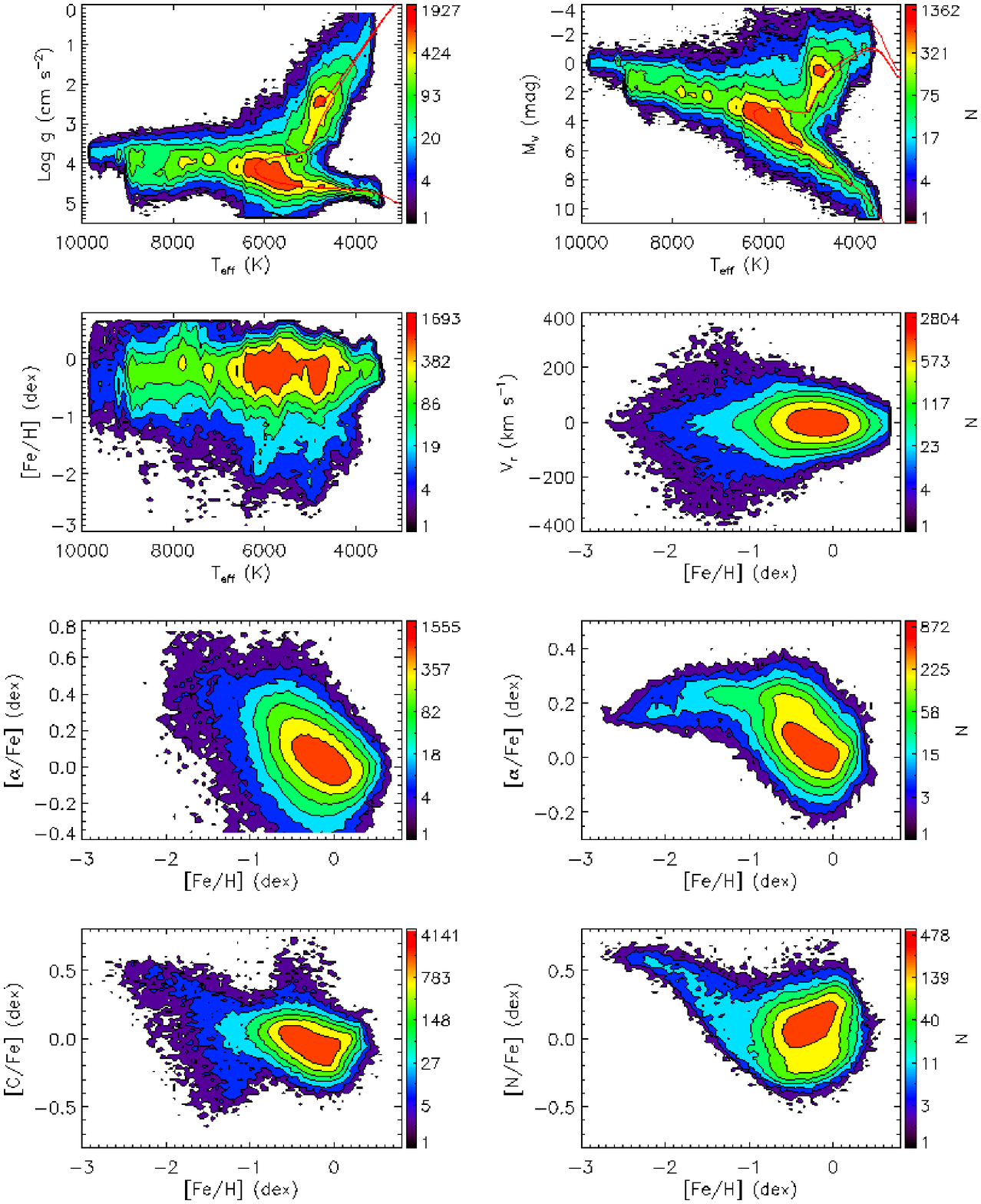}
\caption{Contours of colour-coded stellar number density distributions in various planes of  
recommended parameters for the whole sample of stars in the value-added catalogues. 
Stellar isochrones in the $T_{\rm eff}$ -- log\,$g$ and $T_{\rm eff}$ -- ${\rm M}_V$ 
plots are from \citet{Rosenfield2016} and have a solar metallicity and an age of 4.5\,Gyr. 
The left and right panels of third row are [Fe/H] -- [$\alpha$/Fe] plots  
showing results for dwarfs and giants, respectively.}
\label{Fig9}
\end{figure*}

\begin{figure*}
\centering
\includegraphics[width=180mm]{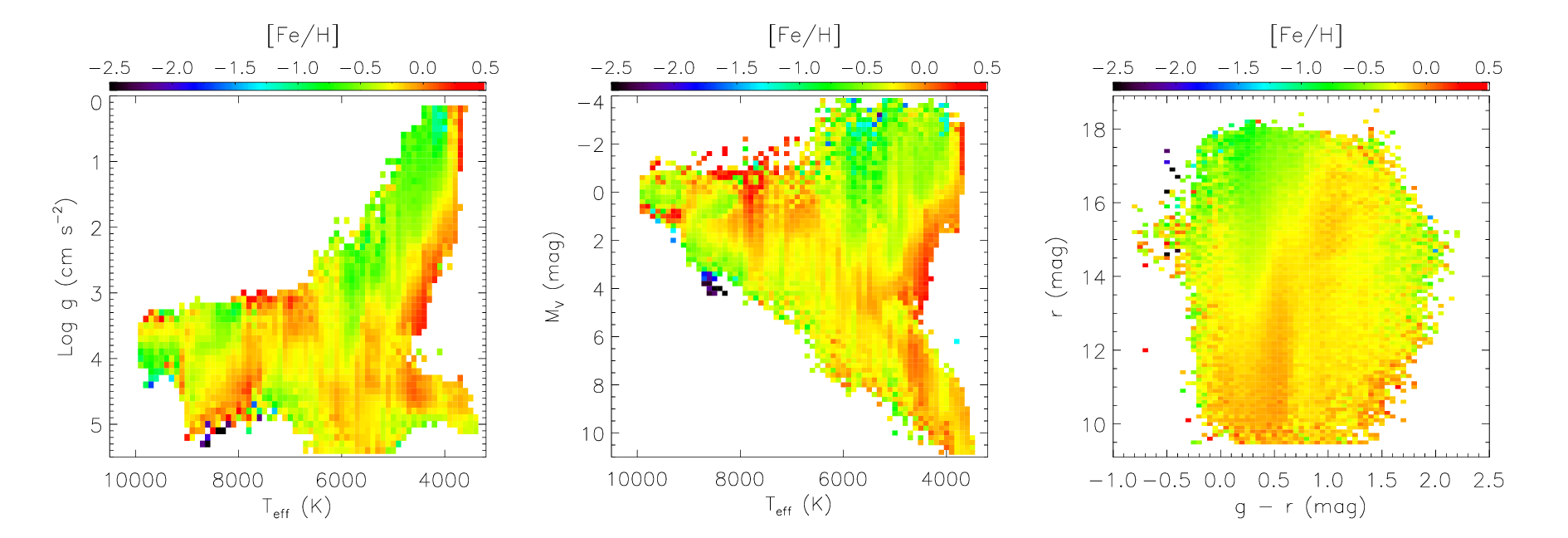}
\caption{Distributions of sample stars in $T_{\rm eff}$ -- log\,$g$, $T_{\rm eff}$ -- ${\rm M}_V$ 
and ($g-r$, $r$) diagrams colour-coded by median metallicity, defined as the median value 
of stars in binsof 100\,K$\times$0.1\,dex in $T_{\rm eff}$ -- log\,$g$, 100\,K$\times$0.2\,mag in 
$T_{\rm eff}$ -- ${\rm M}_V$, and 0.05$\times$0.1\,mag in ($g-r$, $r$) plane, respectively, 
for the three panels from left to right. }
\label{Fig10}
\end{figure*}

Since different methods have been employed to deduce parameters 
$T_{\rm eff}$, log\,$g$, [Fe/H] and [$\alpha$/Fe], one needs to choose 
between parameter values yielded by different methods for the specific problems 
that she/he may want to address. For this purpose, it is necessary to get to know 
both the advantages and limits of the individual methods employed. 

For atmospheric parameters estimated with the weighted-mean algorithm, the 
advantage is that the results are found to be relatively insensitive to spectral SNR compared 
to results yielded by other methods. This implies that for low SNRs (e.g. $< 20$), 
results from the weighted-mean method are more robust and should be preferred. 
A disadvantage of the method is that the resultant parameters suffer from some 
systematics, as described above. 
The advantage of parameters estimated with the KPCA method is that they   
are much less affected by systematics, and thus more accurate than the weighted-mean 
values given sufficiently high spectral SNRs. A limitation is that they are  
available for limited ranges of parameters only, and, in addition, suffer from relatively 
large random errors at low SNRs.   

We provide a recommended set of parameters \{$T_{\rm eff}$, log\,$g$, ${\rm M}_V$, 
${\rm M}_{K_{\rm s}}$, [Fe/H],  [$\alpha$/Fe]\} based on the above considerations. Table\,4 lists the 
adopted methods for the recommended parameters and their effective parameter ranges. 
We emphasize that the recommended parameters are not necessarily 
the most accurate/precise ones. For example, although [Fe/H] values
estimated with the KPCA method using the MILES library are adopted as the 
recommended values, those estimated using the LAMOST-APOGEE training set are in fact 
more precise. Values of $T_{\rm eff}$ derived with the weighted-mean method are 
robust, but suffer from clustering effect, and are thus weakly clumped in the parameter 
space on the scale of a few tens Kelvin.
For [$\alpha$/Fe], considering that there are systematic differences 
between the template matching and KPCA results \citep{Xiang+2016}, 
our recommendation will lead to some level of inconsistency between the results 
for giants and dwarfs. For ${\rm M}_V$ and ${\rm M}_{K_{\rm s}}$, absolute magnitudes yielded using 
300 PCs are adopted as the recommend values, but if the targets of interest are mainly 
composed of stars with low spectral SNRs ($<30$), one would better use results from 100 PCs. 
In addition, because a cut of 0.2 is set for the maximum value of the kernel function, $d_g$, 
to safeguard robust parameter estimates with the KPCA method \citep{Xiang+2016}, about 18 per cent 
of the sample stars (mostly with very low SNRs) have no parameter estimates with   
the KPCA method. For those stars, the recommended parameters are not available.
Rather than adopting the recommended values, we encourage users to make 
their own choice based on their specific problems under consideration.

For [M/H], [$\alpha$/M], [C/H] and [N/H], since at the moment there are only one set of 
measurements, i.e. those with the KPCA method, they are simply adopted as the recommended values. 
Note that [M/H], [$\alpha$/M], [C/H] and [N/H] are available only for stars of 
$T_{\rm eff}<5600$\,K and log\,$g<3.8$\,dex. Here 
$T_{\rm eff}$ refers to that estimated with the weighted-mean method, while 
log\,$g$ is either estimated with the weighted-mean method or 
by the KPCA method using the MILES stars as the training set.

Fig.\,9 plots the stellar number density distributions in various planes of recommended parameters. 
In the $T_{\rm eff}$--log\,$g$ and $T_{\rm eff}$--${\rm M}_V$ planes, the stars show two prominent clumps, 
one composed of FGK dwarfs and another of red clump stars. The compact contours 
of red clump stars indicate a high precision of the estimated parameters. Trajectories of stars 
in those planes are well consistent with the theoretical isochrones. Stars of ${\rm [Fe/H]}>-1.0$\,dex dominate 
the sample, indicating that most of the stars belong to the Galactic disk. The fraction of stars with 
${\rm [Fe/H]}<-1.0$\,dex is only 1.8 per cent. Given the large difference between the numbers of 
metal-rich and metal-poor stars, even a small fraction of those metal-rich stars whose [Fe/H] estimates 
may have suffered large uncertainties could seriously contaminate the sample of metal-poor stars. 
This seems to have led to the appearance of a kinematically cold but metal-poor population in the \
[Fe/H] -- $V_{\rm r}$ plane.  One thus needs to be very careful when studying the metal-poor populations 
using this disk-star dominant sample. Comparison of [Fe/H] values derived with the weighted-mean 
and KPCA methods may help identify potential contaminators. Exercise shows that one excludes 
stars whose KPCA [Fe/H] estimates are smaller than the weighted-mean values by 0.3\,dex or 
replace the estimates by the latter values, the artificial populations of cold, metal-poor 
stars in the [Fe/H] -- $V_{\rm r}$ plane is largely disappears.  

Fig.\,9 also shows that more metal-poor stars generally have higher [$\alpha$/Fe] values. 
This is true for both dwarfs and giants, a natural consequence of the Galactic chemical evolution. 
For both dwarfs and giants, [$\alpha$/Fe] values for stars of solar metallicity are close to zero. 
Nevertheless, [$\alpha$/Fe] values of dwarfs and giants show different patterns. Dwarfs cover 
a wider range of [$\alpha$/Fe] values than giants. There are very few giant stars of ${\rm [\alpha/Fe]} < -0.2$\,dex 
or ${\rm [\alpha/Fe]} > 0.3$\,dex. In contrast, there are quite a number of dwarfs of 
${\rm [\alpha/Fe]} < -0.3$\,dex or ${\rm [\alpha/Fe]} >0.4$\,dex. 
Moreover, besides the dominant thin disk sequence in the [$\alpha$/Fe] -- [Fe/H] plane, giants exhibit also 
an extra, weaker sequence of thick disk stars of a [$\alpha$/Fe] value of $\sim$0.2\,dex. 
This thick disk sequence is not seen in dwarfs. The larger random errors (Fig.\,13) of [$\alpha$/Fe] 
estimates for dwarfs are definitively an important reason for the absence of this sequence. 
Also, giants probe a larger volume than dwarfs as they are brighter. Thus the giant sample 
contains more thick disk stars, whereas the dwarf sample is dominated by the local thin disk stars and 
young, hotter stars in the outer thin disk. In addition, and maybe more importantly, since different algorithms, 
spectral templates and wavelength ranges are used for [$\alpha$/Fe] estimation for dwarfs and giants, 
and the results have not been calibrated externally using data for example from 
high resolution spectroscopy, systematic differences and errors may well be hided in the [$\alpha$/Fe] 
estimates of dwarfs and/or giants. A calibration of [$\alpha$/Fe] estimates is an essential 
and urgent task for the future.      

Fig.\,10 plots distributions of stars in the $T_{\rm eff}$ -- log\,$g$,  $T_{\rm eff}$ -- ${\rm M}_V$ 
and ($g-r$, $r$) planes colour-coded by [Fe/H]. The figure shows that amongst the cool stars, as expected, 
metal-rich ones generally have lower temperatures than those metal-poor ones. However, for hot stars, the 
derived metallicities show some strange patterns in the $T_{\rm eff}$ -- log\,$g$ plane that are not fully understood yet. 
A possible cause could be the inadequacy of parameter coverage of the MILES 
spectral library at higher temperatures. A further, detailed examination 
of log\,$g$ estimates utilizing LAMOST-TGAS common stars, in particular for hot stars, is underway. 
On the whole, the trends of [Fe/H] seen in the $T_{\rm eff}$ -- ${\rm M}_V$ and the ($g-r$, $r$) planes 
seem to be reasonable. The brighter side of the ($g-r$, $r$) diagram is dominated by metal-rich stars 
as they represent a local, thin disk sample. On the other hand, at the fainter side of the diagram, one sees 
metal-rich stars are in general redder while the metal-poor ones are bluer. Again, this is what one would expect.
 
\begin{table*}
\centering
\caption{Adopted methods and their effective parameter ranges for stellar parameter determinations.}
\label{}
\begin{tabular}{lll}
\hline
 Parameter              & Method                                           & Effective range            \\ 
\hline
Effective temperature      \\
$T_{\rm eff}$\_1   &  Weighted-mean template matching                                   &  All \\
$T_{\rm eff}$\_2   &  KPCA with the MILES training set                                       &  $T_{\rm eff}\_1 < 10000$\,K \\
$T_{\rm eff}$  (recommended)   &  $T_{\rm eff}$\_1            & All    \\
 & & \\
 Surface gravity   \\
Logg\_1   &  Weighted-mean template matching                                                                 &  All \\
Logg\_2   &  KPCA with the MILES training set                                       &   $T_{\rm eff}\_1 < 10000$\,K, ${\rm Logg\_1} > 3.0$\,dex \\
Logg\_3   &  KPCA with the LAMOST-$Kepler$ training set                      &  $T_{\rm eff}\_1 < 5600$\,K, ${\rm Logg\_1} < 3.8$\,dex or ${\rm Logg\_2} < 3.8$\,dex\\
Logg (recommended)  &  Logg\_1                                                              & $T_{\rm eff}\_1 > 10000$\,K  \\
         &                         Logg\_3                                                              &  $T_{\rm eff}\_1 < 5600$\,K, ${\rm Logg\_1} < 3.8$\,dex or ${\rm Logg\_2} < 3.8$\,dex \\
         &                         Logg\_2                                                              &   Otherwise  \\
 & & \\
 Metallicity  \\
 {[Fe/H]\_1}      &  Weighted-mean template matching                                                          & All   \\
 {[Fe/H]\_2}      & KPCA with the MILES training set                                & $T_{\rm eff}\_1 < 10000$\,K  \\
 {[Fe/H]\_3}      & KPCA with the LAMOST-APOGEE training set          & $T_{\rm eff}\_1 < 5600$\,K, ${\rm Logg\_1} < 3.8$\,dex or ${\rm Logg\_2} < 3.8$\,dex \\
 {[Fe/H]} (recommended)      & [Fe/H]\_1                                                    & $T_{\rm eff}\_1 > 10000$\,K \\
                                          & [Fe/H]\_2                                                    & Otherwise  \\
  & & \\
  $\alpha$-element to iron abundance ratio  \\
  {[$\alpha$/Fe]\_1}               & Template matching  using spectra of  & $4000 < T_{\rm eff}\_1 < 8000$\,K   \\
                                              &  3900 -- 3980, 4400 -- 4600 and 5000 -- 5300\,{\AA}     &   \\
  {[$\alpha$/Fe]\_2}               & Template matching using spectra of   & $4000 < T_{\rm eff}\_1 < 8000$\,K   \\
                                              & 4400 -- 4600 and 5000 -- 5300\,{\AA}                                        & \\
  {[$\alpha$/Fe]\_3}               & KPCA with the LAMOST-APOGEE training set  & $T_{\rm eff}\_1 < 5600$\,K, ${\rm Logg\_1} < 3.8$\,dex or ${\rm Logg\_2} < 3.8$\,dex \\
   {[$\alpha$/Fe]} (recommended)  &  [$\alpha$/Fe]\_3  & $T_{\rm eff}\_1 < 5600$\,K, ${\rm Logg\_1} < 3.8$\,dex or ${\rm Logg\_2} < 3.8$\,dex \\
                                                   &  [$\alpha$/Fe]\_1  & Otherwise \\ 
    & & \\
   Other elemental abundances  \\
  {[M/H]}         & & \\
  {[$\alpha$/M]}      & KPCA with the LAMOST-APOGEE training set  & $T_{\rm eff}\_1 < 5600$\,K, ${\rm Logg\_1} < 3.8$\,dex or ${\rm Logg\_2} < 3.8$\,dex \\ 
  {[C/H]}  \\
  {[N/H]}  \\                    
  & & \\
  Absolute magnitudes             \\
  ${\rm M}_V$\_1    & KPCA with the LAMOST-Hipparcos training set  &  \\  
  ${\rm M}_{K_{\rm s}}$\_1             &  using 300 PCs &  $T_{\rm eff}\_1 < 12000$\,K \\
   ${\rm M}_V$\_2    & KPCA with the LAMOST-Hipparcos training set &  \\  
  ${\rm M}_{K_{\rm s}}$\_2             &  using 100 PCs &  $T_{\rm eff}\_1 < 12000$\,K \\
   ${\rm M}_V$  (recommended)  &  ${\rm M}_V$\_1 &  \\  
  ${\rm M}_{K_{\rm s}}$  (recommended)             & ${\rm M}_{K_{\rm s}}$\_1 &  $T_{\rm eff}\_1 < 12000$\,K \\
\hline
\end{tabular}
\end{table*}

\begin{figure*}
\centering
\includegraphics[width=160mm]{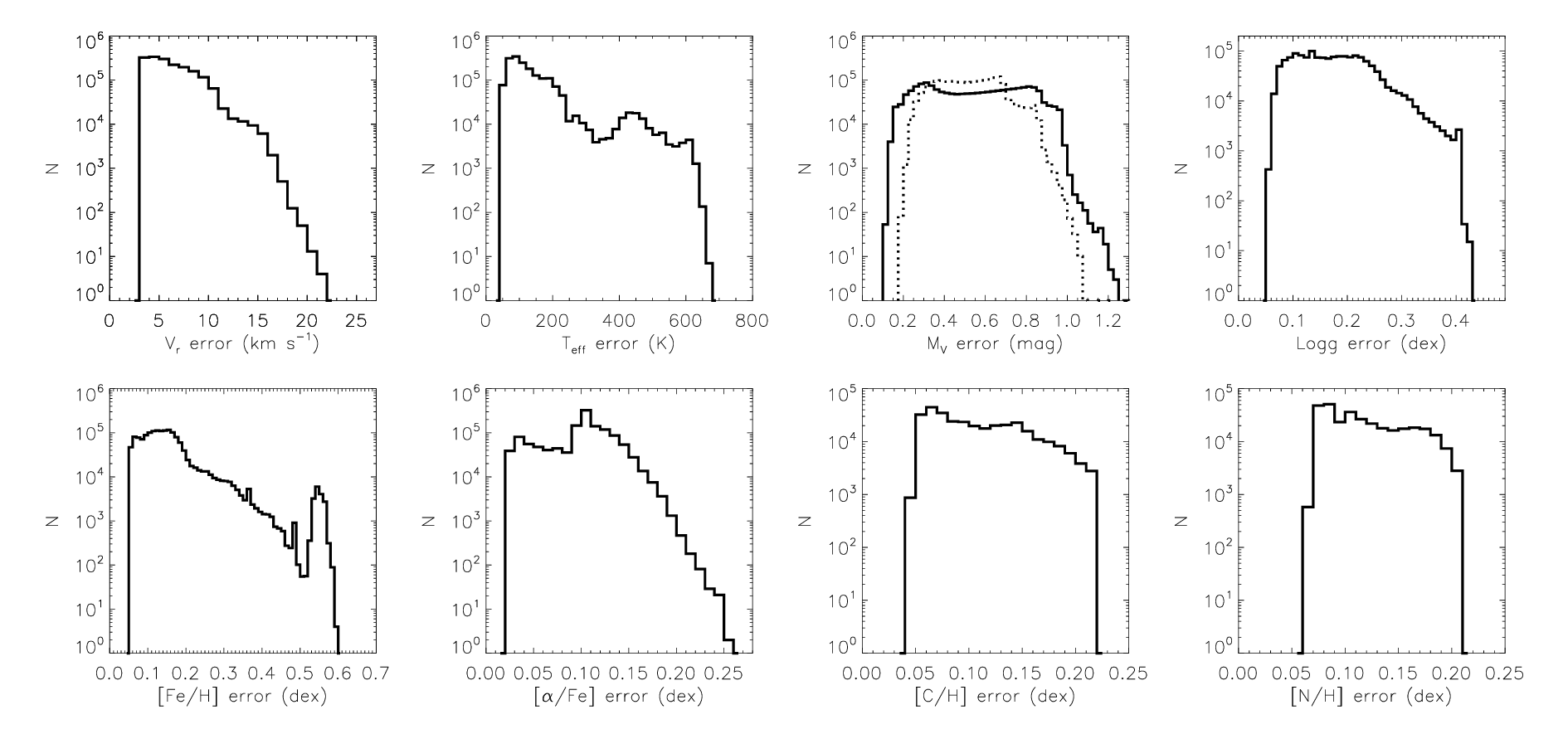}
\caption{Error distributions of recommended parameter estimates. For ${\rm M}_V$, 
error distribution of the other set of ${\rm M}_V$ estimates deduced using 100 PCs is shown by dotted line.}
\label{Fig11}
\end{figure*}
Fig.\,11 plots error distributions of recommended parameter estimates. 
Spectral SNR is the main factor that determines the amount of parameter error, and 
stars of different spectral types have also different parameter errors. A detailed analysis of how the parameter 
errors depends on spectral SNR and stellar parameters can be found in \citet{Xiang+2015a, Xiang+2016}.  
Most of the stars have a radial velocity error of a few km\,s$^{-1}$. However, some of them, 
mostly hot stars of low SNRs, have errors reaching 20\,km\,s$^{-1}$. Temperature errors are 
around 100\,K for most stars, but those of hot ones reach 400 -- 600\,K. 
The large temperature errors for hot stars are mainly caused by the small number and 
large parameter uncertainties of hot stars in the MILES library. Similar trend is seen in [Fe/H] errors, 
where most stars have an error between 0.1 and 0.2\,dex. However, the error distribution exhibits 
a second peak at 0.5--0.6\,dex, mainly contributed by hot stars. 
Errors of log\,$g$ typically have values of 0.1--0.2\,dex, with a small 
fraction reaching 0.3--0.4\,dex. Errors of ${\rm M}_V$ can be as small as 0.2\,mag 
at high spectral SNRs, but reach 0.8 -- 1.0\,mag at low SNRs ($\sim$10). 
${\rm M}_V$ values estimated using 100 PCs have larger errors compared to those 
deduced using 300\,PCs, but they are less sensitive to SNR, and have errors of 0.6 -- 0.8\,mag 
at low SNRs ($\sim$10). Errors of ${\rm M}_{K_{\rm s}}$ have similar trends. 
Errors in [$\alpha$/Fe] show two populations. The lower one is of giants, which 
has values of 0.02--0.09\,dex. The higher one is of dwarfs, and has values of 
0.09--0.25\,dex, peaking around 0.1\,dex. Note that as discussed above, 
since the results are not calibrated by independent, external datasets, there could 
be systematic uncertainties in [$\alpha$/Fe] estimates that have not been included 
in the current error estimates. This is particularly true for giant stars, for which the quoted 
errors are very small. Errors of [C/H] and [N/H] estimates range from 0.05 to 0.20\,dex, 
with most of them smaller than 0.15\,dex.

\begin{figure*}
\centering
\includegraphics[width=160mm]{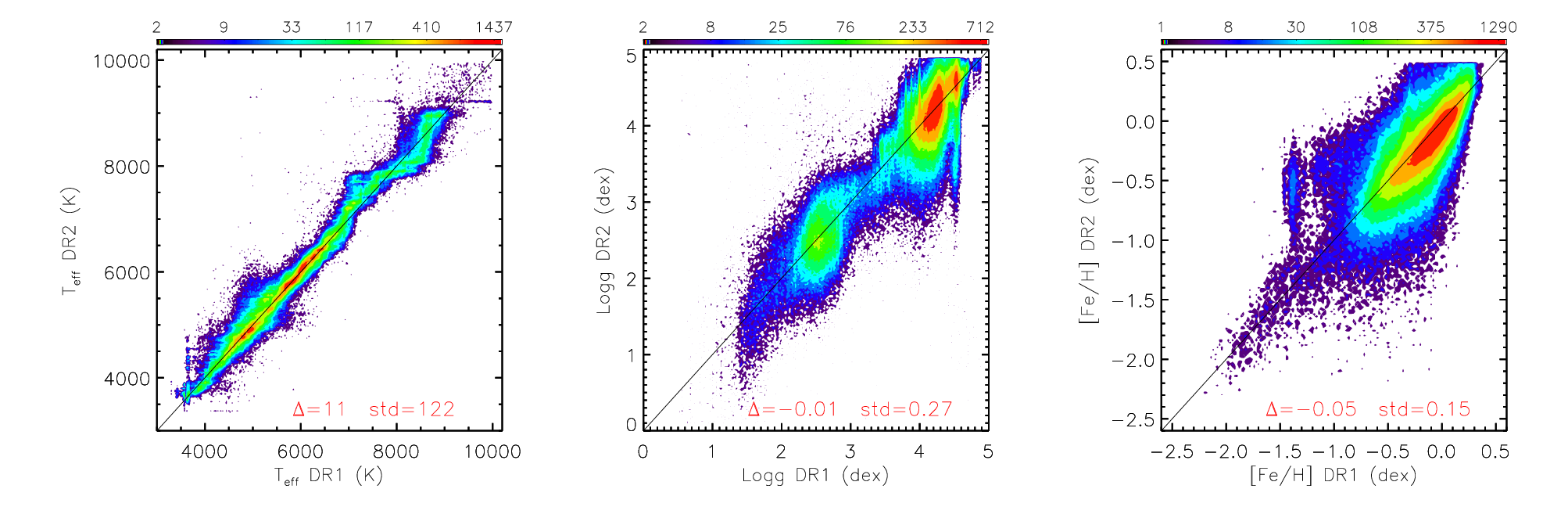}
\caption{Comparison of the new recommended stellar atmospheric parameters with those of 
LSS-GAC DR1. Mean and standard deviations of the parameter differences between DR2 and DR1 
are marked in the plot.}
\label{Fig12}
\end{figure*}
A comparison of recommended stellar atmospheric parameters with LSS-GAC DR1 for the whole 
LSS-GAC DR1 sample is shown in Fig.\,12. The figure shows that $T_{\rm eff}$ estimates from 
the two data releases are quite consistent. For log\,$g$, the plot shows significant patterns. Similar patterns 
are seen by comparing LSS-GAC DR1 log\,$g$ estimates with asteroseismic measurements \citep{Ren+2016}.  
The causes of such patterns are discussed in \citet{Xiang+2016}, and are essentially 
due to the inadequacies of log\,$g$ estimates in LSS-GAC DR1 ones. 
On the whole, new [Fe/H] estimates show good correlation with those of LSS-GAC DR1. The standard deviation 
of the two sets of [Fe/H] estimates of the whole sample is 0.15\,dex. However, the current estimates are  
$0.05$\,dex systematically lower than LSS-GAC DR1. There is also a group of stars whose new [Fe/H] values 
are significantly higher than those of LSS-GAC DR1. In addition, the sharp boundary seen in the LSS-GAC DR1 
results now disappears. Again, the causes of these differences are discussed in \citet{Xiang+2016},  
and are mainly due to the inadequacies of [Fe/H] estimates in LSS-GAC DR1.

\section{repeat observations}

\begin{figure}
\centering
\includegraphics[width=85mm]{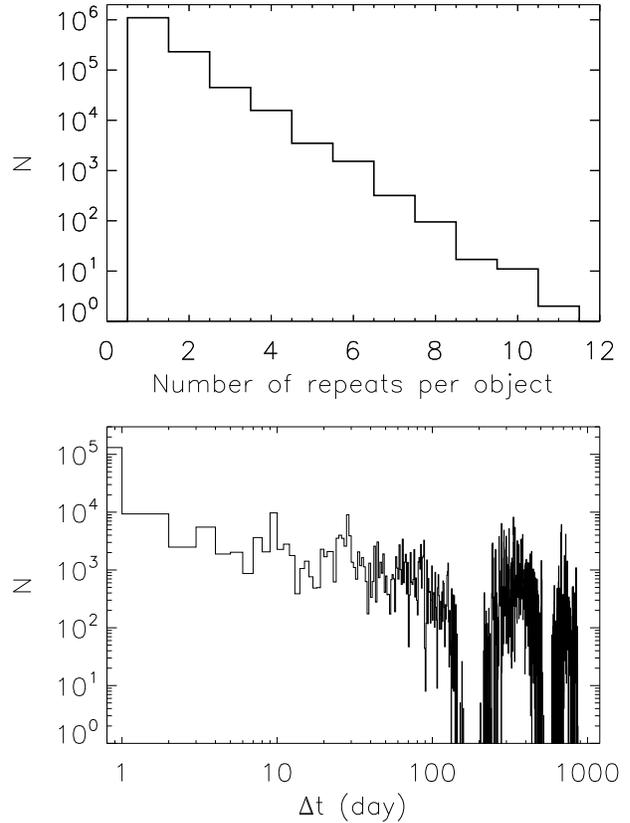}
\caption{$Upper$: Stellar numbers as a function of number of repeat observations per object. 
$Lower$: Distribution of time separation for repeat observations.}
\label{Fig13}
\end{figure}
About 28 per cent of stars in the value-added catalogues have repeat observations. 
The repeat observations are valuable for various purposes, including (1) 
time-domain spectroscopy; (2) robustness check of parameter estimates; (3) 
improvement of stellar parameter estimates by averaging repeat measurements, etc. 
Fig.\,13 plots the distribution of number of stars as a function of number of repeat observations 
per object, as well as the distribution of time separations of repeat observations. 
Stellar number decreases approximately 
linearly with increasing number of repeat observations. There are 95350 unique stars that 
are observed 3 times or more, and 5460 unique stars observed 5 times or more. 
Time separation covers a range from a few hours to nearly 
1000 days. Most repeat observations occurred in the same night, when different plates 
that shared the same central bright stars or adjacent plates that had overlapping fields of views 
were observed. Note that to better present the data, the horizontal axis of the figure has been  
chosen to artificially start from 0.8 days, but the real time separation in the first bin was smaller than 4 hours 
due to hour angle limitations of LAMOST \citep{Cui+2012}. Note also 
that here repeat observations refer to those taken with different plates. In fact, each 
plate observed usually has 2--3 exposures. The current analysis is based on co-added spectra of 
those adjacent individual exposures. Further, independent treatment of the individual exposures 
of each plate observed in the future will significantly increase the repeat observation rate 
in the sense that every star targeted has at least one repeat observation, separated by tens of minutes.

\begin{figure}
\centering
\includegraphics[width=85mm]{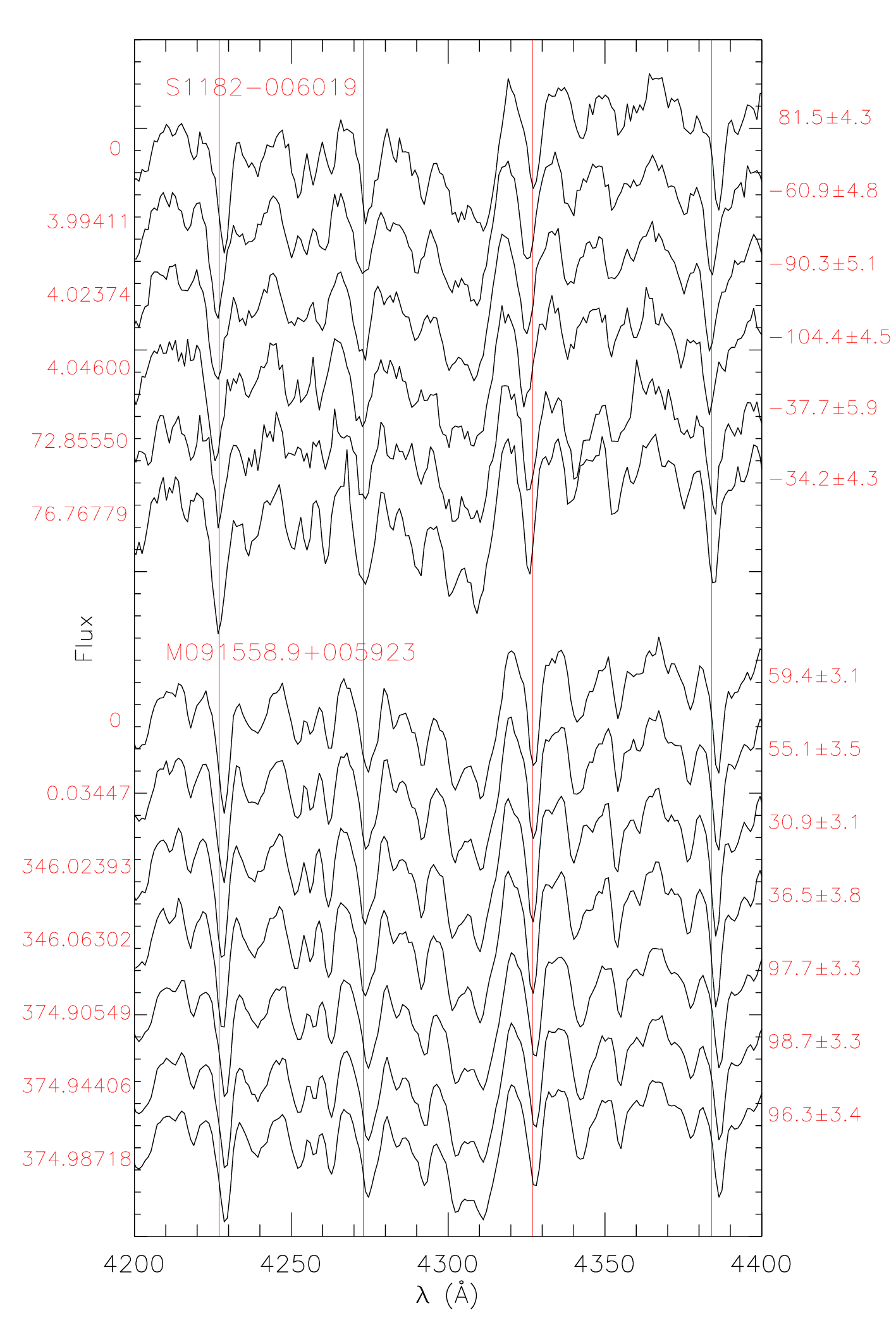}
\caption{Examples of spectra of wavelength range 4200--4400{\AA} of two stars with repeat observations. 
Numbers on the left are JD offsets respect to the first observation, while numbers on the right 
are radial velocity measurements. Object ID from the input catalogues are marked on the plot.}
\label{Fig14}
\end{figure}

Some systematic studies utilizing repeat observations are underway. As an example, Fig.\,14 
shows the spectra in the 3900--4400{\AA} wavelength range of two stars with repeat observations. 
The data clearly show shifts of spectral line positions, thus radial velocity variations. 
The first example has an amplitude of radial velocity variations 
larger than 180\,km\,s$^{-1}$. The offsets of repeat observations from the first one are respectively 
3.99411, 4.02374, 4.04600, 72.85550 and 76.767793 days. Effective temperatures estimated 
from the six observations are respectively 4931, 5192, 5150, 5182, 5507 and 4953\,K, 
${\rm M}_V$ absolute magnitudes 6.10, 5.89, 5.82, 5.43, 5.42 and 6.56\,mag, and [Fe/H] metallicities  
$-0.22$, $-0.33$, $-0.24$, $-0.34$,  $-0.15$ and $-0.07$\,dex. The second example star has a 
radial velocity variation amplitude larger than 60\,km\,s$^{-1}$. The offsets of repeat observations 
from the first one are respectively 0.03447, 346.02393, 346.06302, 374.90549, 374.94406 and 374.98718 days. 
$T_{\rm eff}$ estimates from the seven observations are respectively 5349, 5340, 5046, 5258, 5512, 5512 and 5526\,K, 
the ${\rm M}_V$ absolute magnitudes 4.47, 4.29, 4.62, 4.49, 4.74, 4.79 and 4.76\,mag, and [Fe/H] 
metallicities $-0.11$, $-0.05$, 0.03, 0.00, $-0.04$, $-0.03$ and $-0.04$\,dex. It seems that 
in additions to radial velocity variations, the two stars also exhibit significant variations in effective 
temperature and absolute magnitude. Further analyses are however required to reveal the nature of the variations. 

\begin{figure*}
\centering
\includegraphics[width=160mm]{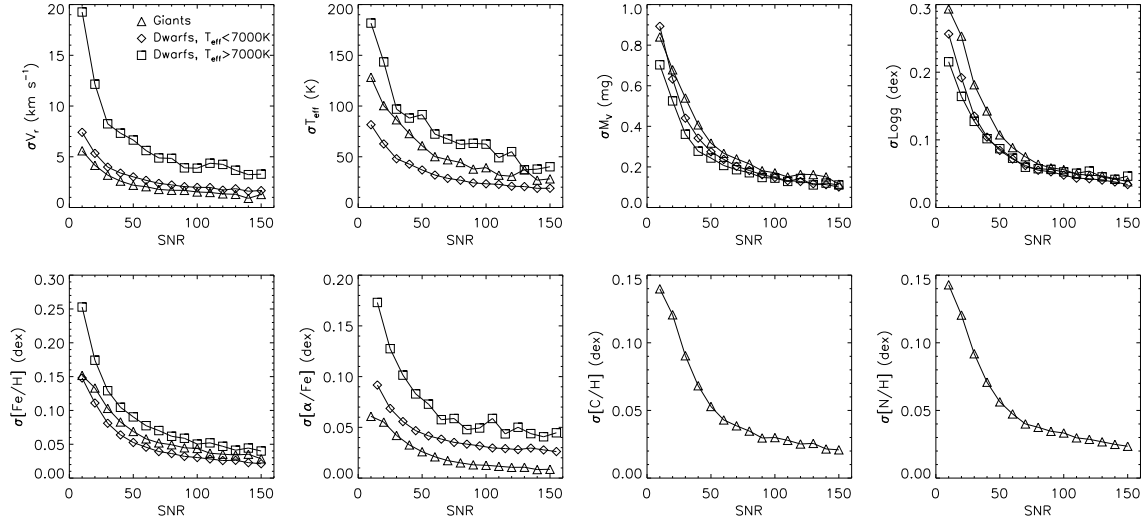}
\caption{Random errors of recommend parameters as a function of spectral SNR 
as deduced by comparing results from repeat observations. Results for giants, hot and cool dwarfs are 
shown separately.}
\label{Fig15}
\end{figure*}

A detailed robustness examination on the LSP3 stellar parameters using repeat observations 
can be found in \citet{Xiang+2015a} for radial velocity measurements and atmospheric parameter 
determinations with the weighted-mean method, and in \citet{Xiang+2016} for atmospheric parameters 
estimated with the KPCA method. Fig.\,15 plots the dispersions of differences of recommended parameters 
deduced from repeat observations that have comparable SNRs (20 per cent) as a function of SNR. 
The dispersions represent random errors of the parameters induced by spectral noises. 
For this purpose, values of dispersions shown in the figure have been divided 
by square root of 2. Results for giants, hot ($T_{\rm eff}$ > 7000\,K) and cool ($T_{\rm eff}$ < 7000\,K) 
dwarfs are shown separately. Generally, hot dwarfs have larger random errors in $V_{\rm r}$, 
$T_{\rm eff}$, [Fe/H] and [$\alpha$/Fe] estimates. For all parameters, the random errors are 
quite sensitive to SNR. Given a SNR higher than 100, random errors of radial velocities 
induced by spectral noises for cool stars can be as small as 1--2\,km\,s$^{-1}$, and 
the values are 3--4\,km\,s$^{-1}$ for hot stars. For a SNR of 10, the errors increase to 5--10\,km\,s$^{-1}$ 
for cool stars and 20\,km\,s$^{-1}$ for hot stars. Note that the final error estimates of $V_{\rm r}$ 
have been set a minimum value of 3\,km\,s$^{-1}$ to account for the potential systematic uncertainties, 
which is $\sim$2\,km\,s$^{-1}$ as estimated via external examinations using the LAMOST-APOGEE 
common stars. As introduced in Section 4.4, for all stellar parameters, these 
random errors and their trends of variations with spectral SNR, $T_{\rm eff}$ and [Fe/H] have been 
incorporated into the final error estimates, which include also uncertainties 
induced by the inadequacies of the parameter estimation methods for both giants and dwarfs.

\section{extinction \& distance}
\subsection{Extinction}
Similar to LSS-GAC DR1 \citep{Yuan+2015a}, various techniques are employed to estimate 
the interstellar reddening toward the individual stars. 
The methods include the star-pair technique, comparing observed colours 
with synthetic ones from stellar model atmospheres, as well as a method based on the stellar colour loci. 

The star-pair method assumes that stars with identical stellar atmospheric parameters 
($T_{\rm eff}$, log\,$g$, [Fe/H]) have the same intrinsic colours. Thus for a reddened star, 
its intrinsic colours can be inferred from its pairs/counterparts with either nil or well known extinction \citep{Yuan+2015a}. 
The method is straightforward, and moreover, free from stellar model atmospheres.
To derive extinction with this method, a control sample with nil or well known 
extinction is inherited from LSS-GAC DR1.
The control sample contains low extinction stars whose $E(B-V)$ values can be well approximated by the reddening 
map of \citet[hereafter SFD98]{Schlegel+1998}. For a given target star, the 
$E(B-V)$ is then derived by fitting $g-r$, $r-i$, $i-J$, $J-H$ and $H-K_{\rm s}$ 
optical to the infrared colours simultaneously by the corresponding values of the pair stars in the control sample, 
assuming the universal extinction coefficients of \citet{Yuan+2013}. When available, $g,r,i$-band 
magnitudes from the XSTPS-GAC survey \citep{Zhang+2013, Zhang+2014, Liu+2014} are used, 
otherwise they are taken from the APASS survey \citep{Munari+2014}, or from the SDSS survey \citep[DR8;][]{Aihara+2011} 
for a few B/M/F plates of high Galactic latitudes. $J,H,K_{\rm s}$ magnitudes are from the 2MASS survey 
\citep{Skrutskie+2006}. There are a few ($\sim$\,3 -- 5 per cent) stars for which different colours 
yield quite different reddenings. For those stars, either the photometry or the stellar atmospheric parameters,  
or both, are problematic. It is found that in such cases, $E(B - V)$ is often overestimated by a significant amount. 
To account for this, if $g-r$ and $r-i$ colours of a given star yield reddening of different
signs and yield values differed by more than 0.2\,mag, the resultant $E(B-V)$ estimate is discarded. 
The corresponding entry in the value-added catalogues are set to $-9$. 
Comparison with the SFD98 reddening map for high Galactic latitude regions of low extinction illustrates 
a typical uncertainty of 0.036\,mag for $E(B-V)$ thus estimated (Fig.\,16), which is similar 
to that of LSS-GAC DR1. 
Note that we use the recommended atmospheric parameters for reddening estimation. 
Different results are expected if different sets of atmospheric parameters are adopted. 
Nevertheless, given the good precision of the current parameter estimates, the differences are small. 
For example, $E(B-V)$ values derived using atmospheric parameters estimated with the weighted-mean 
algorithm differ from the default results by a dispersion of only $\sim$0.01\,mag. 
$E(B-V)$ thus estimated is denoted `EBV\_SP' in the value-added catalogues.

$E(B-V)$ values have also been determined by comparing photometric colours with synthetic ones. 
In doing so, a grid of synthetic colours is first 
constructed by convolving the synthetic spectra of \citet{Castelli+2004} with the 
transmission curves of filters for the photometric systems of SDSS\footnote{http://classic.sdss.org/dr7/instruments/imager/index.html\#filters} \citep{Gunn+1998, Doi+2010} 
and 2MASS\footnote{http://www.ipac.caltech.edu/2mass/releases/second/doc/sec3\_1b1.html} \citep{Milligan+1996}.  
Synthetic colours for a star of given atmospheric parameters $T_{\rm eff}$, 
log\,$g$ and [Fe/H] are then deduced by linearly interpolating the grid. 
$E(B-V)$ is then estimated as the weighted mean of values  
inferred from the individual colours, weighted by the reddening coefficients of \citet{Yuan+2013}. 
As Fig.\,16 shows, $E(B-V)$ values thus derived have a precision of about 
0.03\,mag, but are found to be systematically lower than the values given by either 
the star-pair method or by the SFD98 map, by 0.01 -- 0.02\,mag. 
The underestimation is likely to be caused by the slight different temperature scales 
between the LSP3 and Kurucz atmospheric models. As pointed out above, 
LSP3 temperatures are calibrated to the metallicity-dependent colour-temperature relations  
of \citet{Huang+2015b}, constructed using a sample of stars with $T_{\rm eff}$  
estimated directly from angular diameters measured with interferometry and the trigonometric parallaxes 
from the Hipparcos satellite. It is found that the relations of \citet{Huang+2015b} yield temperatures 
generally $\sim$100\,K lower than earlier results in the literature. 
$E(B-V)$ estimated with this method is denoted `EBV\_MOD' in the catalogues. 

Finally, extinction values deduced from multi-band photometry by fitting the stellar colour loci \citep{Chen+2014} 
are also included in the catalogues. Results from this method 
are currently only available for stars within the footprint and magnitude range of XSTPS-GAC, 
i.e. for stars targeted by the LSS-GAC main and M31/M33 surveys. As shown by \citet{Yuan+2015a}, 
the method yields an $E(B-V)$ precision of about 0.07\,mag and works best 
for dwarf stars. $E(B-V)$ thus estimated is denoted `EBV\_PHOT' in the catalogues.

Considering that the star-pair method is model free, and often achieves a better precision 
compared to other methods, $E(B-V)$ yielded by the method (`EBV\_SP') are generally 
adopted as the recommended values. 
Considering that there are few hot stars in the control sample used by the star-pair method, 
`E(B-V)\_MOD' is adopted as the recommended for stars hotter than 9000\,K. Since the SFD98 map 
is a two dimensional one, it provides an upper limit of $E(B-V)$ for reddening towards the individual 
Galactic stars. Based on this consideration, if the $E(B-V)$ value of a star deduced from the star-pair 
method or from the method based on model atmospheres excess the SFD98 value by 0.1\,mag ($\sim3\sigma$), 
the latter is adopted as the recommended value. The distribution of adopted E(B-V) values 
for all stars in LSS-GAC DR2 are shown in the bottom-right panel of Fig.\,16.  
With $E(B-V)$ determined, values of interstellar extinction in the individual photometric bands 
can be easily computed using the extinction coefficients available from, for example, \citet{Yuan+2013}.     
\begin{figure}
\centering
\includegraphics[width=85mm]{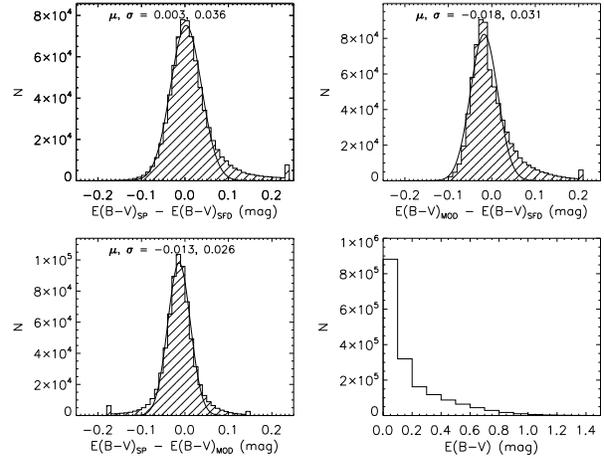}
\caption{Distributions of differences of $E(B-V)$ estimated by different methods or 
from the SFD98 map except for the bottom-right panel where the distribution is for 
the adopted values of $E(B-V)$ for all stars in LSS-GAC DR2. 
$E(B-V)_{\rm SP}$ and $E(B-V)_{\rm MOD}$ denote respectively values derived from 
the star-pair and stellar model atmosphere methods, whereas $E(B-V)_{\rm SFD}$ 
refers to values extracted from the 2-dimensional reddening map of \citet{Schlegel+1998}. 
The comparisons are for stars of $|b|>15^\circ$ and $E(B-V)_{\rm SFD}<0.3$\,mag only.}
\label{Fig16}
\end{figure}

\subsection{Distance}

\begin{figure*}
\centering
\includegraphics[width=160mm]{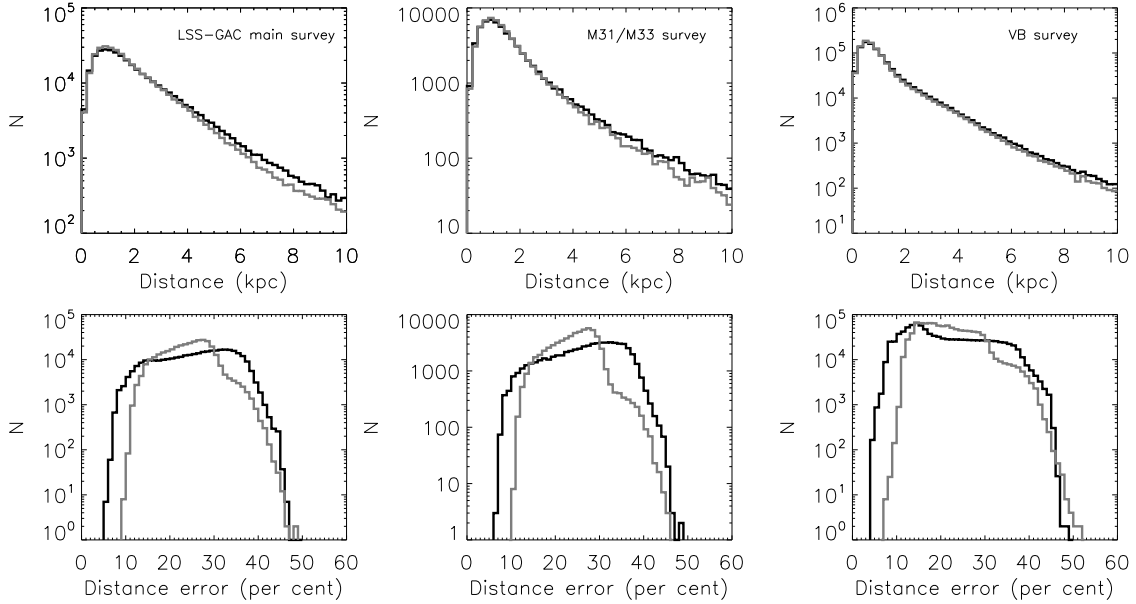}
\caption{Distributions of distances and errors of stars targeted by the 
LSS-GAC main (left), M31/M33 (middle) and VB (right) surveys. Black and grey lines 
represent distances inferred from absolute magnitudes derived from LAMOST spectra 
with the KPCA method using 300 and 100 PCs, respectively. }
\label{Fig17}
\end{figure*}
\begin{figure}
\centering
\includegraphics[width=85mm]{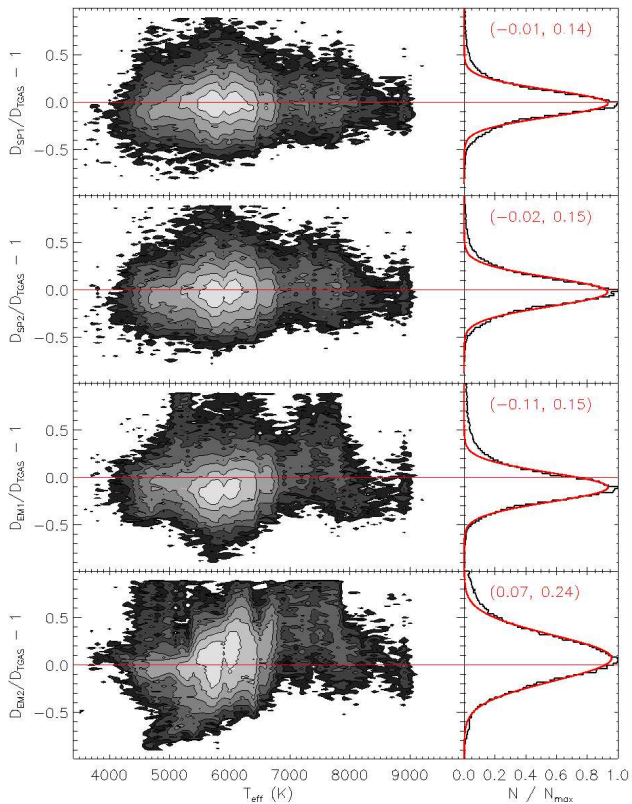}
\caption{Percentage differences of distances between our estimates and those inferred from 
Gaia TGAS parallax as a function of the recommended $T_{\rm eff}$ for a sample of 100,000 
common stars. For up to low, the four panels show results for our distance estimates inferred from absolute 
magnitudes derived from LAMOST spectra using 300 PCs (first panel), 100 PCs (second panel) 
and from absolute magnitudes estimated using empirical relations utilizing the recommended parameters 
(third panel) and the weighted-mean parameters (fourth panel). An error cut of 0.3\,mag 
in the TGAS-based magnitudes is applied.}
\label{Fig18}
\end{figure}

\begin{figure*}
\centering
\includegraphics[width=160mm]{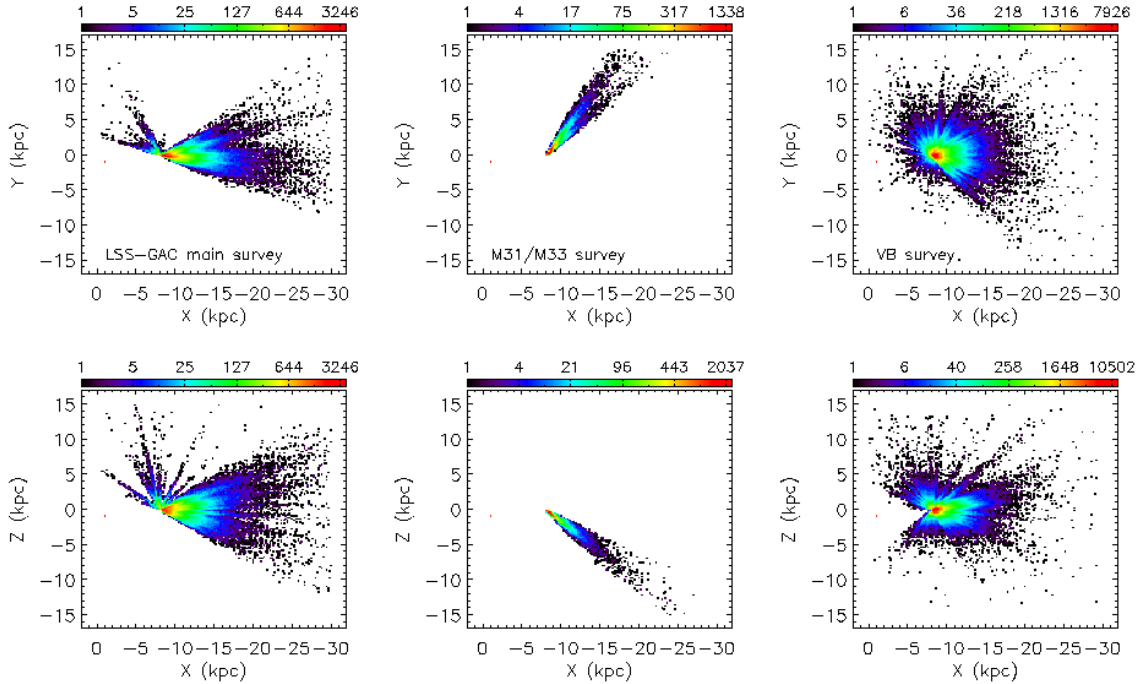}
\caption{Colour-coded stellar density distributions in the Galactic $X-Y$ and $X-Z$ planes for stars 
targeted by the LSS-GAC main (left), M31/M33 (middle) and VB (right) surveys.}
\label{Fig19}
\end{figure*}

Distance to individual stars are also estimated with a variety of methods. 
One is to infer from the absolute magnitudes derived directly from the LAMOST spectra with the KPCA  
method using the LAMOST-Hipparcos training set. 
In doing so, if photometric magnitudes of all $g, r$ and $K_{\rm s}$ bands have an error 
smaller than 0.1\,mag, and are not saturated, then both ${\rm M}_V$ and ${\rm M}_{K_{\rm s}}$  
are used to calculate a weighted-mean distance by taking the inverse squared magnitude 
errors as weights, otherwise only bands with good photometry are utilized to infer the distance.   
Generally, $K_{\rm s}$-band contributes a larger weight as it suffers from much less severe extinction 
than $V-$band. The $V-$band photometry is transformed from $g$ and $r$  
magnitudes using the equation of \citet{Jester+2005}. Errors of absolute magnitudes, extinction 
and photometry are all incorporated in the estimation of distance errors. 
Note that it is found that the estimated ${\rm M}_V$ and ${\rm M}_{K_{\rm s}}$ 
are not independent quantities but largely correlated with each other, so that when estimating 
errors for the weighted-mean distance, we simply adopt the weighted-mean distance errors 
in the individuals bands rather than infer from the reduced variance as done for independent 
measurements. This may have overestimated the distance errors to some extent.
Distance estimated with this method is denoted as `DIST\_SP1' and `DIST\_SP2' in the 
value-added catalogues for results inferred from absolute magnitudes using 300 and 100 PCs, respectively.

Another method to estimate stellar distance is based on empirical relations between 
the absolute magnitudes and stellar atmospheric parameters constructed utilizing the MILES 
stars. A detailed description of the derivation of the empirical relations is presented in 
\citep{Yuan+2015a}. Here since the atmospheric parameters of MILES stars 
have been re-determined/calibrated (cf. \S{4.1}), the relations are updated accordingly. The new 
relations have much smaller residuals --- with dispersions of $g$ (and $r$) band absolute 
magnitudes of only about 0.08, 0.13, 0.15 and 0.20\,mag for OBA stars, FGK dwarfs, 
KM dwarfs and GKM giants, respectively. The corresponding residuals in the 2MASS infrared 
bands are about 0.15, 0.26, 0.06 and 0.26\,mag. The small residuals suggest 
that the method is robust. Errors of absolute magnitudes thus derived for the individual LSS-GAC stars 
are mainly contributed by uncertainties of the stellar atmospheric parameters. 
Here both the recommended atmospheric parameters and atmospheric parameters yielded 
by the weighted-mean algorithm are used to estimate absolute magnitudes. 
To obtain realistic error estimates of absolute magnitudes, we adopt a Monte-Carlo strategy 
to propagate the errors of stellar atmospheric parameters. The distance errors are then estimated 
taking into account contributions by errors in absolute magnitudes, extinction and photometry. 
Distances estimated with this method using the recommended parameters are denoted as 
`DIST\_EM1' in the value-added catalogues, and those using the weighted-mean atmospheric parameters 
are denoted as `DIST\_EM2'. In addition, stellar distances derived from photometry by \citet{Chen+2014} 
via fitting stellar colour loci, if available, are also presented in the value-added catalogues, and are 
denoted by `DIST\_PHOT'.

Estimating distance using absolute magnitudes derived directly from the LAMOST spectra 
does not depend on stellar model atmospheres and stellar atmospheric parameters, 
and thus are expected to suffer from minimal systematics.  
Comparison with distances inferred from the Gaia TGAS parallaxes (Fig.\,8) demonstrates
that the overall systematic errors in our inferred distances are negligible, and the random errors 
can be as small as $\sim$10 per cent given high enough spectral SNRs. 
Uncertainties in absolute magnitude thus distance estimates are sensitive to 
spectral SNRs. For a SNR of 10, the uncertainty increases to 30--40 per cent. 
Fig.\,17 further shows that distance estimates from absolute magnitudes derived using 
both 300 and 100 PCs have essentially negligible (1--2 per cent) systematic errors 
at all $T_{\rm eff}$ in the range of 5000--9000\,K. 
Note that to incorporate stars in a wide temperature range for the comparison, 
especially for the high temperature end, here we have adopted an error cut of 0.3\,mag 
for the TGAS-based magnitudes instead of 0.2\,mag as adopted for Fig.\,8. As a result, 
the dispersion become 14 per cent, which is larger than 12 per cent shown in Fig.\,8. 
While for stars cooler than 5000\,K, our distances seem to be underestimated by 
5--10 per cent. The figure shows also that distances yielded by the empirical relations 
using the recommended parameters are systematically underestimated by about 11 per cent, 
and the systematics show moderate dependence on $T_{\rm eff}$. Such an underestimation 
is probably caused by either hidden systematic errors in the stellar atmospheric parameters, 
especially log\,$g$, or by systematic errors in the empirical relations. 
Note that although the current parameter estimates achieved high precisions, 
potential systematic errors could be existed, which are probably a few tens 
Kelvin in $T_{\rm eff}$, a few per cent to 0.1\,dex in log\,$g$ and a few per cent to 0.1\,dex in [Fe/H]. 
Potential systematic errors for metal-poor stars (${\rm [Fe/H]} < -1.0$\,dex) are likely larger 
and should be treated cautiously. Distances yielded by the empirical relations 
using the weighted-mean parameters are systematically overestimated by 6 per cent, 
and there are strong patterns.

Distances inferred from absolute magnitudes directly determined from the LAMOST spectra 
using 300 PCs (`DIST\_SP1') are currently adopted as the recommended values. It is emphasized 
that users should choose a proper set according to their own problems of interest. 
Fig.\,17 plots the distributions of distances and error estimates for `DIST\_SP1' and `DIST\_SP2' 
for LSS-GAC DR2 stars of the main, M31/M33 and VB surveys, respectively. 
The distances peak at about 1\,kpc, but with a few beyond 10\,kpc. 
For the whole sample, 22, 5 and 2 per cent of the stars have a distance larger than 2, 5 and 
10\,kpc, respectively. The quoted errors have a wide distribution from $\sim$7 to 
$>$40 per cent. For `DIST\_SP1', about 11.5, 25.5, 63.0 and 98.8 per cent of the stars have a quoted  
error smaller than 15, 20 ,30, and 40 per cent, respectively. The numbers are respectively
4.6, 22.9, 87.8 and 99.7 per cent for distances inferred from absolute magnitudes derived using 
100 PCs. Finally, note that as mentioned in Section 4, there are about 18 per cent of stars 
that the KPCA method has failed to provide reliable estimates of absolute magnitudes due to 
low spectral SNRs. For those stars, the recommend distances are not available, 
while as a compromise one may use distance derived with the empirical relations utilizing 
the weighted-mean atmospheric parameters.

Fig.\,19 plots the density distributions of LSS-GAC DR2 stars in the Galactic $X-Y$ and $X-Z$ planes 
for the main, M31/M33 and VB surveys, separately.  
Here the $X$, $Y$ and $Z$ are calcuated assuming $R_\odot=8.0$\,kpc and $Z_\odot=0$\,kpc, 
and the recommended distances are used for the calculation. 
The figure shows that the main survey samples the outer disk well, while the 
VB survey samples the local disk very well.  Some stars are located beyond 20\,kpc in the 
anti-centre direction and some are beyond 10\,kpc in the vertical direction.

\section{Multi-band photometry, proper motion and orbital parameters}
Multi-band photometry from the far-UV to the mid-IR, including magnitudes in $FUV$ and $NUV$ 
bands from the $Galaxy$ $Evolution$ $Explorer$ \citep[GALEX;][]{Martin+2005}, $g$, $r$, $i$ bands 
from the XSTPS-GAC, SDSS and APASS, $J$, $H$, $K_{\rm s}$ bands from 
the 2MASS and $w_1$, $w_2$, $w_3$, $w_4$ bands from the ${\it Wide-field}$
${\it Infrared}$ ${\it Survey}$ ${\it Explorer}$ \citep[WISE;][]{Wright+2010}, are included 
in the value-added catalogues using with a 3 arc-second match radius for source cross-identifications. 

Proper motions for LSS-GAC stars from UCAC4 \citep{Zacharias+2013} 
and PPMXL \citep{Roeser+2010}, as well as those derived by combing the XSTPS-GAC 
and 2MASS catalogues \citep{Yuan+2015a}, are also included in the value-added catalogues. 
 
Given celestial position, distance, radial velocity and proper motions, the stellar orbital parameters, 
including eccentricity, epicentre and pericentre radii, minimum and maximum vertical distances from 
the disk, are computed assuming a simple, axisymmetric Galactic 
potential as presented in Table\,1 of \citet{Gardner+2010}. The Sun is assumed to be located 
at (X, Y, Z) = ($-8$, 0, 0)\,kpc, and the local standard of rest (LSR) has values 
$(U_\odot, V_\odot, W_\odot) =  (7.01, 10.13, 4.95)$\,km\,s$^{-1}$ \citep{Huang+2015c}. 
When calculating the orbital parameters, systematic trends in proper motions with celestial positions 
are corrected for using the relations presented in \citet{Carlin+2013} and in \citet{Huang+2015c} 
for PPMXL and UCAC4 proper motions, respectively. Note that the systematics in proper motions 
have also been found to correlate with colour and magnitudes \citep[e.g.][]{Yuan+2015a, Sun+2015}. 
Such systematics have not been accounted for in the current treatment. 
Fig.\,20 shows the eccentricity distribution of the sample stars.
\begin{figure*}
\centering
\includegraphics[width=160mm]{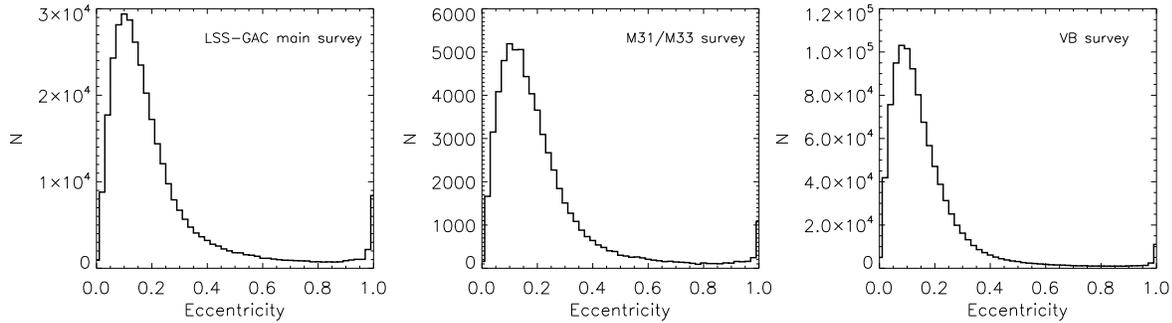}
\caption{Distributions of orbital eccentricities for stars targeted by the LSS-GAC main (left), 
M31/M33 (middle) and the VB (right) surveys.}
\label{Fig20}
\end{figure*}

\section{Format of the value-added catalogues}
Similar to LSS-GAC DR1, this second release of value-added catalogues are stored in three binary 
tables in FITS format. The filenames are `lss-gac\_dr2\_GAC.fits', `lss-gac\_dr2\_M31.fits', `lss-gac\_dr2\_VB.fits', 
corresponding to the LSS-GAC main, M31/M33 and VB surveys, respectively. 
The value-added catalogues include information of the LAMOST observations, stellar parameters 
and error estimates, interstellar reddening, distance, multi-band photometry, 
as well as proper motions and orbital parameters for 1,790,633 observations of 1,393,844 unique stars 
that have a spectral SNR(4650{\AA}) higher than 10. 
A detailed description of the catalogue content is presented in 
Table\,5. Note that compared to LSS-GAC DR1, more informations about the observations and stellar 
parameter estimates are provided for a better description of the data. 
This includes, for example, informations of Julian date at the start of the first exposure, airmass, 
lunar distance, as well as telescope altitude, and so on. 
Stellar type of the best-matching template is also provided, which should be helpful  
for identifying stars of special interest (cf. \S{4.5}). 
The catalogues are accessible from http://lamost973.pku.edu.cn/site/node/4 along with a descriptive
readme.txt file.

\section{Summary}
The current work describes the second release (DR2) of value-added catalogues of LSS-GAC, 
which contain observational information, multi-band photometry, 
radial velocity, stellar atmospheric parameters, elemental abundances, absolute magnitudes, 
extinction, distance, proper motions and orbital parameters for 1,790,633 LAMOST 
spectroscopic observations of 1,393,844 unique stars targeted by the LSS-GAC main, 
M31/M33 and VB surveys between September, 2011 and June, 2014 
that have a spectral SNR (4650{\AA}) higher than 10. 

Spectra used to deliver LSS-GAC DR2 are processed at Peking University 
with the LAMOST 2D pipeline and a spectral flux calibration pipeline specifically developed for LSS-GAC, 
as well as with the LAMOST Stellar Parameter Pipeline at Peking University (LSP3). 
Compared to LSS-GAC DR1, several improvements in stellar parameter determinations 
have been carried out, including (1) More than 260 new spectral templates 
observed with the YAO 2.4\,m and NAOC 2.16\,m telescopes have been added 
to the MILES library used by LSP3; (2) Atmospheric parameters 
of LSP3 template stars have been re-determined/calibrated to reduce the systematic and 
random errors; (3) Stellar atmospheric parameters $T_{\rm eff}$, log\,$g$, [Fe/H], [M/H], 
[$\alpha$/Fe], [$\alpha$/M], absolute magnitudes ${\rm M}_V$, ${\rm M}_{K_{\rm s}}$ and 
elemental abundances [C/H] and [N/H] are deduced from LAMOST spectra with 
a KPCA-based multivariate regression method. The method yields robust results  
 for spectra of good SNRs ($>50$) -- estimates of log\,$g$ are found to be accurate to 0.1\,dex 
 for giants and late-type dwarfs, and estimates of ${\rm M}_V$ and ${\rm M}_{K_{\rm s}}$ are  
accurate to 0.3\,mag, corresponding to a distance accuracy of about 15 per cent. 
Distances inferred from absolute magnitudes determined directly from LAMOST spectra 
using the LAMOST-Hipparcos training set are free from stellar model atmospheres and 
stellar atmospheric parameters and are thus expected to have minimal systematics.  
Estimates of [M/H], [Fe/H], [C/H] and [N/H] for giant stars are shown to have a precision 
better than 0.1\,dex, and those of [$\alpha$/Fe] and [$\alpha$/M] better than 0.05\,dex;   
(4) Estimates of [$\alpha$/Fe] by template matching with the KURUCZ synthetic spectra 
are also provided for both dwarfs and giants; (5) More descriptive critical flags and information  
about the observations and stellar parameter determination are provided to help  
better understand the data and find, for instance, stars of special interest, or to mask 
stars with poorly determined parameters;  
(6) Realistic errors are estimated for atmospheric parameters, 
distance and other inferred quantities (e.g. 3-dimensional positions and velocities). 
Parameter errors are estimated in detail by taking in account of contributions from 
all identified sources, including spectral noise, 
inadequacy of the methods and templates adopted for the analysis, 
and errors propagated from other relevant parameters. 
Errors are assigned to the individual determinations for each star based on the spectral SNR and 
stellar atmospheric parameters.   

Besides the large number of stars, the data sets cover a large and almost contiguous
volume of the Galactic disk and halo, and have a relatively simple target selection function. 
They are thus expected to serve as a unique and valuable asset to study the structure and 
evolution of the Galaxy, especially the disk and the solar neigbourhood. In addition, 
about 28 per cent 
measurements in the value-added catalogues are duplicate results of the same targets. 
The data sets are therefore also useful for time-domain spectroscopic studies.

\vspace{7mm} \noindent {\bf Acknowledgments}{
This work is supported by National Key Basic Research Program of China 2014CB845700 
and Joint Funds of the National Natural Science Foundation of China (Grant No. U1531244 and U1331120).  
Guoshoujing Telescope (the Large Sky Area Multi-Object Fiber
Spectroscopic Telescope LAMOST) is a National Major Scientific
Project built by the Chinese Academy of Sciences. Funding for
the project has been provided by the National Development and
Reform Commission. LAMOST is operated and managed by the National
Astronomical Observatories, Chinese Academy of Sciences. 
We acknowledge the support of the staff of the Xinglong 2.16m telescope 
and the Lijiang 2.4m telescope. This work was partially Supported by the 
Open Project Program of the Key Laboratory of Optical Astronomy, 
National Astronomical Observatories, Chinese Academy of Sciences.
H.-B. Yuan is supported by NSFC grant No. 11443006, 
No. 11603002 and Beijing Normal University grant No. 310232102.
This work has made use of data products from the Galaxy Evolution
Explorer (GALEX), Xuyi Schmidt Telescope Photometric Survey of the
Galactic Anticentre (XSTPS-GAC), Sloan Digital Sky Survey (SDSS), 
AAVSO Photometric All-Sky Survey (APASS), Two Micron All Sky Survey (2MASS), 
Wide-field Infrared Survey Explorer (WISE), NASA/IPAC Infrared Science Archive 
and the cross-match service provided by CDS, Strasbourg. 
We acknowledge the valuable suggestions from the anonymous referee.}
\bibliographystyle{mn2e}
\bibliography{lss_gac_dr2}

\begin{thebibliography}{65}
\expandafter\ifx\csname natexlab\endcsname\relax\def\natexlab#1{#1}\fi

\bibitem[{{Aihara} {et~al}\mbox{.}(2011){Aihara}, {Allende Prieto}, {An},
  {Anderson}, {Aubourg}, {Balbinot}, {Beers}, {Berlind}, {Bickerton},
  {Bizyaev}, {Blanton}, {Bochanski}, {Bolton}, {Bovy}, {Brandt}, {Brinkmann},
  {Brown}, {Brownstein}, {Busca}, {Campbell}, {Carr}, {Chen}, {Chiappini},
  {Comparat}, {Connolly}, {Cortes}, {Croft}, {Cuesta}, {da Costa}, {Davenport},
  {Dawson}, {Dhital}, {Ealet}, {Ebelke}, {Edmondson}, {Eisenstein},
  {Escoffier}, {Esposito}, {Evans}, {Fan}, {Femen{\'{\i}}a Castell{\'a}},
  {Font-Ribera}, {Frinchaboy}, {Ge}, {Gillespie}, {Gilmore}, {Gonz{\'a}lez
  Hern{\'a}ndez}, {Gott}, {Gould}, {Grebel}, {Gunn}, {Hamilton}, {Harding},
  {Harris}, {Hawley}, {Hearty}, {Ho}, {Hogg}, {Holtzman}, {Honscheid}, {Inada},
  {Ivans}, {Jiang}, {Johnson}, {Jordan}, {Jordan}, {Kazin}, {Kirkby}, {Klaene},
  {Knapp}, {Kneib}, {Kochanek}, {Koesterke}, {Kollmeier}, {Kron}, {Lampeitl},
  {Lang}, {Le Goff}, {Lee}, {Lin}, {Long}, {Loomis}, {Lucatello}, {Lundgren},
  {Lupton}, {Ma}, {MacDonald}, {Mahadevan}, {Maia}, {Makler}, {Malanushenko},
  {Malanushenko}, {Mandelbaum}, {Maraston}, {Margala}, {Masters}, {McBride},
  {McGehee}, {McGreer}, {M{\'e}nard}, {Miralda-Escud{\'e}}, {Morrison},
  {Mullally}, {Muna}, {Munn}, {Murayama}, {Myers}, {Naugle}, {Neto}, {Nguyen},
  {Nichol}, {O'Connell}, {Ogando}, {Olmstead}, {Oravetz}, {Padmanabhan},
  {Palanque-Delabrouille}, {Pan}, {Pandey}, {P{\^a}ris}, {Percival},
  {Petitjean}, {Pfaffenberger}, {Pforr}, {Phleps}, {Pichon}, {Pieri}, {Prada},
  {Price-Whelan}, {Raddick}, {Ramos}, {Reyl{\'e}}, {Rich}, {Richards}, {Rix},
  {Robin}, {Rocha-Pinto}, {Rockosi}, {Roe}, {Rollinde}, {Ross}, {Ross},
  {Rossetto}, {S{\'a}nchez}, {Sayres}, {Schlegel}, {Schlesinger}, {Schmidt},
  {Schneider}, {Sheldon}, {Shu}, {Simmerer}, {Simmons}, {Sivarani}, {Snedden},
  {Sobeck}, {Steinmetz}, {Strauss}, {Szalay}, {Tanaka}, {Thakar}, {Thomas},
  {Tinker}, {Tofflemire}, {Tojeiro}, {Tremonti}, {Vandenberg}, {Vargas
  Maga{\~n}a}, {Verde}, {Vogt}, {Wake}, {Wang}, {Weaver}, {Weinberg}, {White},
  {White}, {Yanny}, {Yasuda}, {Yeche}, \& {Zehavi}}]{Aihara+2011}
{Aihara} H. {et~al.}, 2011, \apjs, 193, 29

\bibitem[{{Anderson} \& {Francis}(2012)}]{Anderson+2012}
{Anderson} E., {Francis} C., 2012, Astronomy Letters, 38, 331

\bibitem[{{Carlin} {et~al}\mbox{.}(2013){Carlin}, {DeLaunay}, {Newberg},
  {Deng}, {Gole}, {Grabowski}, {Jin}, {Liu}, {Liu}, {Luo}, {Yuan}, {Zhang},
  {Zhao}, \& {Zhao}}]{Carlin+2013}
{Carlin} J.~L. {et~al.}, 2013, \apjl, 777, L5

\bibitem[{{Carlin} {et~al}\mbox{.}(2015){Carlin}, {Liu}, {Newberg}, {Beers},
  {Chen}, {Deng}, {Guhathakurta}, {Hou}, {Hou}, {L{\'e}pine}, {Li}, {Luo},
  {Smith}, {Wu}, {Yang}, {Yanny}, {Zhang}, \& {Zheng}}]{Carlin+2015}
{Carlin} J.~L. {et~al.}, 2015, \aj, 150, 4

\bibitem[{{Castelli} \& {Kurucz}(2004)}]{Castelli+2004}
{Castelli} F., {Kurucz} R.~L., 2004, ArXiv Astrophysics e-prints

\bibitem[{{Cenarro} {et~al}\mbox{.}(2007){Cenarro}, {Peletier},
  {S{\'a}nchez-Bl{\'a}zquez}, {Selam}, {Toloba}, {Cardiel},
  {Falc{\'o}n-Barroso}, {Gorgas}, {Jim{\'e}nez-Vicente}, \&
  {Vazdekis}}]{Cenarro+2007}
{Cenarro} A.~J. {et~al.}, 2007, \mnras, 374, 664

\bibitem[{{Chen} {et~al}\mbox{.}(2014){Chen}, {Liu}, {Yuan}, {Zhang},
  {Schultheis}, {Jiang}, {Huang}, {Xiang}, {Zhao}, {Yao}, \& {Lu}}]{Chen+2014}
{Chen} B.-Q. {et~al.}, 2014, \mnras, 443, 1192

\bibitem[{{Cui} {et~al}\mbox{.}(2012){Cui}, {Zhao}, {Chu}, {Li}, {Li}, {Zhang},
  {Su}, {Yao}, {Wang}, {Xing}, {Li}, {Zhu}, {Wang}, {Gu}, {Luo}, {Xu}, {Zhang},
  {Liu}, {Zhang}, {Yang}, {Cao}, {Chen}, {Chen}, {Chen}, {Chen}, {Chu}, {Feng},
  {Gong}, {Hou}, {Hu}, {Hu}, {Hu}, {Jia}, {Jiang}, {Jiang}, {Jiang}, {Jin},
  {Li}, {Li}, {Li}, {Liu}, {Liu}, {Lu}, {Mao}, {Men}, {Qi}, {Qi}, {Shi},
  {Tang}, {Tao}, {Wang}, {Wang}, {Wang}, {Wang}, {Wang}, {Wang}, {Wang},
  {Wang}, {Wang}, {Wang}, {Wang}, {Wang}, {Xu}, {Xu}, {Yang}, {Yu}, {Yuan},
  {Yuan}, {Zhai}, {Zhang}, {Zhang}, {Zhang}, {Zhao}, {Zhou}, {Zhou}, {Zhu}, \&
  {Zou}}]{Cui+2012}
{Cui} X.-Q. {et~al.}, 2012, Research in Astronomy and Astrophysics, 12, 1197

\bibitem[{{Deng} {et~al}\mbox{.}(2012){Deng}, {Newberg}, {Liu}, {Carlin},
  {Beers}, {Chen}, {Chen}, {Christlieb}, {Grillmair}, {Guhathakurta}, {Han},
  {Hou}, {Lee}, {L{\'e}pine}, {Li}, {Liu}, {Pan}, {Sellwood}, {Wang}, {Wang},
  {Yang}, {Yanny}, {Zhang}, {Zhang}, {Zheng}, \& {Zhu}}]{Deng+2012}
{Deng} L.-C. {et~al.}, 2012, Research in Astronomy and Astrophysics, 12, 735

\bibitem[{{Doi} {et~al}\mbox{.}(2010){Doi}, {Tanaka}, {Fukugita}, {Gunn},
  {Yasuda}, {Ivezi{\'c}}, {Brinkmann}, {de Haars}, {Kleinman}, {Krzesinski}, \&
  {French Leger}}]{Doi+2010}
{Doi} M. {et~al.}, 2010, \aj, 139, 1628

\bibitem[{{Dotter} {et~al}\mbox{.}(2008){Dotter}, {Chaboyer}, {Jevremovi{\'c}},
  {Kostov}, {Baron}, \& {Ferguson}}]{Dotter+2008}
{Dotter} A., {Chaboyer} B., {Jevremovi{\'c}} D., {Kostov} V., {Baron} E.,
  {Ferguson} J.~W., 2008, \apjs, 178, 89

\bibitem[{{Falc{\'o}n-Barroso} {et~al}\mbox{.}(2011){Falc{\'o}n-Barroso},
  {S{\'a}nchez-Bl{\'a}zquez}, {Vazdekis}, {Ricciardelli}, {Cardiel}, {Cenarro},
  {Gorgas}, \& {Peletier}}]{Falcon-Barroso+2011}
{Falc{\'o}n-Barroso} J., {S{\'a}nchez-Bl{\'a}zquez} P., {Vazdekis} A.,
  {Ricciardelli} E., {Cardiel} N., {Cenarro} A.~J., {Gorgas} J., {Peletier}
  R.~F., 2011, \aap, 532, A95

\bibitem[{{Gao} {et~al}\mbox{.}(2015){Gao}, {Zhang}, {Xiang}, {Huang}, {Liu},
  {Luo}, {Zhang}, {Wu}, {Zhang}, {Li}, \& {Du}}]{Gao+2015}
{Gao} H. {et~al.}, 2015, Research in Astronomy and Astrophysics, 15, 2204

\bibitem[{{Garc{\'{\i}}a P{\'e}rez} {et~al}\mbox{.}(2015){Garc{\'{\i}}a
  P{\'e}rez}, {Allende Prieto}, {Holtzman}, {Shetrone}, {M{\'e}sz{\'a}ros},
  {Bizyaev}, {Carrera}, {Cunha}, {Garc{\'{\i}}a-Hern{\'a}ndez}, {Johnson},
  {Majewski}, {Nidever}, {Schiavon}, {Shane}, {Smith}, {Sobeck}, {Troup},
  {Zamora}, {Bovy}, {Eisenstein}, {Feuillet}, {Frinchaboy}, {Hayden}, {Hearty},
  {Nguyen}, {O'Connell}, {Pinsonneault}, {Weinberg}, {Wilson}, \&
  {Zasowski}}]{Garcia_Perez+2015}
{Garc{\'{\i}}a P{\'e}rez} A.~E. {et~al.}, 2015, ArXiv e-prints

\bibitem[{{Gardner} \& {Flynn}(2010)}]{Gardner+2010}
{Gardner} E., {Flynn} C., 2010, \mnras, 405, 545

\bibitem[{{Gray}(1999)}]{Gray+1999}
{Gray} R.~O., 1999, {SPECTRUM: A stellar spectral synthesis program}.
  Astrophysics Source Code Library

\bibitem[{{Gunn} {et~al}\mbox{.}(1998){Gunn}, {Carr}, {Rockosi}, {Sekiguchi},
  {Berry}, {Elms}, {de Haas}, {Ivezi{\'c}}, {Knapp}, {Lupton}, {Pauls},
  {Simcoe}, {Hirsch}, {Sanford}, {Wang}, {York}, {Harris}, {Annis}, {Bartozek},
  {Boroski}, {Bakken}, {Haldeman}, {Kent}, {Holm}, {Holmgren}, {Petravick},
  {Prosapio}, {Rechenmacher}, {Doi}, {Fukugita}, {Shimasaku}, {Okada}, {Hull},
  {Siegmund}, {Mannery}, {Blouke}, {Heidtman}, {Schneider}, {Lucinio}, \&
  {Brinkman}}]{Gunn+1998}
{Gunn} J.~E. {et~al.}, 1998, \aj, 116, 3040

\bibitem[{{Ho} {et~al}\mbox{.}(2016){Ho}, {Ness}, {Hogg}, {Rix}, {Liu}, {Yang},
  {Zhang}, {Hou}, \& {Wang}}]{Ho+2016}
{Ho} A.~Y.~Q. {et~al.}, 2016, ArXiv e-prints

\bibitem[{{Holtzman} {et~al}\mbox{.}(2015){Holtzman}, {Shetrone}, {Johnson},
  {Allende Prieto}, {Anders}, {Andrews}, {Beers}, {Bizyaev}, {Blanton}, {Bovy},
  {Carrera}, {Chojnowski}, {Cunha}, {Eisenstein}, {Feuillet}, {Frinchaboy},
  {Galbraith-Frew}, {Garc{\'{\i}}a P{\'e}rez}, {Garc{\'{\i}}a-Hern{\'a}ndez},
  {Hasselquist}, {Hayden}, {Hearty}, {Ivans}, {Majewski}, {Martell},
  {Meszaros}, {Muna}, {Nidever}, {Nguyen}, {O Connell}, {Pan}, {Pinsonneault},
  {Robin}, {Schiavon}, {Shane}, {Sobeck}, {Smith}, {Troup}, {Weinberg},
  {Wilson}, {Wood-Vasey}, {Zamora}, \& {Zasowski}}]{Holtzman+2015}
{Holtzman} J.~A. {et~al.}, 2015, \aj, 150, 148

\bibitem[{{Huang} {et~al}\mbox{.}(2015{\natexlab{a}}){Huang}, {Liu}, {Yuan},
  {Xiang}, {Chen}, \& {Zhang}}]{Huang+2015b}
{Huang} Y., {Liu} X.-W., {Yuan} H.-B., {Xiang} M.-S., {Chen} B.-Q., {Zhang}
  H.-W., 2015{\natexlab{a}}, \mnras, 454, 2863

\bibitem[{{Huang} {et~al}\mbox{.}(2015{\natexlab{b}}){Huang}, {Liu}, {Yuan},
  {Xiang}, {Huo}, {Chen}, {Zhang}, \& {Hou}}]{Huang+2015c}
{Huang} Y., {Liu} X.-W., {Yuan} H.-B., {Xiang} M.-S., {Huo} Z.-Y., {Chen}
  B.-Q., {Zhang} Y., {Hou} Y.-H., 2015{\natexlab{b}}, \mnras, 449, 162

\bibitem[{{Huang} {et~al}\mbox{.}(2015{\natexlab{c}}){Huang}, {Liu}, {Zhang},
  {Yuan}, {Xiang}, {Chen}, {Ren}, {Sun}, {Wang}, {Zhang}, {Hou}, {Wang}, \&
  {Yang}}]{Huang+2015a}
{Huang} Y. {et~al.}, 2015{\natexlab{c}}, Research in Astronomy and
  Astrophysics, 15, 1240

\bibitem[{{Jester} {et~al}\mbox{.}(2005){Jester}, {Schneider}, {Richards},
  {Green}, {Schmidt}, {Hall}, {Strauss}, {Vanden Berk}, {Stoughton}, {Gunn},
  {Brinkmann}, {Kent}, {Smith}, {Tucker}, \& {Yanny}}]{Jester+2005}
{Jester} S. {et~al.}, 2005, \aj, 130, 873

\bibitem[{{Jofr{\'e}} {et~al}\mbox{.}(2015){Jofr{\'e}}, {Heiter}, {Soubiran},
  {Blanco-Cuaresma}, {Masseron}, {Nordlander}, {Chemin}, {Worley}, {Van Eck},
  {Hourihane}, {Gilmore}, {Adibekyan}, {Bergemann}, {Cantat-Gaudin},
  {Delgado-Mena}, {Gonz{\'a}lez Hern{\'a}ndez}, {Guiglion}, {Lardo}, {de
  Laverny}, {Lind}, {Magrini}, {Mikolaitis}, {Montes}, {Pancino},
  {Recio-Blanco}, {Sordo}, {Sousa}, {Tabernero}, \& {Vallenari}}]{Jofre+2015}
{Jofr{\'e}} P. {et~al.}, 2015, \aap, 582, A81

\bibitem[{{Lee} {et~al}\mbox{.}(2011){Lee}, {Beers}, {Allende Prieto}, {Lai},
  {Rockosi}, {Morrison}, {Johnson}, {An}, {Sivarani}, \& {Yanny}}]{Lee+2011b}
{Lee} Y.~S. {et~al.}, 2011, \aj, 141, 90

\bibitem[{{Lee} {et~al}\mbox{.}(2015){Lee}, {Beers}, {Carlin}, {Newberg},
  {Hou}, {Li}, {Luo}, {Wu}, {Yang}, {Zhang}, {Zhang}, \& {Zhang}}]{Lee+2015}
{Lee} Y.~S. {et~al.}, 2015, \aj, 150, 187

\bibitem[{{Li} {et~al}\mbox{.}(2016){Li}, {Han}, {Xiang}, {Shi}, {Zhao}, {Liu},
  {Zhang}, {Yuan}, {Ci}, {Zhang}, {Wang}, {Huang}, {Zhang}, {Hou}, {Wang}, \&
  {Cao}}]{Liji+2016}
{Li} J. {et~al.}, 2016, Research in Astronomy and Astrophysics, 16, 010

\bibitem[{{Lindegren} {et~al}\mbox{.}(2016){Lindegren}, {Lammers}, {Bastian},
  {Hern{\'a}ndez}, {Klioner}, {Hobbs}, {Bombrun}, {Michalik}, {Ramos-Lerate},
  {Butkevich}, {Comoretto}, {Joliet}, {Holl}, {Hutton}, {Parsons},
  {Steidelm{\"u}ller}, {Abbas}, {Altmann}, {Andrei}, {Anton}, {Bach},
  {Barache}, {Becciani}, {Berthier}, {Bianchi}, {Biermann}, {Bouquillon},
  {Bourda}, {Br{\"u}semeister}, {Bucciarelli}, {Busonero}, {Carlucci},
  {Casta{\~n}eda}, {Charlot}, {Clotet}, {Crosta}, {Davidson}, {de Felice},
  {Drimmel}, {Fabricius}, {Fienga}, {Figueras}, {Fraile}, {Gai}, {Garralda},
  {Geyer}, {Gonz{\'a}lez-Vidal}, {Guerra}, {Hambly}, {Hauser}, {Jordan},
  {Lattanzi}, {Lenhardt}, {Liao}, {L{\"o}ffler}, {McMillan}, {Mignard}, {Mora},
  {Morbidelli}, {Portell}, {Riva}, {Sarasso}, {Serraller}, {Siddiqui}, {Smart},
  {Spagna}, {Stampa}, {Steele}, {Taris}, {Torra}, {van Reeven}, {Vecchiato},
  {Zschocke}, {de Bruijne}, {Gracia}, {Raison}, {Lister}, {Marchant},
  {Messineo}, {Soffel}, {Osorio}, {de Torres}, \& {O'Mullane}}]{Lindegren2016}
{Lindegren} L. {et~al.}, 2016, ArXiv e-prints

\bibitem[{{Liu} {et~al}\mbox{.}(2015){Liu}, {Fang}, {Wu}, {Deng}, {Wang},
  {Wang}, {Fu}, {Hou}, {Li}, \& {Zhang}}]{Liuchao+2015}
{Liu} C. {et~al.}, 2015, \apj, 807, 4

\bibitem[{{Liu} {et~al}\mbox{.}(2014){Liu}, {Yuan}, {Huo}, {Deng}, {Hou},
  {Zhao}, {Zhao}, {Shi}, {Luo}, {Xiang}, {Zhang}, {Huang}, \&
  {Zhang}}]{Liu+2014}
{Liu} X.-W. {et~al.}, 2014, in IAU Symposium, Vol. 298, IAU Symposium,
  {Feltzing} S., {Zhao} G., {Walton} N.~A., {Whitelock} P., eds., pp. 310--321

\bibitem[{{Luo} {et~al}\mbox{.}(2012){Luo}, {Zhang}, {Zhao}, {Zhao}, {Cui},
  {Li}, {Chu}, {Shi}, {Wang}, {Zhang}, {Bai}, {Chen}, {Wang}, {Guo}, {Chen},
  {Du}, {Kong}, {Lei}, {Li}, {Song}, {Wu}, {Zhang}, {Zhou}, {Zuo}, {Du}, {He},
  {Hou}, {Dong}, {Li}, {Li}, {Li}, {Song}, {Tian}, {Wang}, {Wu}, {Yang},
  {Yuan}, {Cao}, {Chen}, {Chen}, {Chen}, {Chu}, {Feng}, {Gong}, {Gu}, {Hou},
  {Huo}, {Hu}, {Hu}, {Hu}, {Jia}, {Jiang}, {Jiang}, {Jiang}, {Jin}, {Li}, {Li},
  {Li}, {Li}, {Li}, {Liu}, {Liu}, {Liu}, {Lu}, {Lu}, {Luo}, {Mao}, {Men}, {Ni},
  {Qi}, {Qi}, {Shi}, {Su}, {Sun}, {Su}, {Tang}, {Tao}, {Tu}, {Wang}, {Wang},
  {Wang}, {Wang}, {Wang}, {Wang}, {Wang}, {Wang}, {Wang}, {Wang}, {Wang},
  {Wang}, {Wang}, {Wang}, {Wei}, {Xue}, {Xing}, {Xu}, {Xu}, {Xu}, {Yang},
  {Yang}, {Yao}, {Yu}, {Yuan}, {Zhai}, {Zhang}, {Zhang}, {Zhang}, {Zhang},
  {Zhang}, {Zhang}, {Zhao}, {Zhou}, {Zhu}, {Zhu}, \& {Zou}}]{Luo+2012}
{Luo} A.-L. {et~al.}, 2012, Research in Astronomy and Astrophysics, 12, 1243

\bibitem[{{Luo} {et~al}\mbox{.}(2015){Luo}, {Zhao}, {Zhao}, {Deng}, {Liu},
  {Jing}, {Wang}, {Zhang}, {Shi}, {Cui}, {Chu}, {Li}, {Bai}, {Wu}, {Cai},
  {Cao}, {Cao}, {Carlin}, {Chen}, {Chen}, {Chen}, {Chen}, {Chen}, {Chen},
  {Chen}, {Christlieb}, {Chu}, {Cui}, {Dong}, {Du}, {Fan}, {Feng}, {Fu}, {Gao},
  {Gong}, {Gu}, {Guo}, {Han}, {He}, {Hou}, {Hou}, {Hou}, {Hu}, {Hu}, {Hu},
  {Huo}, {Jia}, {Jiang}, {Jiang}, {Jiang}, {Jin}, {Kong}, {Kong}, {Lei}, {Li},
  {Li}, {Li}, {Li}, {Li}, {Li}, {Li}, {Li}, {Li}, {Li}, {Li}, {Li}, {Liang},
  {Lin}, {Liu}, {Liu}, {Liu}, {Liu}, {Lu}, {Luo}, {Mao}, {Newberg}, {Ni}, {Qi},
  {Qi}, {Shen}, {Shi}, {Song}, {Song}, {Su}, {Su}, {Tang}, {Tao}, {Tian},
  {Wang}, {Wang}, {Wang}, {Wang}, {Wang}, {Wang}, {Wang}, {Wang}, {Wang},
  {Wang}, {Wang}, {Wang}, {Wang}, {Wang}, {Wang}, {Wang}, {Wang}, {Wang},
  {Wang}, {Wang}, {Wei}, {Wei}, {Wu}, {Wu}, {Wu}, {Wu}, {Xing}, {Xu}, {Xu},
  {Xu}, {Yan}, {Yang}, {Yang}, {Yang}, {Yang}, {Yao}, {Yu}, {Yuan}, {Yuan},
  {Yuan}, {Yuan}, {Zhai}, {Zhang}, {Zhang}, {Zhang}, {Zhang}, {Zhang}, {Zhang},
  {Zhang}, {Zhang}, {Zhao}, {Zhou}, {Zhou}, {Zhu}, {Zhu}, {Zou}, \&
  {Zuo}}]{Luo+2015}
{Luo} A.-L. {et~al.}, 2015, Research in Astronomy and Astrophysics, 15, 1095

\bibitem[{{Majewski} {et~al}\mbox{.}(2010){Majewski}, {Wilson}, {Hearty},
  {Schiavon}, \& {Skrutskie}}]{Majewski+2010}
{Majewski} S.~R., {Wilson} J.~C., {Hearty} F., {Schiavon} R.~R., {Skrutskie}
  M.~F., 2010, in IAU Symposium, Vol. 265, IAU Symposium, {Cunha} K., {Spite}
  M., {Barbuy} B., eds., pp. 480--481

\bibitem[{{Martin} {et~al}\mbox{.}(2005){Martin}, {Fanson}, {Schiminovich},
  {Morrissey}, {Friedman}, {Barlow}, {Conrow}, {Grange}, {Jelinsky},
  {Milliard}, {Siegmund}, {Bianchi}, {Byun}, {Donas}, {Forster}, {Heckman},
  {Lee}, {Madore}, {Malina}, {Neff}, {Rich}, {Small}, {Surber}, {Szalay},
  {Welsh}, \& {Wyder}}]{Martin+2005}
{Martin} D.~C. {et~al.}, 2005, \apjl, 619, L1

\bibitem[{{Milligan}, {Cranton} \& {Skrutskie}(1996){Milligan}, {Cranton}, \&
  {Skrutskie}}]{Milligan+1996}
{Milligan} S., {Cranton} B.~W., {Skrutskie} M.~F., 1996, in \procspie, Vol.
  2863, Current Developments in Optical Design and Engineering VI, {Fischer}
  R.~E., {Smith} W.~J., eds., pp. 2--13

\bibitem[{{Munari} {et~al}\mbox{.}(2014){Munari}, {Henden}, {Frigo}, {Zwitter},
  {Bienaym{\'e}}, {Bland-Hawthorn}, {Boeche}, {Freeman}, {Gibson}, {Gilmore},
  {Grebel}, {Helmi}, {Kordopatis}, {Levine}, {Navarro}, {Parker}, {Reid},
  {Seabroke}, {Siebert}, {Siviero}, {Smith}, {Steinmetz}, {Templeton},
  {Terrell}, {Welch}, {Williams}, \& {Wyse}}]{Munari+2014}
{Munari} U. {et~al.}, 2014, \aj, 148, 81

\bibitem[{{Munari} {et~al}\mbox{.}(2005){Munari}, {Sordo}, {Castelli}, \&
  {Zwitter}}]{Munari+2005}
{Munari} U., {Sordo} R., {Castelli} F., {Zwitter} T., 2005, \aap, 442, 1127

\bibitem[{{Perryman} {et~al}\mbox{.}(2001){Perryman}, {de Boer}, {Gilmore},
  {H{\o}g}, {Lattanzi}, {Lindegren}, {Luri}, {Mignard}, {Pace}, \& {de
  Zeeuw}}]{Perryman+2001}
{Perryman} M.~A.~C. {et~al.}, 2001, \aap, 369, 339

\bibitem[{{Perryman} {et~al}\mbox{.}(1997){Perryman}, {Lindegren},
  {Kovalevsky}, {Hoeg}, {Bastian}, {Bernacca}, {Cr{\'e}z{\'e}}, {Donati},
  {Grenon}, {Grewing}, {van Leeuwen}, {van der Marel}, {Mignard}, {Murray}, {Le
  Poole}, {Schrijver}, {Turon}, {Arenou}, {Froeschl{\'e}}, \&
  {Petersen}}]{Perryman+1997}
{Perryman} M.~A.~C. {et~al.}, 1997, \aap, 323, L49

\bibitem[{{Rebassa-Mansergas} {et~al}\mbox{.}(2015){Rebassa-Mansergas}, {Liu},
  {Cojocaru}, {Yuan}, {Torres}, {Garc{\'{\i}}a-Berro}, {Xiang}, {Huang},
  {Koester}, {Hou}, {Li}, \& {Zhang}}]{Rebassa-Mansergas+2015}
{Rebassa-Mansergas} A. {et~al.}, 2015, \mnras, 450, 743

\bibitem[{{Ren} {et~al}\mbox{.}(2016){Ren}, {Liu}, {Xiang}, {Huang}, {Hekker},
  {Wang}, {Yuan}, {Rebassa-Mansergas}, {Chen}, {Sun}, {Zhang}, {Huo}, {Zhang},
  {Zhang}, {Hou}, \& {Wang}}]{Ren+2016}
{Ren} J.-J. {et~al.}, 2016, Research in Astronomy and Astrophysics, 16, 009

\bibitem[{{Roeser}, {Demleitner} \& {Schilbach}(2010){Roeser}, {Demleitner}, \&
  {Schilbach}}]{Roeser+2010}
{Roeser} S., {Demleitner} M., {Schilbach} E., 2010, \aj, 139, 2440

\bibitem[{{Rosenfield} {et~al}\mbox{.}(2016){Rosenfield}, {Marigo}, {Girardi},
  {Dalcanton}, {Bressan}, {Williams}, \& {Dolphin}}]{Rosenfield2016}
{Rosenfield} P., {Marigo} P., {Girardi} L., {Dalcanton} J.~J., {Bressan} A.,
  {Williams} B.~F., {Dolphin} A., 2016, \apj, 822, 73

\bibitem[{{S{\'a}nchez-Bl{\'a}zquez}
  {et~al}\mbox{.}(2006){S{\'a}nchez-Bl{\'a}zquez}, {Peletier},
  {Jim{\'e}nez-Vicente}, {Cardiel}, {Cenarro}, {Falc{\'o}n-Barroso}, {Gorgas},
  {Selam}, \& {Vazdekis}}]{Sanchez-Blazquez+2006}
{S{\'a}nchez-Bl{\'a}zquez} P. {et~al.}, 2006, \mnras, 371, 703

\bibitem[{{Schlegel}, {Finkbeiner} \& {Davis}(1998){Schlegel}, {Finkbeiner}, \&
  {Davis}}]{Schlegel+1998}
{Schlegel} D.~J., {Finkbeiner} D.~P., {Davis} M., 1998, \apj, 500, 525

\bibitem[{{Skrutskie} {et~al}\mbox{.}(2006){Skrutskie}, {Cutri}, {Stiening},
  {Weinberg}, {Schneider}, {Carpenter}, {Beichman}, {Capps}, {Chester},
  {Elias}, {Huchra}, {Liebert}, {Lonsdale}, {Monet}, {Price}, {Seitzer},
  {Jarrett}, {Kirkpatrick}, {Gizis}, {Howard}, {Evans}, {Fowler}, {Fullmer},
  {Hurt}, {Light}, {Kopan}, {Marsh}, {McCallon}, {Tam}, {Van Dyk}, \&
  {Wheelock}}]{Skrutskie+2006}
{Skrutskie} M.~F. {et~al.}, 2006, \aj, 131, 1163

\bibitem[{{Soubiran} {et~al}\mbox{.}(2010){Soubiran}, {Le Campion}, {Cayrel de
  Strobel}, \& {Caillo}}]{Soubiran+2010}
{Soubiran} C., {Le Campion} J.-F., {Cayrel de Strobel} G., {Caillo} A., 2010,
  \aap, 515, A111

\bibitem[{{Sun} {et~al}\mbox{.}(2015){Sun}, {Liu}, {Huang}, {Yuan}, {Xiang},
  {Zhang}, {Chen}, {Ren}, {Wang}, {Zhang}, {Hou}, {Wang}, \& {Yang}}]{Sun+2015}
{Sun} N.-C. {et~al.}, 2015, Research in Astronomy and Astrophysics, 15, 1342

\bibitem[{{Wang} {et~al}\mbox{.}(2016{\natexlab{a}}){Wang}, {Shi}, {Zhao},
  {Zhang}, {Huo}, {Zhang}, {Chen}, {Wu}, {Zhang}, \& {Hou}}]{WangJL+2016}
{Wang} J. {et~al.}, 2016{\natexlab{a}}, \mnras, 456, 672

\bibitem[{{Wang} {et~al}\mbox{.}(2016{\natexlab{b}}){Wang}, {Wang}, {Wu},
  {Zhao}, {Li}, {Luo}, {Liu}, {Zhang}, {Hou}, {Wang}, \& {Cao}}]{Wang+2016}
{Wang} L. {et~al.}, 2016{\natexlab{b}}, \aj, 152, 6

\bibitem[{{Wenger} {et~al}\mbox{.}(2000){Wenger}, {Ochsenbein}, {Egret},
  {Dubois}, {Bonnarel}, {Borde}, {Genova}, {Jasniewicz}, {Lalo{\"e}},
  {Lesteven}, \& {Monier}}]{Wenger+2000}
{Wenger} M. {et~al.}, 2000, \aaps, 143, 9

\bibitem[{{Wright} {et~al}\mbox{.}(2010){Wright}, {Eisenhardt}, {Mainzer},
  {Ressler}, {Cutri}, {Jarrett}, {Kirkpatrick}, {Padgett}, {McMillan},
  {Skrutskie}, {Stanford}, {Cohen}, {Walker}, {Mather}, {Leisawitz}, {Gautier},
  {McLean}, {Benford}, {Lonsdale}, {Blain}, {Mendez}, {Irace}, {Duval}, {Liu},
  {Royer}, {Heinrichsen}, {Howard}, {Shannon}, {Kendall}, {Walsh}, {Larsen},
  {Cardon}, {Schick}, {Schwalm}, {Abid}, {Fabinsky}, {Naes}, \&
  {Tsai}}]{Wright+2010}
{Wright} E.~L. {et~al.}, 2010, \aj, 140, 1868

\bibitem[{{Wu} {et~al}\mbox{.}(2014){Wu}, {Du}, {Luo}, {Zhao}, \&
  {Yuan}}]{Wu+2014}
{Wu} Y., {Du} B., {Luo} A., {Zhao} Y., {Yuan} H., 2014, in IAU Symposium, Vol.
  306, Statistical Challenges in 21st Century Cosmology, {Heavens} A., {Starck}
  J.-L., {Krone-Martins} A., eds., pp. 340--342

\bibitem[{{Wu} {et~al}\mbox{.}(2011){Wu}, {Luo}, {Li}, {Shi}, {Prugniel},
  {Liang}, {Zhao}, {Zhang}, {Bai}, {Wei}, {Dong}, {Zhang}, \& {Chen}}]{Wu+2011}
{Wu} Y. {et~al.}, 2011, Research in Astronomy and Astrophysics, 11, 924

\bibitem[{{Xiang} {et~al}\mbox{.}(2017){Xiang}, {Liu}, {Shi}, {Yuan}, {Huang},
  {Luo}, {Zhang}, {Zhao}, {Zhang}, {Ren}, {Chen}, {Wang}, {Li}, {Huo}, {Zhang},
  {Wang}, {Zhang}, {Hou}, \& {Wang}}]{Xiang+2016}
{Xiang} M.-S. {et~al.}, 2017, \mnras, 464, 3657

\bibitem[{{Xiang} {et~al}\mbox{.}(2015{\natexlab{a}}){Xiang}, {Liu}, {Yuan},
  {Huang}, {Huo}, {Zhang}, {Chen}, {Zhang}, {Sun}, {Wang}, {Zhao}, {Shi},
  {Luo}, {Li}, {Wu}, {Bai}, {Zhang}, {Hou}, {Yuan}, {Li}, \&
  {Wei}}]{Xiang+2015a}
{Xiang} M.~S. {et~al.}, 2015{\natexlab{a}}, \mnras, 448, 822

\bibitem[{{Xiang} {et~al}\mbox{.}(2015{\natexlab{b}}){Xiang}, {Liu}, {Yuan},
  {Huang}, {Wang}, {Ren}, {Chen}, {Sun}, {Zhang}, {Huo}, \&
  {Rebassa-Mansergas}}]{Xiang+2015c}
{Xiang} M.-S. {et~al.}, 2015{\natexlab{b}}, Research in Astronomy and
  Astrophysics, 15, 1209

\bibitem[{{Xiang} {et~al}\mbox{.}(2015{\natexlab{c}}){Xiang}, {Liu}, {Yuan},
  {Huo}, {Huang}, {Zheng}, {Zhang}, {Chen}, {Zhang}, {Sun}, {Wang}, {Zhao},
  {Shi}, {Luo}, {Li}, {Bai}, {Zhang}, {Hou}, {Yuan}, \& {Li}}]{Xiang+2015b}
{Xiang} M.~S. {et~al.}, 2015{\natexlab{c}}, \mnras, 448, 90

\bibitem[{{York} {et~al}\mbox{.}(2000){York}, {Adelman}, {Anderson},
  {Anderson}, {Annis}, {Bahcall}, {Bakken}, {Barkhouser}, {Bastian}, {Berman},
  {Boroski}, {Bracker}, {Briegel}, {Briggs}, {Brinkmann}, {Brunner}, {Burles},
  {Carey}, {Carr}, {Castander}, {Chen}, {Colestock}, {Connolly}, {Crocker},
  {Csabai}, {Czarapata}, {Davis}, {Doi}, {Dombeck}, {Eisenstein}, {Ellman},
  {Elms}, {Evans}, {Fan}, {Federwitz}, {Fiscelli}, {Friedman}, {Frieman},
  {Fukugita}, {Gillespie}, {Gunn}, {Gurbani}, {de Haas}, {Haldeman}, {Harris},
  {Hayes}, {Heckman}, {Hennessy}, {Hindsley}, {Holm}, {Holmgren}, {Huang},
  {Hull}, {Husby}, {Ichikawa}, {Ichikawa}, {Ivezi{\'c}}, {Kent}, {Kim},
  {Kinney}, {Klaene}, {Kleinman}, {Kleinman}, {Knapp}, {Korienek}, {Kron},
  {Kunszt}, {Lamb}, {Lee}, {Leger}, {Limmongkol}, {Lindenmeyer}, {Long},
  {Loomis}, {Loveday}, {Lucinio}, {Lupton}, {MacKinnon}, {Mannery}, {Mantsch},
  {Margon}, {McGehee}, {McKay}, {Meiksin}, {Merelli}, {Monet}, {Munn},
  {Narayanan}, {Nash}, {Neilsen}, {Neswold}, {Newberg}, {Nichol}, {Nicinski},
  {Nonino}, {Okada}, {Okamura}, {Ostriker}, {Owen}, {Pauls}, {Peoples},
  {Peterson}, {Petravick}, {Pier}, {Pope}, {Pordes}, {Prosapio},
  {Rechenmacher}, {Quinn}, {Richards}, {Richmond}, {Rivetta}, {Rockosi},
  {Ruthmansdorfer}, {Sandford}, {Schlegel}, {Schneider}, {Sekiguchi}, {Sergey},
  {Shimasaku}, {Siegmund}, {Smee}, {Smith}, {Snedden}, {Stone}, {Stoughton},
  {Strauss}, {Stubbs}, {SubbaRao}, {Szalay}, {Szapudi}, {Szokoly}, {Thakar},
  {Tremonti}, {Tucker}, {Uomoto}, {Vanden Berk}, {Vogeley}, {Waddell}, {Wang},
  {Watanabe}, {Weinberg}, {Yanny}, {Yasuda}, \& {SDSS
  Collaboration}}]{York+2000}
{York} D.~G. {et~al.}, 2000, \aj, 120, 1579

\bibitem[{{Yuan} {et~al}\mbox{.}(2015){Yuan}, {Liu}, {Huo}, {Xiang}, {Huang},
  {Chen}, {Zhang}, {Sun}, {Wang}, {Zhang}, {Zhao}, {Luo}, {Shi}, {Li}, {Yuan},
  {Dong}, {Li}, {Hou}, \& {Zhang}}]{Yuan+2015a}
{Yuan} H.-B. {et~al.}, 2015, \mnras, 448, 855

\bibitem[{{Yuan}, {Liu} \& {Xiang}(2013){Yuan}, {Liu}, \& {Xiang}}]{Yuan+2013}
{Yuan} H.~B., {Liu} X.~W., {Xiang} M.~S., 2013, \mnras, 430, 2188

\bibitem[{{Zacharias} {et~al}\mbox{.}(2013){Zacharias}, {Finch}, {Girard},
  {Henden}, {Bartlett}, {Monet}, \& {Zacharias}}]{Zacharias+2013}
{Zacharias} N., {Finch} C.~T., {Girard} T.~M., {Henden} A., {Bartlett} J.~L.,
  {Monet} D.~G., {Zacharias} M.~I., 2013, \aj, 145, 44

\bibitem[{{Zhang} {et~al}\mbox{.}(2013){Zhang}, {Liu}, {Yuan}, {Zhao}, {Yao},
  {Zhang}, \& {Xiang}}]{Zhang+2013}
{Zhang} H.-H., {Liu} X.-W., {Yuan} H.-B., {Zhao} H.-B., {Yao} J.-S., {Zhang}
  H.-W., {Xiang} M.-S., 2013, Research in Astronomy and Astrophysics, 13, 490

\bibitem[{{Zhang} {et~al}\mbox{.}(2014){Zhang}, {Liu}, {Yuan}, {Zhao}, {Yao},
  {Zhang}, {Xiang}, \& {Huang}}]{Zhang+2014}
{Zhang} H.-H., {Liu} X.-W., {Yuan} H.-B., {Zhao} H.-B., {Yao} J.-S., {Zhang}
  H.-W., {Xiang} M.-S., {Huang} Y., 2014, Research in Astronomy and
  Astrophysics, 14, 456

\bibitem[{{Zhao} {et~al}\mbox{.}(2012){Zhao}, {Zhao}, {Chu}, {Jing}, \&
  {Deng}}]{Zhao+2012}
{Zhao} G., {Zhao} Y.-H., {Chu} Y.-Q., {Jing} Y.-P., {Deng} L.-C., 2012,
  Research in Astronomy and Astrophysics, 12, 723

\end{thebibliography}

\begin{table*}
\begin{minipage}[]{160mm} \centering  
\caption{Description of the second release of LSS-GAC value-added catalogues.}
\label{}
\begin{tabular}{lll}
\hline
 Col.              &  Name                                           &  Description            \\ 
\hline
1   &  spec\_id                                                                  &  LAMOST unique spectral ID, in format of date-plateid-spectrographid-fibreid \\
2   &  date                  &  Date of observation  \\
3   & plate                  &  LAMOST plate ID, not necessarily unique  \\
4   & sp\_id                 &  LAMOST spectrograph ID, ranging from 1 to 16  \\
5   & fiber\_id              &  LAMOST fibre ID for a given spectrograph, ranging from 1 to 250  \\
6   & objid                  &  Object ID in the input catalogues  \\
7   & objtype              & Initial object type from the survey input catalogues  \\
8 & ra                       & Right ascension of J2000.0 ($\circ$)  \\
9 & dec                    & Declination of J2000.0 ($\circ$)  \\
10 & l                         & Galactic longitude ($\circ$)  \\
11 & b                        & Galactic latitude ($\circ$)  \\
12 & jd                       & Julian days at the middle time of the first exposure  \\
13 & moonphase       & Moon phase at the middle time of the first exposure \\
14 & monodis            & Moon distance at the middle time of the first exposure \\
15 & airmass             & Airmass at the middle time of the first exposure \\
16 & alt                      & Altitude (in degrees) of the telescope at the middle time of the first exposure \\
17 & az                      & Azimuth angle (in degrees) of the telescope at the middle time of the first exposure \\
18 & ha                      &  Hour angle (in degrees) of the telescope at the middle time of the first exposure \\
19 & badfiber           & Is it a bad fibre? (yes -- 1,  no -- 0)  \\
20 & satflag               & Is the spectrum saturated? (yes -- 1,  no -- 0) \\
21 & brightflag           & Are there bright stars (snr\_b > 300) in the nearby 5 fibres (yes -- 1, no -- 0) \\  
22 & brightsnr            & Largest snr\_b in the nearby 5 fibres \\
23 & object\_sky\_ratio & Object to sky flux ratio \\ 
24  & snr\_b                 & S/N(4650\AA) per pixel   \\
25  & snr\_r                  & S/N(7450\AA) per pixel  \\
26 & uqflag                & Uniqueness flag. If the target has been observed n times, then uqflag runs from 1 to n, with 1 denoting \\
&   &  the spectrum of the highest snr\_b  \\
27 & vr                       &  Radial velocity yielded by LSP3, after corrected for a systematic offset of $-$3.1 km\,s$^{-1}$ (km\,s$^{-1}$)  \\
28 & err\_vr                & Error of radial velocity yielded by LSP3 (km\,s$^{-1}$)   \\
29 & teff                     & Recommended $T_{\rm eff}$\,(K)   \\
30 & err\_teff             & Error of the recommended $T_{\rm eff}$ \\
31 & teff\_from          & Which method is the recommended $T_{\rm eff}$ adopted \\
32 & logg                     & Recommended log\,$g$~(cm\,s$^{-2}$)    \\
33 & err\_logg             & Error of the recommended log\,$g$ \\
34 & logg\_from          & Which method is the recommended log\,$g$ adopted \\
35 & mv                     & Recommended absolute magnitude in $V$-band  \\
36 & err\_mv              & Error of ${\rm M}_V$  \\
37 & mv\_from          & Which method is the recommended ${\rm M}_V$ adopted \\
38 & mk                     & Recommended absolute magnitude in $K_{\rm s}$-band  \\
39 & err\_mk             & Error of ${\rm M}_{K_{\rm s}}$  \\
40 & mk\_from          & Which method is the recommended ${\rm M}_{K_{\rm s}}$ adopted \\
41 & feh                     & Recommended [Fe/H]~(dex)   \\
42 & err\_feh             & Error of the recommended [Fe/H] \\
43 & feh\_from          & Which method is the recommended [Fe/H] adopted \\
44 & afe                    & Recommended [$\alpha$/Fe]~(dex)  \\
45 & err\_afe             & Error of the recommended [$\alpha$/Fe] \\
46 & afe\_from          & Which method is the recommended [$\alpha$/Fe] adopted \\
47 & am                    & $\alpha$-element to metal abundance ratio [$\alpha$/M] estimated by training the LAMOST-APOGEE stars (dex) \\
48 & err\_am             & Error of [$\alpha$/M]  \\
49 & mh                    & Metal abundance [M/H] estimated by training the LAMOST-APOGEE stars (dex) \\
50 & err\_mh             & Error of [M/H]  \\
51 & ch                     & Carbon abundance [C/H] estimated by training the LAMOST-APOGEE stars (dex) \\
52 & err\_mh             & Error of [C/H]  \\
53 & nh                     & Nitrogen abundance [N/H] estimated by training the LAMOST-APOGEE stars (dex) \\
54 & err\_nh             & Error of [N/H]  \\
55 & ebv                     & Recommended $E(B-V)$  \\
56 & dist                    &  Recommended distance (pc) \\
57 & err\_dist             &  Error of the recommended distance \\
58 & dist\_from          &   Which method is the recommended distance adopted \\
59 & vr\_peak\_corr\_coeff & Peak correlation coefficient for radial velocity estimation \\
60 & vr\_flag                & Flag to describe value of `vr\_peak\_corr\_coeff', $n$ means the vr\_peak\_corr\_coeff \\
& & is smaller than the median value of the sample stars with the same snr\_b by $n\times$MAD \\       
\hline
\end{tabular}
\end{minipage}
\end{table*}

\begin{table*}
\begin{minipage}[]{160mm} \centering 
\caption{\it -- continued}
\label{}
\begin{tabular}{lll}
\hline
 Col.              &  Name                                           &  Description            \\ 
\hline
61 & teff1                   & $T_{\rm eff}$ estimated with the weighted-mean algorithm (K)    \\
62 & err\_teff1             & Error of teff1          \\   
63 & teff2                   & $T_{\rm eff}$ estimated with the KPCA method (K)    \\
64 & err\_teff2             & Error of teff2         \\   
65 & logg1                   & Log\,$g$ estimated with the weighted-mean algorithm (cm\,s$^{-2}$)    \\
66 & err\_logg1             & Error of logg1          \\    
67 & logg2                   & Log\,$g$ estimated with the KPCA method using MILES stars as training set (cm\,s$^{-2}$)    \\
68 & err\_logg2             & Error of logg2          \\    
69 & logg3                   & Log\,$g$ estimated with the KPCA method using LAMOST-$Kepler$ stars as training set (cm\,s$^{-2}$)    \\
70 & err\_logg3             & Error of logg3          \\    
71 & mv1                     & ${\rm M}_V$ estimated with the KPCA method taking LAMOST-Hipparcos stars as training set using 300 PCs (mag) \\
72 & err\_mv1              & Error of mv1  \\
73 & mv2                     & ${\rm M}_V$ estimated with the KPCA method taking LAMOST-Hipparcos stars as training set using 100 PCs (mag) \\
74 & err\_mv2              & Error of mv2  \\
75 & mk1                     & ${\rm M}_{K_{\rm s}}$ estimated with the KPCA method taking LAMOST-Hipparcos stars as training set using 300 PCs (mag) \\
76 & err\_mk1             & Error of mk1  \\
77 & mk2                     & ${\rm M}_{K_{\rm s}}$ estimated with the KPCA method taking LAMOST-Hipparcos stars as training set using 100 PCs (mag) \\
78 & err\_mk2             & Error of mk2  \\
79 & feh1                     & [Fe/H] estimated with the LSP3 weighted mean algorithm (dex)    \\
80 & err\_feh1               & Error of feh1          \\    
81 & feh2                     & [Fe/H] estimated with the LSP3 KPCA method by training the MILES stars (dex)    \\
82 & err\_feh2               & Error of feh2          \\    
83 & feh3                     & [Fe/H] estimated with the LSP3 KPCA method by training the LAMOST-APOGEE stars (dex)    \\
84 & err\_feh3               & Error of feh3          \\ 
85 & afe1                     & [$\alpha$/Fe] estimated by template matching with synthetic spectra using spectral segments \\
      &                             &  of 3900 -- 3980\,{\AA}, 4400 -- 4600\,{\AA} and 5000 -- 5300\,{\AA} (dex)    \\
86 & err\_afe1               & Error of afe1          \\    
87 & afe2                     & [$\alpha$/Fe] estimated by template matching with synthetic spectra using spectral segments \\
      &                            &  of 4400 -- 4600\,{\AA} and 5000 -- 5300\,{\AA} (dex)     \\
88 & err\_afe2               & Error of afe2          \\ 
89 & afe3                     & [$\alpha$/Fe] estimated with the LSP3 KPCA method using the LAMOST-APOGEE training set (dex)    \\
90 & err\_afe3               & Error of afe3          \\    
91 & besttemp\_chi2     & Name of the best-matching template star with $\chi^2$ algorithm \\
92 & besttemp\_corr     & Name of the best-matching template star with correlation algorithm \\
93 & typeflag\_chi2       & Stellar types of the best-matching template star with $\chi^2$ algorithm\\
94 & typeflag\_corr       & Stellar types of the best-matching template star with correlation algorithm\\
95 & min\_chi2              & Minimum $\chi^2$ yielded by LSP3 from template matching with the MILES library \\
96 & chi2\_flag              & The `min\_chi2' is smaller than the median value of the sample stars in the same SNR and \\
& & atmospheric parameter bin by $n\times$MAD \\ 
97 & dg\_miles\_tm       & Maximum value of the kernel function for estimation of $T_{\rm eff}$ and [Fe/H] with KPCA method \\
& & via training the MILES stars \\
98 & dg\_miles\_g         & Maximum value of the kernel function for estimation of log\,$g$ with KPCA method via training \\
& & the MILES stars \\
99 & dg\_lm\_kep          & Maximum value of the kernel function for estimation of log\,$g$ with KPCA method via training \\
& & the LAMOST-$Kepler$ stars \\
100 & dg\_lm\_hip          & Maximum value of the kernel function for estimation of absolute magnitudes with KPCA method \\
& & via training the LAMOST-Hipparcos stars \\
101 & dg\_lm\_apo          & Maximum value of the kernel function for estimation of abundances with KPCA method via training \\
& & the LAMOST-APOGEE stars \\
102 & fuv                     & GALEX FUV band magnitude (mag)   \\
103 & err\_fuv               & Error of fuv  \\
104 & nuv                    & GALEX NUV band magnitude (mag)  \\
105 & err\_nuv              & Error of nuv  \\
106 & g                        & XSTPS-GAC g-band magnitude (mag)  \\
107 & err\_g                  & Error of g  \\
108 & r                         & XSTPS-GAC r-band magnitude (mag)  \\
109 & err\_r                   & Error of r  \\
110 & i                         & XSTPS-GAC i-band magnitude (mag)  \\
111 & err\_i                   & Error of i  \\
112 & J                         & 2MASS J-band magnitude (mag)  \\
113 & err\_J                   & Error of J  \\
114 & H                        & 2MASS H-band magnitude (mag)  \\
115 & err\_H                 &Error of H  \\
116 & Ks                        & 2MASS Ks-band magnitude (mag)  \\
117 & err\_Ks                &Error of Ks  \\
\hline
\end{tabular}
\end{minipage}
\end{table*}

\begin{table*}
\begin{minipage}[]{160mm} \centering 
\caption{\it -- continued}
\label{}
\begin{tabular}{lll}
\hline
 Col.              &  Name                                           &  Description            \\ 
\hline
118 & w1                     & WISE W1-band magnitude (mag)  \\
119 & err\_w1             &Error of w1  \\
120 & w2                     & WISE W2-band magnitude (mag)  \\
121 & err\_w2              &Error of w2  \\
122 & w3                     & WISE W3-band magnitude (mag)  \\
123 & err\_w3              &Error of w3  \\
124 & w4                     & WISE W4-band magnitude (mag)  \\
125 & err\_w4              &Error of w4  \\
126 & ph\_qual\_2mass &2MASS photometric quality flag  \\
127 & ph\_qual\_wise &WISE photometric quality flag  \\
128 & var\_flag\_wise &WISE variability flag  \\
129 & ext\_flag\_wise &WISE extended source flag  \\
130 & cc\_flag\_wise &WISE contamination and confusion flag  \\
131 & ebv\_sfd &  $E(B-V)$ from the SFD extinction map (mag)  \\
132 & ebv\_sp &  $E(B-V)$ derived from the star pair method (mag)  \\
133 & err\_ebv\_sp & Error of ebv\_sp \\
134 & ebv\_mod &  $E(B-V)$ derived by comparing the observed and synthetic model atmosphere colours (mag)  \\
135 & ebv\_phot &  $E(B-V)$ derived by fitting multiband photometry to the empirical stellar loci (Chen et al. 2014) (mag)  \\
136 & dist\_em1 &  Distance derived using the empirical relations of absolute magnitudes as a function of stellar atmospheric \\ 
     & & parameters as yielded by the MILES stars using the recommended atmospheric parameters\\
137 & err\_dist\_em1  & Error of  dist\_em1   \\   
138 & dist\_em2 &  Distance derived using the empirical relations of absolute magnitudes as a function of stellar atmospheric \\ 
     & & parameters as yielded by the MILES stars using the weighted-mean atmospheric parameters\\
139 & err\_dist\_em2  & Error of  dist\_em2   \\   
140 & dist\_sp1 &  Distance inferred from absolute magnitudes estimated with KPCA method using 300 PCs (pc)  \\
141 & err\_dist\_sp1 & Error of dist\_sp1 \\
142 & dist\_sp2 &  Distance inferred from absolute magnitudes estimated with KPCA method using 100 PCs (pc)  \\
143 & err\_dist\_sp2 & Error of dist\_sp2 \\
144 & dist\_phot &  Distance derived from the photometric parallax method (Chen et al. 2014) (pc)  \\
145 & err\_dist\_phot & Error of dist\_phot \\
146 & pmra\_ppmxl &  Proper motion in RA from the PPMXL catalogues (mas yr$^{-1}$)  \\
147 & epmra\_ppmxl &   Error of pmra\_ppmxl  \\
148 & pmdec\_ppmxl &  Proper motion in Dec. from the PPMXL catalogues (mas yr$^{-1}$)  \\
149 & epmdec\_ppmxl & Error of pmdec\_ppmxl \\
150 & pmra\_ucac4 &  Proper motion in RA from the UCAC4 catalogues (mas yr$^{-1}$)  \\
151 & epmra\_ucac4 &  Error of pmra\_ucac4  \\
152 & pmdec\_ucac4 &  Proper motion in Dec. from the UCAC4 catalogues (mas yr$^{-1}$)  \\
153 & epmdec\_ucac4 &  Error of pmdec\_ucac4  \\
154 & pmra\_xuyi2mass &  Proper motion in RA derived by comparing the XSTPS-GAC and 2MASS positions (mas yr$^{-1}$)  \\
155 & pmdec\_xuyi2mass &  Proper motion in Dec. derived by comparing the XSTPS-GAC and 2MASS positions (mas yr$^{-1}$)  \\
156 & x &       x coordinate in a Galactocentric Cartesian reference system, positive towards the Galactic Centre (kpc)  \\
157 & err\_x   & Error of x \\
158 & y &       y coordinate in a Galactocentric Cartesian reference system, positive in the direction of disk rotation (kpc)  \\
159 & err\_y   & Error of y \\
160 & z &       z coordinate in a Galactocentric Cartesian reference system, positive towards the North Galactic Pole (kpc)  \\
161 & err\_z   & Error of z \\
162 & u[2] &   Galactic space velocities in the x direction computed from dist\_em and the corrected PPMXL and UCAC4 \\
     & & proper motions, respectively, positive towards the Galactic Centre (km\,s$^{-1}$)  \\
163 & err\_u[2]  & Error of u[2] \\
164 & v[2]  &  Galactic space velocities in the y direction computed from dist\_em and the corrected PPMXL and UCAC4 \\
     & & proper motions, respectively, positive in the direction of Galactic rotation (km\,s$^{-1}$)  \\
165 & err\_v[2]  & Error of v[2] \\
166 & w[2]  & Galactic space velocities in the z direction computed from dist\_em and the corrected PPMXL and UCAC4 \\
     & & proper motions, respectively, positive towards the North Galactic Pole (km\,s$^{-1}$)  \\
\hline
\end{tabular}
\end{minipage}
\end{table*}

\begin{table*}
\begin{minipage}[]{160mm} \centering 
\caption{\it -- continued}
\label{}
\begin{tabular}{lll}
\hline
 Col.              &  Name                                           &  Description            \\ 
\hline
167 & err\_w[2]  & Error of w[2] \\
168 & v\_r[2] &  Galactic space velocities in the r direction in a Galactocentric cylindrical polar coordinate system, \\
     & & computed from dist\_em and the corrected PPMXL and UCAC4 proper motions, respectively (km\,s$^{-1}$)  \\
169 & err\_v\_r[2]  & Error of v\_r[2] \\     
170 & v\_phi[2] &  Galactic space velocities in the $\Phi$ direction in a Galactocentric cylindrical polar coordinate system, \\
     & & computed from dist\_em and the corrected PPMXL and UCAC4 proper motions, respectively, \\
     & & positive in the direction of counter disk rotation (km\,s$^{-1}$)  \\
171 & err\_v\_phi[2]  & Error of v\_phi[2] \\     
172 & v\_z[2] &  Galactic space velocities in the z direction in a Galactocentric cylindrical polar coordinate system, \\
     & & computed from dist\_em and the corrected PPMXL and UCAC4 proper motions, respectively, \\
     & & positive towards the North Galactic Pole (km\,s$^{-1}$)  \\
173 & err\_v\_z[2]  & Error of v\_z[2] \\
174 & e[2] &   Orbital eccentricities computed from dist\_em and the corrected PPMXL and UCAC4 proper motions, respectively \\
175 & Rapo[2] &   Maximum Galactic radii reached by the orbits computed with PPMXL and UCAC4 proper motions, respectively (kpc)  \\
176 & Rperi[2] &   Minimum Galactic radii reached by the orbits computed with PPMXL and UCAC4 proper motions, respectively (kpc)  \\
177 & zmin[2]   & Minimum height reached by the orbits computed with PPMXL and UCAC4 proper motions, respectively (kpc)  \\
178 & zmax[2]  &  Maximum height reached by the orbits computed with PPMXL and UCAC4 proper motions, respectively (kpc)  \\   
\hline
\end{tabular}
\end{minipage}
\end{table*}

\label{lastpage}

\end{document}